\documentclass[twocolumn, times]{aastex62}
\usepackage[T1]{fontenc}
\usepackage{bm}
\newcommand{\wise}{\textit{WISE}}
\newcommand{\spitzer}{\textit{Spitzer}}

\newcommand{\mstar}{$M_\star$}

\newcommand{\ssfr}{$\left< \mathrm{sSFR} \right>_8$} 
\newcommand{\radcut}{$r = 13 \, \mathrm{kpc}$}
\newcommand{\av}{$A_V$}
\newcommand{\medianav}{$\tilde A_V$}
\newcommand{\meanav}{$\left< A_V \right>$}
\newcommand{\fred}{$f_{red}$}
\newcommand{\dav}{$\mathrm{d}A_V$}
\newcommand{\ml}{$M_\star/L$}
\newcommand{\logml}{$\log (M_\star/L)$}
\newcommand{\gi}{$g-i$}
\newcommand{\wcolor}{$\mathrm{W}1-\mathrm{W}2$}
\newcommand{\mli}{$M_\star/L_i$}
\newcommand{\logmli}{$\log (M_\star/L_i)$}
\newcommand{\mlw}{$M_\star/L_{\mathrm{W}1}$}
\newcommand{\logmlw}{$\log (M_\star/L_{\mathrm{W}1})$}
\newcommand{\mformedcmd}{$M^\mathrm{CMD}_{\star, \mathrm{formed}}$}
\newcommand{\mcmd}{$M^\mathrm{CMD}_\star$}
\newcommand{\mlcmd}{$M^\mathrm{CMD}_\star/L_\mathrm{obs}$}
\newcommand{\logmlcmd}{$\log (M^\mathrm{CMD}_\star/L_\mathrm{obs})$}
\newcommand{\mlicmd}{$M^\mathrm{CMD}_\star/L_{i, \mathrm{obs}}$}
\newcommand{\logmlicmd}{$\log (M^\mathrm{CMD}_\star/L_{i, \mathrm{obs}})$}
\newcommand{\mlwcmd}{$M^\mathrm{CMD}_\star/L_{\mathrm{W}1, \mathrm{obs}}$}
\newcommand{\logmlwcmd}{$\log (M^\mathrm{CMD}_\star/L_{\mathrm{W}1, \mathrm{obs}})$}
\defcitealias{williams17}{W17}
\defcitealias{lewis15}{L15}
\defcitealias{dalcanton15}{D15}

\newcommand{\uw}{University of Washington Astronomy Department, Box 351580, Seattle, WA 98195-1580, USA}
\newcommand{\rutgers}{Rutgers University, Department of Physics and Astronomy, 136 Frelinghuysen Road, Piscataway, NJ 08854, USA}
\newcommand{\mich}{Department of Astronomy, University of Michigan, 1085 S. University Avenue, Ann Arbor, MI 48109-1107, USA}
\newcommand{\raytheon}{Raytheon, 1151 E. Hermans Road, Tucson, AZ 85706, USA}
\newcommand{\stsci}{Space Telescope Science Institute, 3700 San Martin Drive, Baltimore, MD 21218, USA}
\shorttitle{Stellar Mass-to-Light Ratios in M31}
\shortauthors{Telford et al.}

\begin{document}
\correspondingauthor{Grace Telford}
\email{grace.telford@rutgers.edu}

\title{Mass-to-Light Ratios of Spatially Resolved Stellar Populations in M31}

\author[0000-0003-4122-7749]{O. Grace Telford}
\affiliation{\uw}
\affiliation{\rutgers}
\author[0000-0002-1264-2006]{Julianne J. Dalcanton}
\affiliation{\uw}
\author[0000-0002-7502-0597]{Benjamin F. Williams}
\affiliation{\uw}
\author[0000-0002-5564-9873]{Eric F. Bell}
\affiliation{\mich}
\author[0000-0001-8416-4093]{Andrew E. Dolphin}
\affiliation{\raytheon}
\author[0000-0001-7531-9815]{Meredith J. Durbin}
\affiliation{\uw}
\author[0000-0003-1680-1884]{Yumi Choi}
\affiliation{\stsci}

\begin{abstract}

A galaxy's stellar mass-to-light ratio (\ml{}) is a useful tool for
converting luminosity to stellar mass (\mstar{}). However, the
practical utility of \ml{} inferred from stellar population synthesis
(SPS) models is limited by mismatches between the real and assumed
models for star formation history (SFH) and dust geometry, both of
which vary within galaxies. Here, we measure spatial variations in
\ml{} and their dependence on color, star formation history,
and dust across the disk of M31, using a map of \mcmd{}
derived from color-magnitude diagrams of resolved stars in the
Panchromatic Hubble Andromeda Treasury (PHAT) survey. First, we find
comparable scatter in \ml{} for the optical and mid-IR, contrary to
the common idea that \ml{} is less variable in the
IR. Second, we confirm that \ml{} is correlated with color for both the
optical and mid-IR and report color vs. \ml{} relations (CMLRs) in M31 for filters used in the Sloan Digital Sky Survey (SDSS) and \textit{Widefield Infrared Survey Explorer} (\wise{}). Third, we show that the CMLR residuals correlate with
recent SFH, such that quiescent regions are offset to higher \ml{}
than star-forming regions at a fixed color. The mid-IR CMLR, however,
is not linear due to the high scatter of \ml{} in star-forming
regions. Finally, we find a flatter optical CMLR than any SPS-based
CMLRs in the literature. We show this is an effect of dust geometry,
which is typically neglected but should be
accounted for when using optical data to map \mstar{}.

\end{abstract}



\section{Introduction\label{sec:intro}}

\subsection{Stellar Mass Inference Techniques and Challenges}\label{sec:mstar_overview}

Stellar mass (\mstar{}) is a key galaxy property, essential to our  understanding of how galaxies assemble and evolve. Many scaling relations used to calibrate galaxy formation models depend on \mstar{}: e.g., the star-forming main sequence \citep{brinchmann04, speagle14}, the stellar mass-halo mass relation \citep{behroozi10, moster10}, and the mass-metallicity relation \citep{tremonti04}. Galaxy formation models are commonly tuned to reproduce these observed relationships \citep{somerville15}, making the implicit assumption that the ``observed'' galaxy properties that are used to construct them have been inferred accurately.

Stellar population synthesis (SPS) models are the most common tool used to infer the stellar mass-to-light ratio (\ml{}) from the light emitted by galaxies \citep{tinsley80, walcher11, conroy13}. The many available SPS codes all combine models of stellar evolution with a stellar spectral library and assumed parameterization of the star formation history (SFH) to predict the total light output by stars. This stellar emission is then attenuated using a simple dust model, typically assuming a uniform foreground screen, and sometimes allowing extra extinction toward young stellar populations. Though these models are powerful and have enabled rapid progress in our understanding of galaxy evolution, they are limited by necessary assumptions and simplifications. The possible biases in \mstar{} inferred from SPS models are a popular topic in the literature \citep[e.g.,][]{kannappan07, pforr12, roediger15}, underscoring the importance of accurate \mstar{} measurements to many aspects of galaxy science.

There remains inconsistency at the factor of 2 level among various \mstar{} inference techniques: using SPS-based color vs. \ml{} relations (CMLRs), fitting SPS models to optical spectra or ultraviolet (UV) through infrared (IR) spectral energy distributions (SEDs), and dynamical modeling \citep[e.g.,][]{de-jong07, de-lucia14, mcgaugh15}. Every method of inferring \mstar{} is subject to systematic uncertainty, so the task at hand is to identify which systematics can bring \mstar{} measurements from different codes and data into agreement. 

In the optical, many CMLR predictions exist in the literature, each fit to libraries of SPS models generated with different codes and assumptions. The priors inherent in constructing SPS model libraries, particularly the imposed form of the SFH and treatment of dust, affect the slope and normalization of the best-fit CMLRs. These relations all have different slopes and normalizations, and it is not clear which most accurately captures the behavior of real galaxies. 

We are primarily interested in testing the performance of SPS-predicted CMLRs, for two key reasons. First, predicted CMLRs are a convenient tool for comparing different SPS codes. The choices and priors used to construct the various model libraries to which CMLRs are fit are essentially summarized by the predicted relationship between color and \ml{}. By checking the performance of predicted CMLRs, we are implicitly checking the performance of the SPS models themselves. Second, the ability to infer \mstar{} robustly from a single color is extremely valuable for maximizing the potential of large photometric surveys. Though more sophisticated \mstar{} inference techniques exist (e.g., Bayesian SED fitting with flexible SFHs; \citealt{leja17}), they require expensive multi-wavelength observations, and this will always limit the size of galaxy samples to which those techniques can be applied.

In the mid-IR, $3.6 \, \mu$m emission has long been used as a \mstar{} tracer. However, emission from hot dust and young stellar populations can strongly affect mid-IR \ml{} and colors, making the application of a constant \ml{} at $3.4-3.6 \, \mu$m inappropriate for galaxies (and regions within them) that are not dominated by old stellar populations \citep[e.g.,][]{querejeta15}. Furthermore, SPS models are less well-understood in this wavelength regime than in the optical due to different approaches to modeling luminous stellar evolutionary phases (e.g., thermally pulsing asymptotic giant branch (TP-AGB), red supergiants (RSGs), and red He-burning (RHeB) stars). Due to both of these issues, SPS models cannot at present be reliably used to calibrate CMLRs appropriate for star-forming galaxies in the mid-IR. These issues are of pressing concern, as an improved understanding of how infrared light traces \mstar{} will be necessary to interpret future observations (e.g., from the \textit{James Webb Space Telescope}).

\subsection{This Work: Combining CMD-Based \mstar{} and Observed Surface Brightness in M31}\label{sec:this_work}

M31 is the closest massive spiral within which the \textit{Hubble Space Telescope} (\textit{HST}) can resolve individual stars. The Panchromatic Hubble Andromeda Treasury survey (PHAT; \citealt{dalcanton12}) obtained high quality \textit{HST} imaging across 1/3 of the galaxy, covering the northern disk out to $\sim 20 \,\mathrm{kpc}$. PHAT obtained UV through near-IR (NIR) photometry for over 100 million stars \citep{williams14}, enabling spatially resolved measurements of SFHs and dust properties via modeling the distribution of stars in color-magnitude diagrams (CMDs). 

\citet{williams17} \citepalias[hereafter][]{williams17} measured the ancient SFH of M31's disk from optical and NIR CMDs, which can then be integrated with time to produce a map of the total stellar mass formed (\mformedcmd{}) within the PHAT footprint. \textit{This map of \mformedcmd{} was inferred with no constraints from integrated light, and is therefore an independent measurement from what would be obtained by fitting SPS models to the observed SED across the disk of M31} \citep[for an example of the latter strategy, see][]{sick15}.

Several imaging surveys have also mapped M31 in optical, near-IR, and mid-IR filters.  Surface brightness maps covering the PHAT footprint have been produced by the Sloan Digital Sky Survey \citep[SDSS;][]{york00}, the Two Micron All-Sky Survey \citep[2MASS;][]{skrutskie06}, the \textit{Spitzer Space Telescope} \citep{werner04}, and the \textit{Widefield Infrared Survey Explorer} \citep[\wise{};][]{wright10}. In recognition of M31's value in calibrating analyses of more distant galaxies, higher-quality optical-NIR surface photometry was recently obtained by the ANDRomeda Optical and Infrared Disk Survey \citep[ANDROIDS; ][]{sick14}. This wealth of available data for M31 make this galaxy a unique target in which to test the standard SPS-based methods for measuring a galaxy's \mstar{} from its observed brightness and colors.

Our main goal in this work is to map \ml{} ratios in M31 using a technique that is complementary to SPS modeling of integrated light. We combine the \mcmd{}, i.e., the stellar mass calculated from SFHs inferred by modeling resolved stellar populations, with the observed surface brightness of M31 to construct \mlcmd{}. We use this notation to signify that our measurements in M31 are fundamentally different from the \ml{} predictions of SPS models.

The CMD-based SFHs within the PHAT footprint in M31 were determined in $83''\times83''$ regions \citepalias{williams17}, which we call ``SFH pixels'' throughout the text. SFH pixels have a deprojected physical size of $0.3\times1.4$ kpc, for a total area of 0.42 kpc$^2$. We must therefore match the spatial resolution of the observed surface brightness maps to the SFH pixel size to calculate \mlcmd{}. We also measure the observed colors in matched areas and construct CMLRs in M31 (specifically, linear relations between \mlcmd{} and observed colors), which we then compare against other CMLRs in the literature.

We characterize the M31 CMLRs and compare the slopes of these best-fit relations for M31 to other CMLRs reported in the literature. In the optical, we assess whether the recent SFH or dust attenuation may bias the \mstar{} inferred from integrated light using SPS-based CMLRs that adopt simple dust and SFH parameterizations. In the mid-IR, we explore the impact of recent SFH and dust emission on $3.4 \, \mu$m \mlcmd{} and assess whether \wcolor{} color is a useful tool for improving \mstar{} estimates for galaxies with ongoing, low-level star formation.
 
This paper is organized as follows. Section~\ref{sec:phat_data_products} describes the spatially resolved SFHs and dust maps from modeling PHAT CMDs and our calculation of \mcmd{}. Section~\ref{sec:obs} presents the observed surface brightness and color maps matched to the resolution of the $83''\times83''$ SFH pixels, and maps of the \mlcmd{} in the optical and mid-IR.  In Section~\ref{sec:cmlrs}, we characterize the distribution of SFH pixels in color-\ml{} space and compare the best-fit CMLRs in M31 to those previously reported in the literature. In Section~\ref{sec:cmlr_scatter}, we analyze the impact of recent SFH and dust content and geometry on the slope, normalization, and scatter in the M31 CMLRs. Finally, we discuss the implications of this study for best practices when estimating \mstar{} in Section~\ref{sec:discussion} and summarize our key results in Section~\ref{sec:conclusions}. Finally, we present CMLRs fit to the observed colors and \mlcmd{} in M31 for other SDSS filter combinations in Appendix~\ref{sec:appendix}. We assume a distance of 785 kpc to M31 \citep{mcconnachie05}. All \ml{} are in Solar units, i.e., $(M_\star/M_\odot) / (L_\mathrm{filter}/L_{\mathrm{filter, }\odot})$, and all magnitudes are in the AB system. 


\section{PHAT Data Products: Spatially Resolved SFHs, Stellar Mass, and Dust}
\label{sec:phat_data_products} 

Here, we describe the data products from the PHAT survey \citep{dalcanton12} used in this work: ancient SFHs from \citetalias{williams17} used to determine \mcmd{}, recent SFHs from \citet{lewis15} \citepalias[hereafter][]{lewis15} used to determine the average star formation rate (SFR) over the past 100 Myr, and dust maps from \citet{dalcanton15} \citepalias[hereafter][]{dalcanton15} used to constrain the extinction and dust geometry. All of these were inferred from modeling optical and/or NIR CMDs constructed from the PHAT resolved-star photometry \citep{williams14}. We describe our method for mapping the present-day \mcmd{} distribution using the \citetalias{williams17} ancient SFHs; this map is combined with observed surface brightness maps to calculate \mlcmd{} in Section~\ref{sec:ml_cmd}. The recent SFHs and dust properties are used to analyze the sensitivity of \mlcmd{} to dust geometry and ongoing star formation in Section~\ref{sec:cmlr_scatter}.

\subsection{CMD-Based \mstar{} from Ancient SFHs}
\label{sec:stellar_mass}

\subsubsection{Ancient SFHs}
\label{sec:ancient_sfhs}

\begin{figure*}[!ht]
\minipage{0.23\textwidth}
  \includegraphics[width=\linewidth]{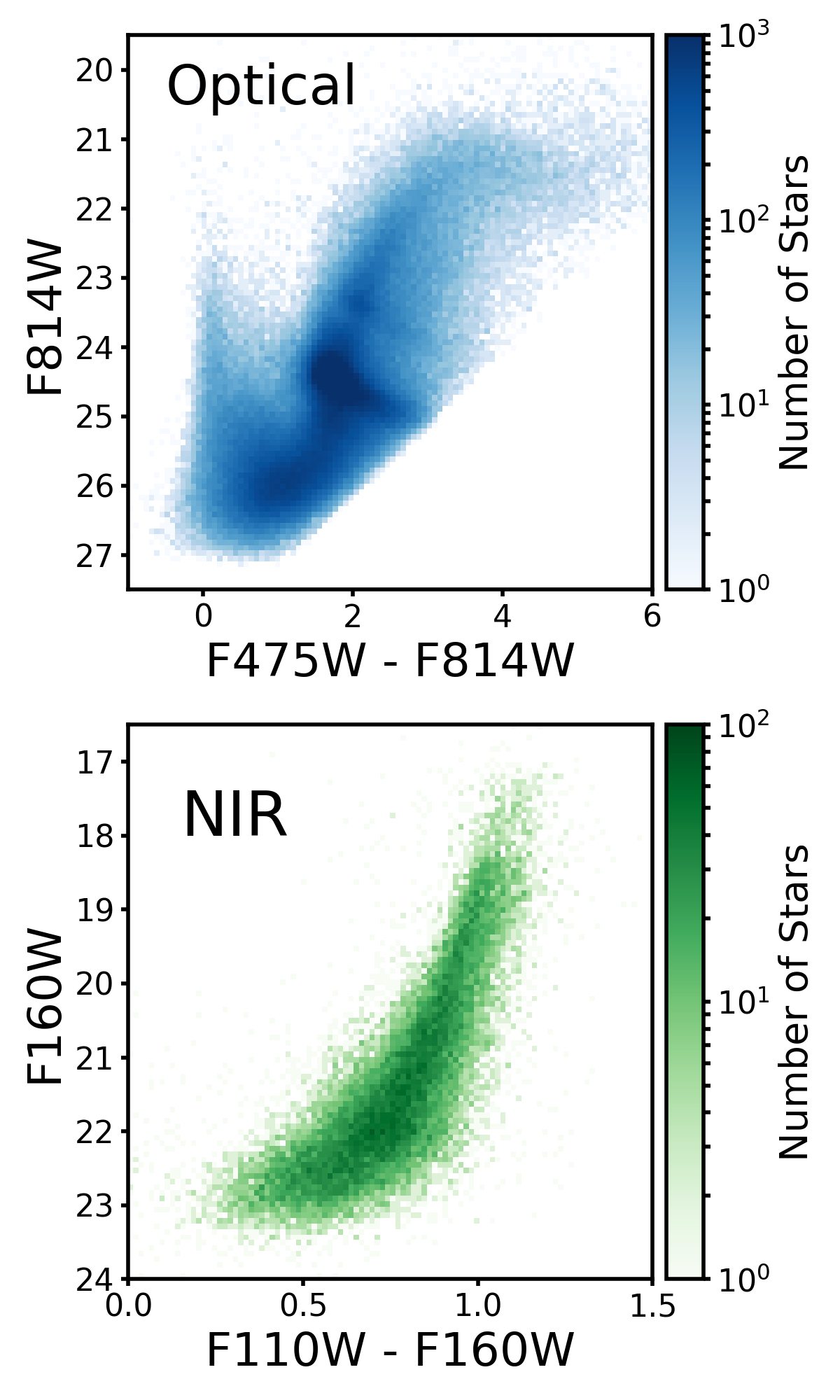}
\endminipage\hfill
\minipage{0.38\textwidth}
  \includegraphics[width=\linewidth]{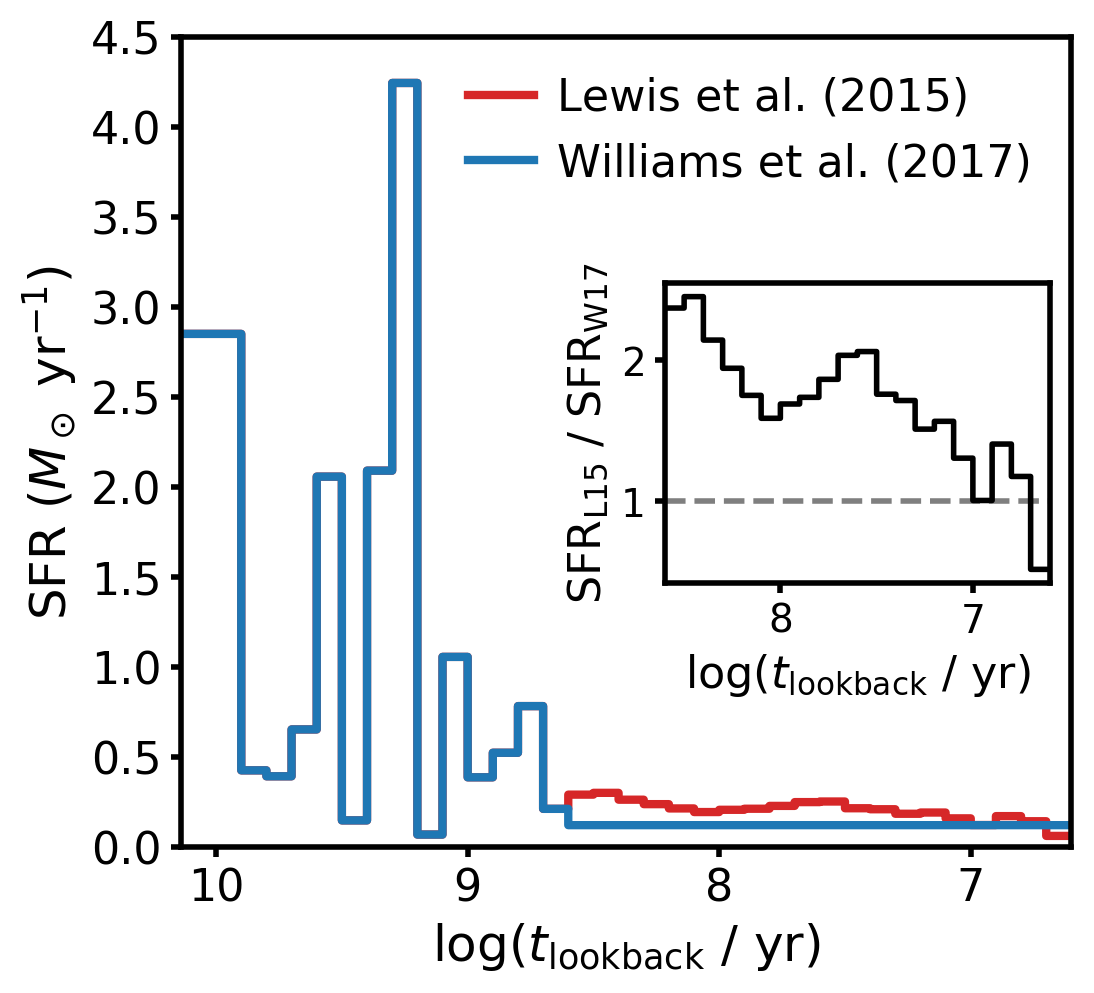}
\endminipage\hfill
\minipage{0.38\textwidth}
  \includegraphics[width=\linewidth]{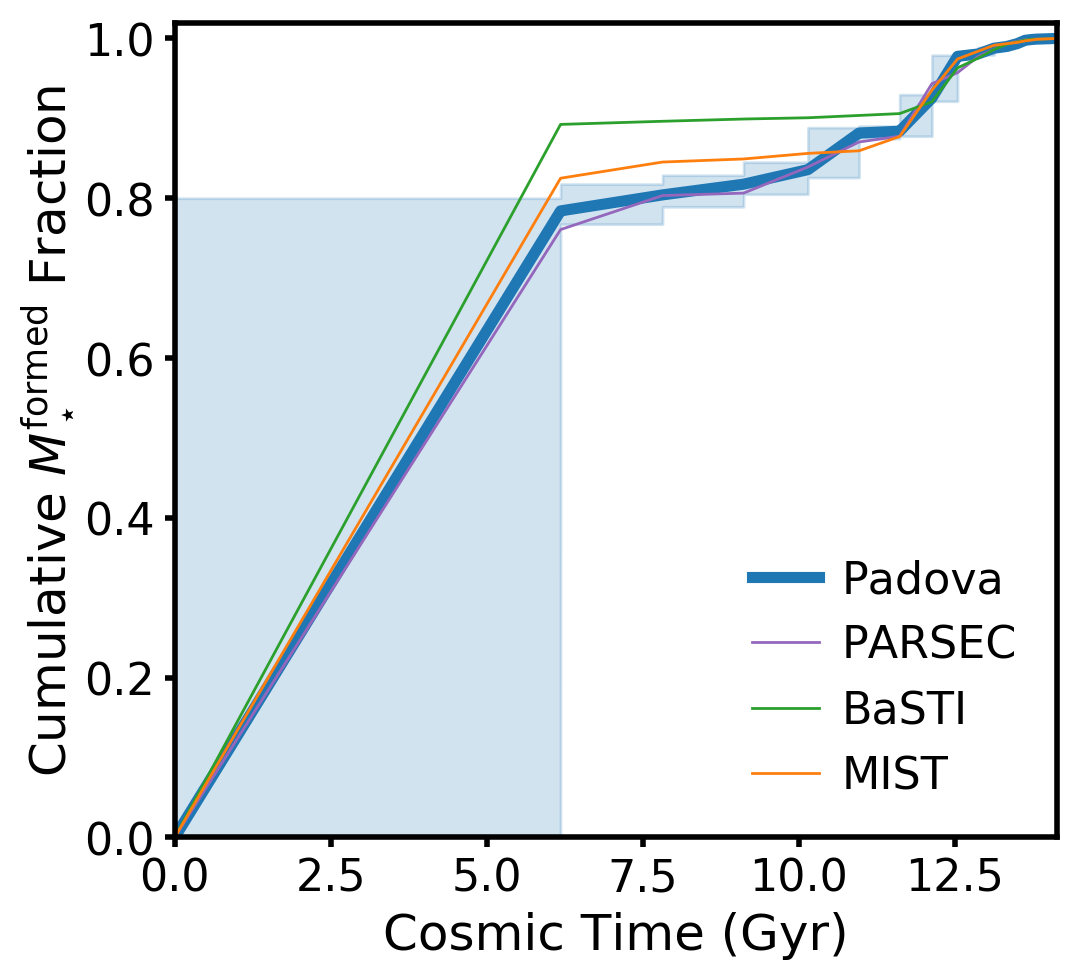}
\endminipage\hfill
\caption{\textbf{\textsc{The ancient and recent SFHs inferred from PHAT CMDs}}. Left: Example optical (top, blue) and NIR (bottom, green) color-magnitude diagrams constructed from PHAT resolved star photometry. Both CMDs come from the same field (Brick 13, Field 1), near the center of the PHAT footprint and representative of typical stellar densities (and therefore photometric quality). Center: Comparison between the recent SFHs from \citetalias{lewis15} (red line, Section~\ref{sec:recent_sfhs}), who optimized their CMD modeling for ages $\leq 500 \,\mathrm{Myr}$, and from \citetalias{williams17} (blue line, Section~\ref{sec:ancient_sfhs}), whose modeling was appropriate for older populations. SFR is plotted against logarithmic lookback time; in both the center and right panels, the present day is at the right of the plot. The inset shows the ratio of \citetalias{lewis15} SFH to the constant value from \citetalias{williams17} over the last 500 Myr, and demonstrates that the \citetalias{lewis15} modeling recovers a higher SFR averaged over the past 100 Myr. Right: The cumulative fraction of \mstar{} formed up to a given time over the \citetalias{williams17} analysis area. The fiducial Padova SFH is shown as the thick blue line, with random uncertainties shown as blue shading. The thin lines show the best-fit SFH for different stellar evolutionary models: PARSEC (purple), BaSTI (green), and MIST (orange).
\label{fig:sfhs}}
\end{figure*}

\citetalias{williams17} measured spatially resolved, ancient SFHs by modeling optical-NIR CMDs using \texttt{MATCH} \citep{dolphin02, dolphin12, dolphin13}. The CMDs were constructed from \textit{HST} resolved-star photometry in the F475W and F814W filters on the Advanced Camera for Surveys (ACS) and the F110W and F160W filters on Wide Field Camera 3 (WFC3). In the left panel of Figure~\ref{fig:sfhs}, we show example optical (top) and NIR (bottom) CMDs for the same region near the center of the PHAT footprint (Brick 13, Field 1), chosen to be representative of typical stellar densities. \citetalias{williams17} derived the ancient SFHs within 826 SFH pixels ($83''\times83''$ regions). We refer the reader to \citetalias{williams17} for details, but summarize here the key modeling choices. 

\citetalias{williams17} determined SFHs in logarithmic age bins, with 0.1 dex resolution from $\log(t/\mathrm{yr}) = 8.5-9.9$, and wider time bins at the oldest and youngest ages: $\log(t/\mathrm{yr})=6.6-8.5$ and $\log(t/\mathrm{yr})={9.9-10.15}$. The total SFH over the region analyzed in this paper is shown as the blue line in the center panel of Figure~\ref{fig:sfhs}. The CMD modeling assumes a \citet{kroupa01} initial mass function (IMF), a binary fraction of 30\%, and a mean metal enrichment history fixed to a physically motivated and iteratively tuned model that imposes exponentially decreasing enrichment rates. The adopted enrichment histories vary with radius such that the inner regions of the disk enrich earlier. The oldest main sequence turnoffs cannot be resolved in the PHAT CMDs (due to photometric depth and crowding; see the left panel of Figure~\ref{fig:sfhs}), so the enrichment history cannot be reliably inferred from the data. Fixing the age-metallicity relation in the CMD modeling avoids introducing biases due to unphysical fluctuations in metallicity, at the cost of not being able to account for variations in metallicity not captured by the adopted model.

The presence of dust affects age- and metallicity-sensitive CMD features in a way that is degenerate with real variation in SFH and enrichment history. To appropriately model the effects of dust attenuation, \citetalias{williams17} implemented a sophisticated dust model in \texttt{MATCH}, following the model used by \citetalias{dalcanton15} to map the dust distribution in M31 (described in Section~\ref{sec:dust} below). In this model, some fraction of the stars is assumed to lie behind the dust layer, and those reddened stars experience attenuation described by a lognormal distribution in \av{}. This model is appropriate for old stellar populations with a large scale height relative to that of the dust layer, but not for young stars embedded in the dust layer. \citetalias{williams17} therefore allowed for stars younger than a transition age to experience more dust attenuation. The lognormal dust model parameters in the \citetalias{williams17} CMD modeling were fixed using the best-fit parameters in the higher-resolution dust maps from \citetalias{dalcanton15} and values of the transition age and ratio of old star and dust scale heights were optimized through extensive testing (see \citetalias{williams17} for details on this procedure). A uniform foreground dust component was also included in the \citetalias{williams17} dust model, and was fit independently in each SFH pixel.

Both random and systematic uncertainties contribute to the total uncertainty in the derived SFHs. \texttt{MATCH} computes random uncertainties that capture the effects of photometric quality and number of stars in the CMD. To quantify the systematic uncertainty, \citetalias{williams17} repeated their SFH measurements using four different stellar evolutionary tracks: Padova \citep{marigo08, girardi10}, PARSEC \citep{bressan12}, BaSTI \citep{pietrinferni04, cassisi06, pietrinferni13}, and MIST \citep{choi16}. We adopt the results using the Padova models as our fiducial ancient SFHs in this work for consistency with the derivation of the recent SFHs (described in Section~\ref{sec:recent_sfhs} below). 

We calculate the total stellar mass formed (\mformedcmd{}) by integrating the SFH to the present day, as shown in the right panel of Figure~\ref{fig:sfhs}. The solid blue line shows the cumulative fraction of \mstar{} formed within the area of PHAT analyzed by \citetalias{williams17} up to a given time since the Big Bang (i.e., the present day is at the right of the plot). The random uncertainties are shown as the shaded envelope, and account for covariance between adjacent bins. The thin lines show the results using the PARSEC (purple), BaSTI (green), and MIST (orange) models; the spread in these captures the systematic uncertainty due to model choice. Across these different SFHs, the total \mformedcmd{} varies from 2\% lower to 10\% higher than the Padova SFH, while the range in \mformedcmd{} due to random uncertainties is typically $\pm8$\%.

\subsubsection{Returned Mass Fraction}
\label{sec:present_mass}

Because of mass loss during stellar evolution, the total mass formed is larger than the present-day stellar mass, \mstar{}. The fraction of formed stellar mass lost is known as the returned fraction, $R$, and depends on SFH, metallicity, and the IMF. Most stellar mass loss happens quickly after star formation as massive stars end their lives. However, intermediate mass stars also experience substantial mass loss during the asymptotic giant branch (AGB) phase, extending mass loss over many Gyr timescales. The initial metallicity affects the rate of mass loss over a star's lifetime, and the IMF dictates the relative abundance of high-mass and low-mass stars (which do not return any mass to the ISM). The choice of IMF and isochrone set are the dominant factors that determine $R$. 

To compute $R$ over the CMD-based SFHs in M31, we use the Flexible Stellar Population Synthesis\footnote{\url{https://github.com/cconroy20/fsps}, commit hash 3656df5} (\texttt{FSPS}, v3.0, \citealt{conroy09, conroy10a}) package and its Python wrapper, \texttt{python-fsps}\footnote{\url{https://github.com/dfm/python-fsps}, commit hash 8361d60} \citep[v0.3.0,][]{foreman-mackey14}. This is a computationally convenient tool that allows us to account for the effects of the SFH and enrichment history on $R$. We use Padova stellar evolutionary models \citep{marigo07, marigo08} and a \citet{kroupa01} IMF, consistent with the modeling choices made in deriving the CMD-based SFHs. For each SFH pixel, we input the CMD-based SFHs (SFR in $M_\odot \mathrm{yr}^{-1}$ vs.\ $t$ in Gyr since the galaxy formed) and the enrichment histories enforced in the \citetalias{williams17} modeling (the average stellar metal mass fraction $Z$ vs.\ $t$) to \texttt{FSPS}. \texttt{FSPS} calculates the mass lost due to stellar evolution from the Padova isochrones, integrating over the SFHs and accounting for metallicity-dependent variation in mass loss rates, and returns the present-day \mstar{} in each SFH pixel. The \texttt{FSPS}-based stellar masses include the mass contribution of stellar remnants. We have verified that calculating $R$ self-consistently using the evolutionary models in \texttt{MATCH}, which do not tabulate remnant masses, agrees with \texttt{FSPS}-based $R$ calculations that exclude the mass in stellar remnants.

A caveat is that the CMD-based SFHs were inferred assuming that 30\% of stars have a binary companion, where the companion stellar mass is drawn from the uniform distribution [0, $m_\mathrm{primary}$], requiring that the companion is less massive than the primary star. This changes the ``effective'' IMF slightly from the \citet{kroupa01} primary star IMF, driving a steeper slope at the low-mass ($m_\star < 0.5\, M_\odot$) end. The extra low-mass stars will drive $R$ lower for the primary + binary star population than would be calculated from the \citet{kroupa01} IMF, meaning the present-day stellar mass will be higher at the few percent level. The result is a small, SFH-independent \mstar{} normalization change that does not affect our results, as we discuss further in Section~\ref{sec:mass_norm}.

Even with this detailed approach that accounts for variation in SFH and metallicity across the PHAT footprint, we find little variation in $R$. The median is $R=39\%$, and the full range of $R$ spans just $2.5\%$ (corresponding to an 0.03 dex range in $\log M_\star$). This small range reflects that SSPs complete their mass loss within a few gigayears, and thus in cases like M31 where 78\% of the  stellar mass formed $\geq 8$ Gyr ago, variation in the recent SFH has little impact on the total $R$.

\subsubsection{Scaling $M_\star$ Measurements to a Common IMF}
\label{sec:mass_norm}

\begin{table}
\caption{Conversion from Various IMFs to \citet{kroupa01}} 
\label{tab:imfs}
\begin{center}
\begin{tabular}{lc}
IMF & $\log(M^\mathrm{Kroupa}_\star/M^\mathrm{IMF}_\star)$
\\ \hline
\citet{bell01} ``diet Salpeter'' & $-0.07$ \\
\citet{chabrier03} & $+0.03$ \\
\texttt{MATCH} \citet{kroupa01}, no \mstar{} limits & $-0.12$
\end{tabular}
\end{center}
\tablecomments{The ``diet Salpeter'' IMF is $0.15$ dex lighter than a \citet{salpeter55} IMF, which \citet{gallazzi08} calculated is $0.25$ dex heavier than a \citet{chabrier03} IMF. \citet{salim07} calculated the scaling between \citet{chabrier03} and \citet{kroupa01} IMFs.}
\end{table}

All \mstar{} inferences require an adopted IMF, but the IMF itself is uncertain due to both the difficulty of measuring the IMF from observations and to its possible variation within and/or among galaxies (due to, e.g., changes in star formation intensity, metallicity, or redshift). There are several IMFs that are commonly adopted in the literature, causing systematic differences in \mstar{} at the $\sim0.25$ dex level. In Section~\ref{sec:lit_cmlrs}, we compare our \mlcmd{} in M31 to \ml{} in the literature that adopted different IMFs. We scale all literature \ml{} to a common \citet{kroupa01} IMF using the constant scale factors in Table~\ref{tab:imfs}. We emphasize that our goal is to ensure that all \mstar{} measurements are \textit{on the same scale}; we do not assert that our chosen \mstar{} scale is the truth.

Our \mcmd{} in M31 were constructed from ancient SFHs inferred by \texttt{MATCH} assuming a \citet{kroupa01} IMF. The code models the number of stars in different regions of the CMD, where the relative number of stars of different masses is determined by the IMF slope. The normalization of the IMF, i.e., the number of stars formed per unit stellar mass formed, then dictates the conversion between number of stars in the CMD and the SFR in each age bin.

Most SPS models adopt physically motivated low- and high- mass cutoffs on the IMF, typically $0.1 - 100 \, M_\odot$. In contrast, \texttt{MATCH} integrates over all possible stellar masses ($0 - \infty \, M_\odot$) when calculating the IMF normalization, essentially allowing stellar mass to populate the very low and very high mass extremes of the IMF. This choice does not affect the modeled distribution of stars in the CMD, but does result in a lower IMF normalization: fewer stars of any given mass are formed per unit star formation. The SFHs, and therefore \mformedcmd{}, output by \texttt{MATCH} are therefore systematically higher than would be inferred for an IMF with stellar mass limits. 

To account for the differences described above, we make our \mlcmd{} consistent with SPS-based \ml{} in the literature by subtracting $0.12$ dex from our stellar evolution-corrected \mcmd{}, where $-0.12$ dex is the ratio between the \citet{kroupa01} IMF normalization calculated with mass cutoffs of $0.1 - 100 \, M_\odot$ and that calculated with no mass cutoffs. The magnitude of this correction is well within the factor of $\sim$2 uncertainty in \mstar{} that is commonly acknowledged in the literature \citep[e.g.,][]{courteau14, mcgaugh15} and discussed further in Section~\ref{sec:mass_uncertainty}. We use our \texttt{FSPS} models (Section~\ref{sec:present_mass}) to confirm that predicted luminosity maps in the SDSS and \wise{} filters better match the observed brightness maps after scaling down the \texttt{MATCH} SFHs input into the SPS models.

\subsubsection{Map of the $M_\star$ Distribution in M31}
\label{sec:mass_map}

\begin{figure}[]
\begin{centering}
  \includegraphics[width=\linewidth]{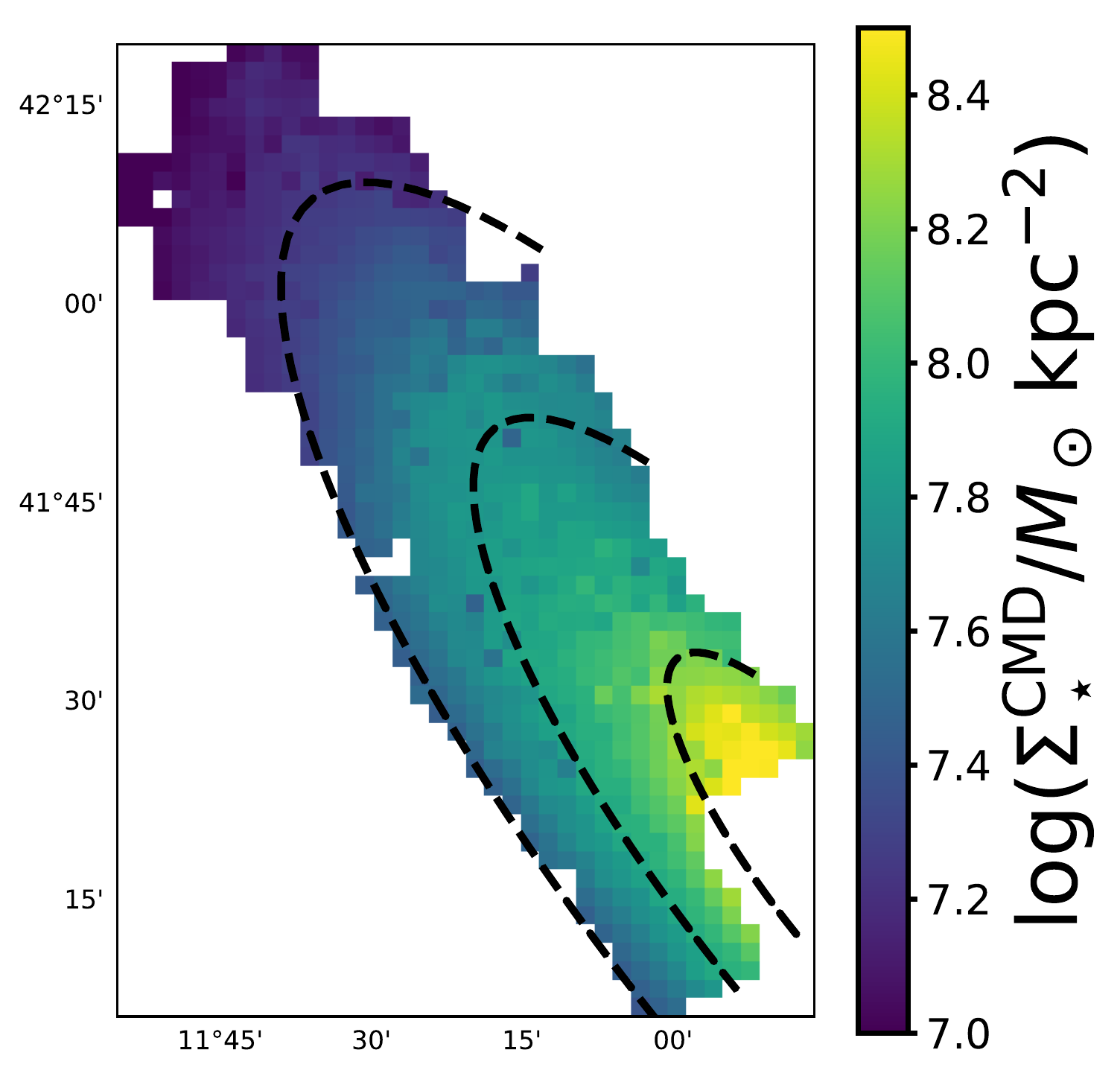}
\caption{\textbf{\textsc{Stellar mass map from CMD-based SFHs.}} Map of the present-day stellar mass surface density, $\Sigma^\mathrm{CMD}_\star$, within the PHAT footprint.  The dashed black lines show radii of 5, 10, and 15 kpc for reference. Yellow, higher $\Sigma^\mathrm{CMD}_\star$ SFH pixels lie near the center of M31, while darker, lower-density regions lie in the outer disk. The distribution of \mcmd{} is overall smooth, though some pixel-to-pixel fluctuations are obvious.
\label{fig:mass_map}}
\end{centering}
\end{figure}

After applying the corrections for mass recycling and the IMF stellar mass limits, we calculate a present-day stellar mass surface density ($\Sigma^\mathrm{CMD}_\star$) map using a SFH pixel area of $0.42\,\mathrm{kpc^2}$. The resulting map is shown in Figure~\ref{fig:mass_map}, with dashed black lines at radii of 5, 10, and 15 kpc for reference. Overall, the profile of $\Sigma^\mathrm{CMD}_\star$ is quite smooth, decreasing with radial distance from the center of M31, even though no smoothness in the stellar mass profile was enforced in the CMD fitting procedure. There are a few SFH pixels that deviate from a smooth profile, visible as stronger color contrasts between neighboring pixels. However, the total number of these pixels is quite small, with 41 SFH pixels (5.3\%) deviating by more than 20\% from a median-smoothed version of the map. We prefer the un-smoothed version of the map because the radial decline $\Sigma^\mathrm{CMD}_\star$ biases the median-smoothed $\Sigma^\mathrm{CMD}_\star$ in the SFH pixels at the edges of the map. We combine this $\Sigma^\mathrm{CMD}_\star$ map with the observed surface brightness maps of M31 to map \mlcmd{} across the PHAT footprint in Section~\ref{sec:ml_cmd}.

\subsection{Recent SFHs}
\label{sec:recent_sfhs}

\begin{figure*}[!htp]
\begin{centering}
\minipage{0.33\textwidth}
  \includegraphics[width=\linewidth]{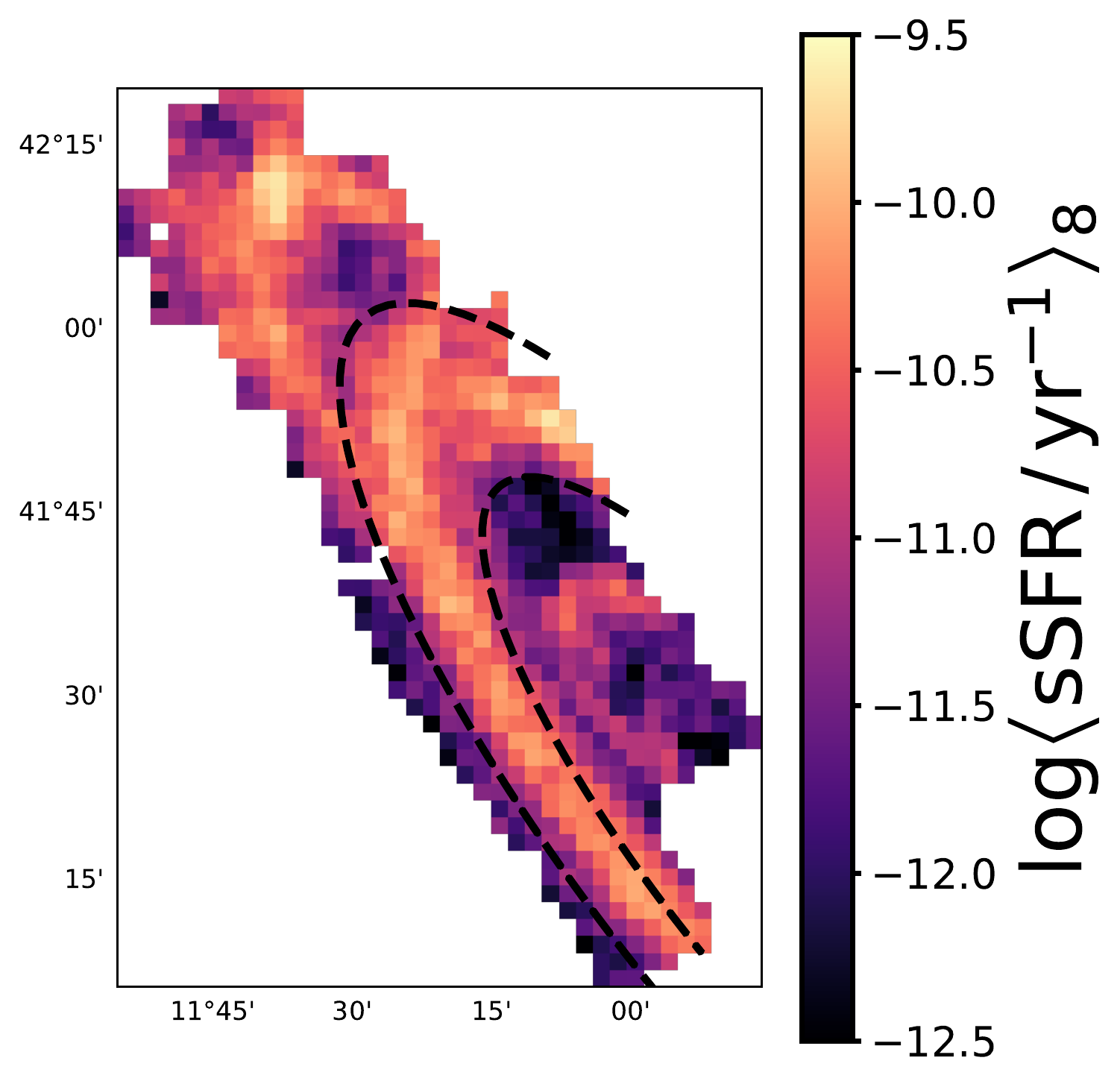}
\endminipage\hfill
\minipage{0.33\textwidth}
  \includegraphics[width=\linewidth]{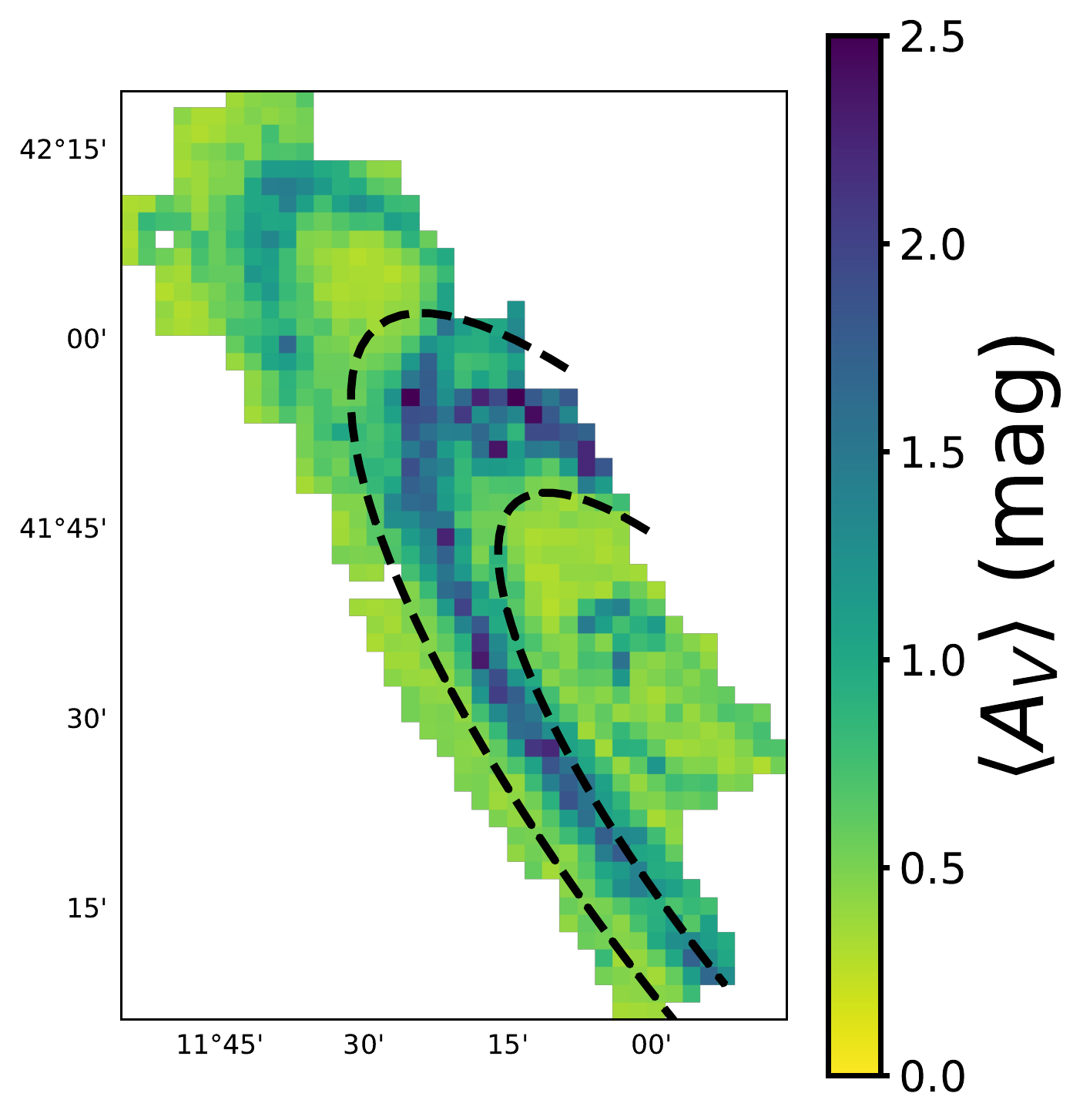}
\endminipage\hfill
\minipage{0.33\textwidth}
  \includegraphics[width=\linewidth]{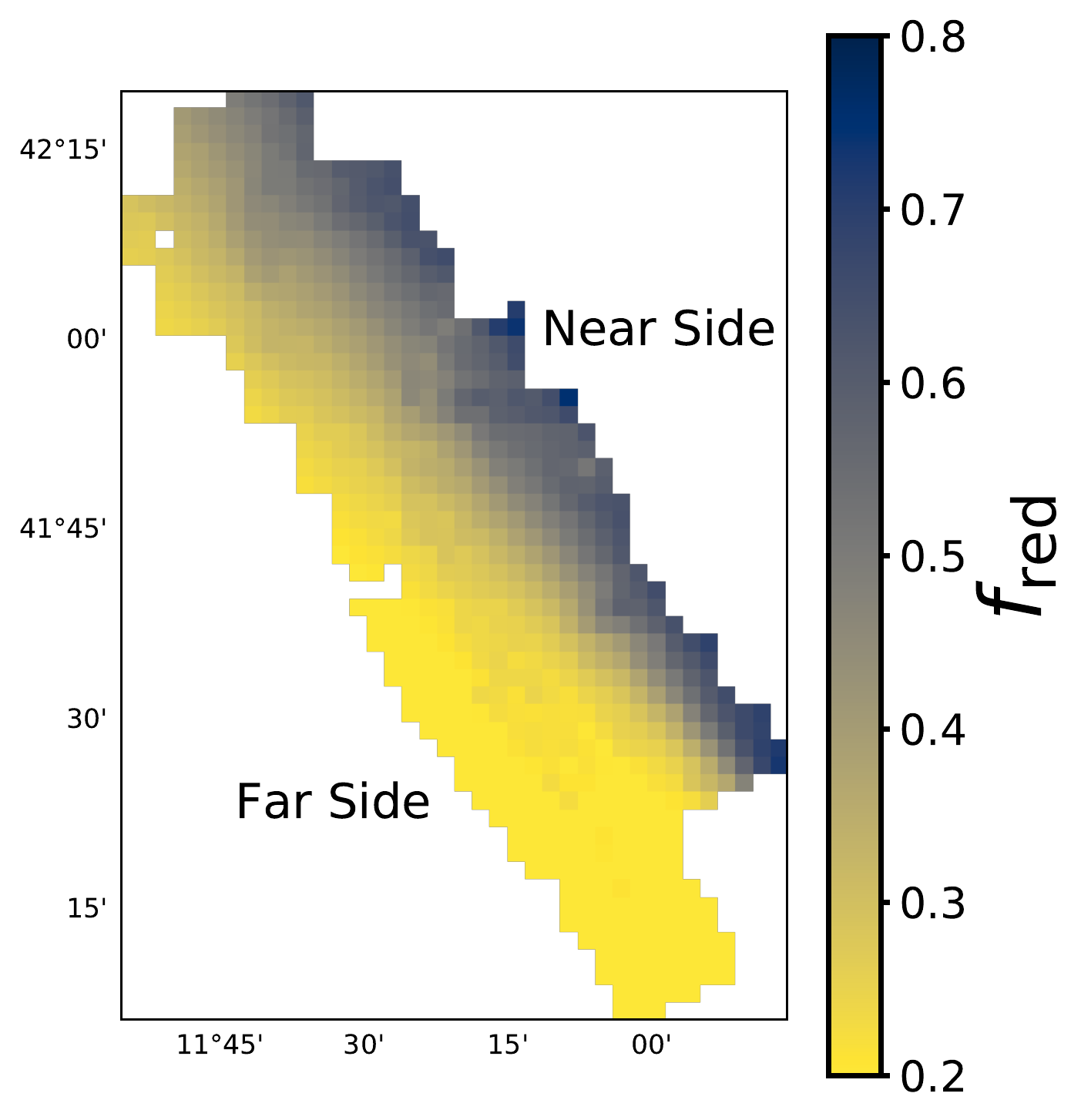}
\endminipage\hfill
\caption{\textbf{\textsc{Maps of recent star formation and dust parameters inferred from PHAT CMDs.}} From left to right, we show maps of the PHAT footprint color-coded by: average specific SFR over the past $10^8\,\mathrm{yr}$ (Section~\ref{sec:recent_sfhs}); average dust extinction \meanav{}; and average \fred{} (Section~\ref{sec:dust}). Dashed black lines are overplotted at radii of 9 and 13 kpc in the left and center panels to roughly bound the star-forming 10 kpc ring; we use these lines of constant radius for reference later in the paper. Ongoing star formation is highly correlated with dust content, while the fraction of old stars behind the dust layer is azimuthally dependent (such that lower \fred{} regions are on the far side of the disk). 
\label{fig:useful_maps}}
\end{centering}
\end{figure*}

Recent star formation is known to affect $M_\star/L$, especially in the optical, so we require a robust and spatially resolved SFR measurement to assess its impact on \mlcmd{} (see Section~\ref{sec:cmlr_recent_sfh} below).  The \citetalias{williams17} SFHs capture the bulk of star formation that has occurred over M31's lifetime. However, because the dust model and time binning used in their CMD fitting were optimized to infer the SFH for old stellar populations, the SFR measured over recent times is likely not as reliable as one optimized to fit the young main sequence.

To evaluate the effect of recent star formation, we use results from \citetalias{lewis15}, who modeled the main sequence in the PHAT optical CMDs (using the F475W and F814W filters on ACS) to recover the recent SFH at ages $\leq 500$ Myr. The assumptions made in this modeling were largely consistent with those in \citetalias{williams17}: a \citet{kroupa01} IMF, Padova isochrones, a binary fraction of 0.35, and stellar metallicity increasing with time. Compared to \citetalias{williams17}, the recent SFHs have higher temporal resolution at young ages, using logarithmic age bins of 0.1 dex width from $\log(t/\mathrm{yr}) = 6.6-9.9$. The recent SFHs were also determined with higher spatial resolution than the ancient SFHs, in regions of $24''\times27''$ ($100\times400$ pc, deprojected).

Because young, main sequence stars are well-mixed with the dust in star-forming regions, \citetalias{lewis15} used a different dust model optimization than that in \citetalias{williams17}. They optimized a uniform distribution between \av{} and \av{} + \dav{}, where \av{} is a foreground extinction and \dav{} is the differential extinction within the pixel. The differential extinction component allows the model to account for the broadening of the main sequence feature in the optical CMDs due to individual stars lying behind different total amounts of dust along the line of sight to the observer. This dust model is only appropriate for young stellar populations, so evolved stars (e.g., the red giant branch) were excluded from the \citetalias{lewis15} CMD modeling.

We use \texttt{Montage}\footnote{\url{http://montage.ipac.caltech.edu}} (\citealt{berriman03, jacob10}; version 5.0) to spatially align the two sets of SFHs.  \texttt{Montage} uses a flux-conserving algorithm to compute the exact overlap between input and output pixels and appropriately redistribute the \citetalias{lewis15} SFHs into the larger $83''\times83''$ SFH pixels. \citetalias{lewis15} excluded more of the high-density central disk from their analysis than \citetalias{williams17}, so we use only the regions that are fully covered by both the ancient and recent SFH maps in our analysis (see Figure~\ref{fig:mosaics}). This leaves us with a sample of 778 SFH pixels that we use throughout the paper. 

The center panel of Figure~\ref{fig:sfhs} shows the total (i.e., summed over all 778 SFH pixels) ancient and recent SFHs, where SFR is plotted as a function of logarithmic lookback time, such that the present is at the right of the plot. The \citetalias{williams17} ancient SFH (used to calculate \mstar{}) is shown in blue, and the \citetalias{lewis15} recent SFH (used to measure the recent SFR) is shown in red. The inset shows the ratio of the \citetalias{lewis15} SFH to the \citetalias{williams17} SFH over $6.6 < \log{(t_\mathrm{lookback}\,/\,\mathrm{yr})} < 8.6$. This comparison illustrates that the \citetalias{lewis15} SFH is more detailed and better captures the variation in SFR over the past $500\,\mathrm{Myr}$ than the \citetalias{williams17} measurement over the same lookback times. The two measurements do not precisely agree because of \citetalias{williams17}'s larger spatial binning and different dust model optimization.

We use the \citetalias{lewis15} SFHs to measure the average SFR surface density in each SFH pixel over the past 100 Myr, a timescale similar to that probed by UV-based SFR indicators. This SFR surface density map is then divided by the $\Sigma^\mathrm{CMD}_\star$ map presented in Figure~\ref{fig:mass_map} to find the average specific SFR, \ssfr{} (in units of yr$^{-1}$, where the subscript 8 indicates an average over the past 10$^8$ yr). The left panel of Figure~\ref{fig:useful_maps} shows a map of \ssfr{} in the PHAT footprint, with bright yellow regions tracing the well-studied, star-forming rings in M31. We overplot dashed black lines at 9 and 13 kpc to roughly bound the 10 kpc star-forming ring, and use these same lines for visual reference later in the paper. We use these \ssfr{} measurements in Section~\ref{sec:cmlr_recent_sfh} below to assess whether recent star formation drives scatter in \mlcmd{} at a given color.

\subsection{Dust Maps}
\label{sec:dust}

\citetalias[][]{dalcanton15} fit a dust model to the morphology of the red giant branch (RGB) in NIR CMDs within $3.3''\times3.3''$ regions across the PHAT footprint. They adopted a lognormal probability distribution for the $V$-band extinction \av{} (described by a median \medianav{} and dimensionless parameter $\sigma$), and assumed that some ``reddened fraction'' ($0 < f_\mathrm{red} < 1$) of stars in each region lie behind a thin dust layer and experience dust attenuation (so $f_\mathrm{red} = 1$ means that all stars in a given region lie behind the dust layer). Regions with both high \medianav{} and high \fred{} experience the strongest attenuation; if a region has a high dust content but low \fred{}, then most old stars do not experience attenuation and the effect of dust on color and luminosity of the old stellar population is small.

Variation in \fred{} is strongest within galaxies that are highly inclined and have thick stellar disks.  The assumption that stars lie either in front of or behind the dust layer (i.e., are not embedded in the dust layer) is appropriate for evolved stellar populations, which are expected to have a scale height much larger than that of a thin dust layer concentrated near the disk midplane. The dust attenuation experienced by younger stellar populations with a scale height more similar to that of the dust layer is better approximated by the differential extinction model, as used in \citetalias{lewis15}.

We use the mean \meanav{} as a dust surface density tracer, defined in terms of the \citepalias{dalcanton15} model parameters as: 
\begin{equation}
\left<  A_V \right> = \tilde{A}_V \, e^{\sigma^2 / 2}.
\end{equation}
Because the dust parameters are inferred at higher resolution than the \citetalias{williams17} SFHs, we adopt the mean \meanav{} and \fred{} within each SFH pixel as tracers of the dust content and geometry relative to the evolved stellar population.

The center and right panels of Figure~\ref{fig:useful_maps} show maps of \meanav{} and \fred{}, respectively, across the PHAT footprint. The dust mass, as traced by \meanav{}, is co-located with ongoing star formation, while \fred{} has a clear azimuthal dependence such that the far side of the disk has lower \fred{} (yellow colors in the right panel). This is due to the geometry of M31: its high inclination and thick stellar disk produce an increasing \fred{} from the far to near side of the disk, with $f_\mathrm{red}=0.5$ along the major axis. For the case of a face-on disk galaxy, $f_\mathrm{red}=0.5$ everywhere.

The \citetalias{dalcanton15} dust model is more complex and realistic than the uniform foreground dust screen models that are typically adopted in SPS modeling used to infer \ml{} from galaxy colors or SEDs. Our detailed knowledge of the dust distribution in M31, coupled with our \mlcmd{}, enables a unique test of the validity of fitting CMLRs to SPS models that assume simple foreground screen dust models (see Section~\ref{sec:cmlr_dust} below).


\section{Observed Colors and \mlcmd{}\label{sec:obs}}

For our analysis of color-\mlcmd{} relations in M31, we focus on the SDSS (optical) and \wise{} (mid-IR) data to bracket the range of wavelengths commonly used for \mstar{} inference. We do not analyze the public NIR data from 2MASS because that photometry is too shallow for us to reliably measure \mlcmd{}. The optical and mid-IR data are expected to be sensitive to different effects: e.g., dust attenuation strongly affects the optical, but not the mid-IR. Both of these filter sets have been used to image large samples of galaxies, so the insights gained from our analysis will inform the interpretation of results from widely used survey datasets. 

In the optical, we study the relation between \mli{} and \gi{} because this CMLR is reported to yield the most precise \ml{} estimates of the available SDSS filter combinations. In SPS models, the presence of dust moves galaxies along this relation, instead of introducing scatter about the CMLR, so \mli{} is predicted to be accurate within $\sim0.1$ dex \citep{zibetti09, taylor11}. We present best-fit CMLRs in M31 for other SDSS filter combinations in Appendix~\ref{sec:appendix}.

In the mid-IR, both \spitzer{} and \wise{} data are available for M31. Though \spitzer{} 3.6 $\mu$m imaging is often used as a tracer of \mstar{} \citep[e.g.,][]{barmby06, courteau11, eskew12}, we choose to focus on the relation between \mlw{} and \wcolor{} because the wealth of all-sky data from \wise{} is more recent and less well-studied. For nearby spiral galaxies, integrated \spitzer{} 3.6 and 4.5 $\mu$m  fluxes agree to within 5\% with integrated \wise{} W1 and W2 (3.4 and 4.6 $\mu$m) fluxes, respectively \citep{jarrett13}.

In this section, we map the observed brightness in the $g$, $i$, W1, and W2 filters using mosaics of M31 constructed from archival SDSS and \wise{} imaging. We describe our methods of masking foreground stars and calculating the surface brightness and colors at the spatial resolution of the map of $\Sigma^\mathrm{CMD}_\star$ in the PHAT footprint (Figure~\ref{fig:mass_map}). From these resolution-matched maps, we calculate colors and \mlcmd{} across the disk of M31. These measurements are used to construct optical and mid-IR CMLRs for M31 in Section~\ref{sec:m31_cmlrs} below.

\begin{figure*}[!ht]
\minipage{0.5\textwidth}
  \includegraphics[width=\linewidth]{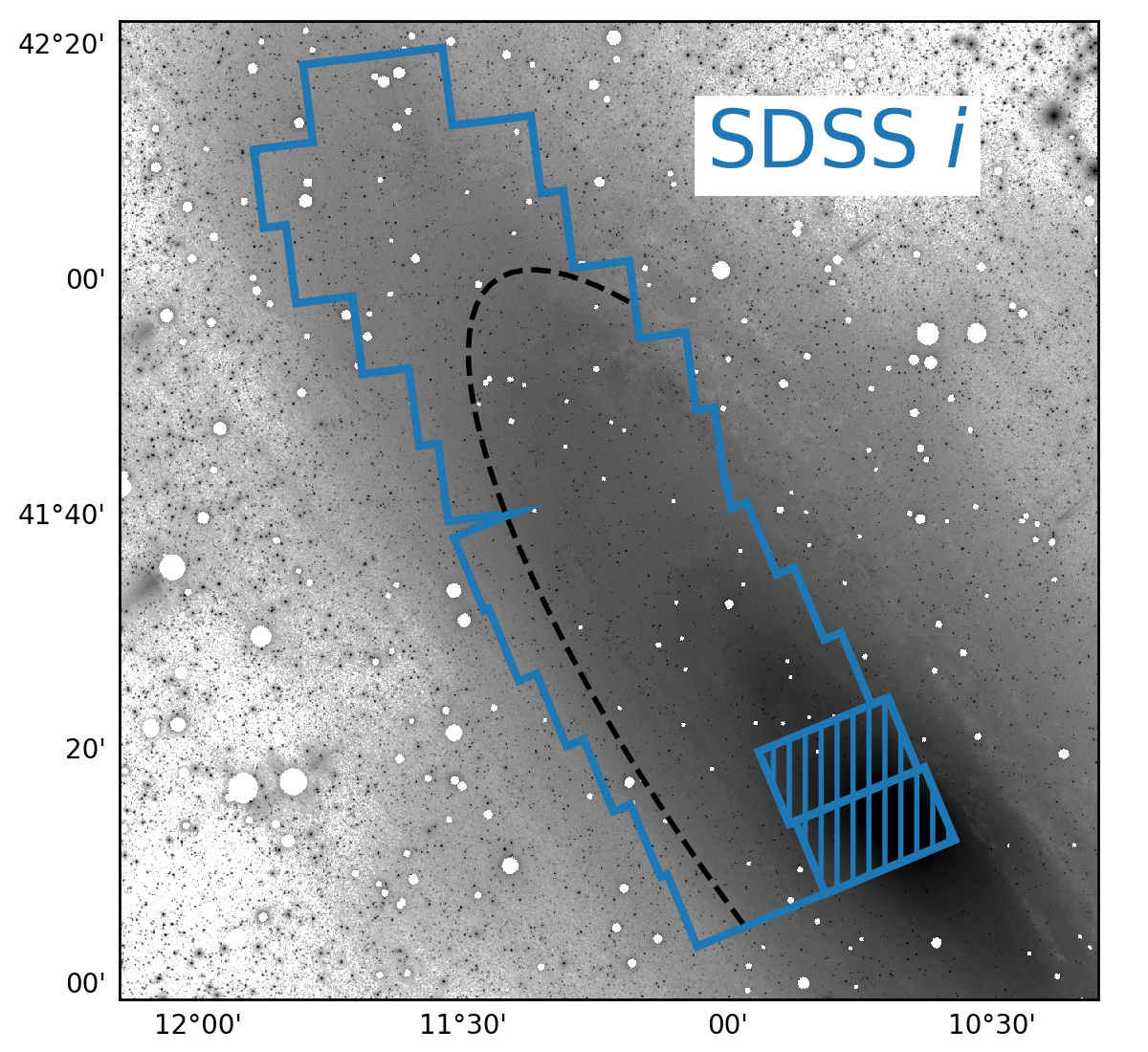}
\endminipage\hfill
\minipage{0.5\textwidth}
  \includegraphics[width=\linewidth]{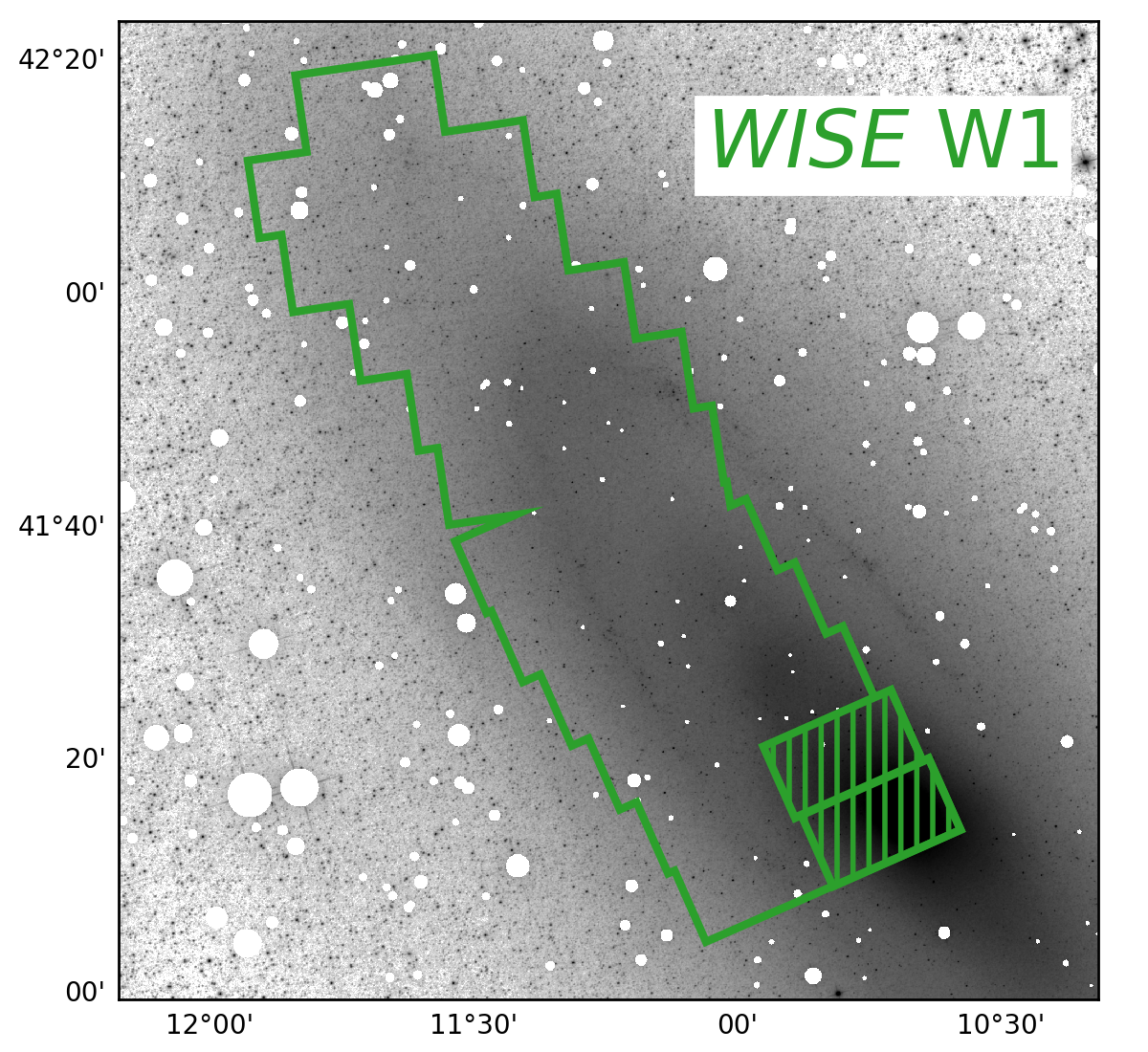}
\endminipage\hfill
\caption{\textbf{\textsc{Optical and mid-IR mosaics of M31 used to map surface brightness within the PHAT footprint.}} Left: SDSS $i$-band mosaic from \citet{tempel11}, with the PHAT footprint shown as the blue outline. The hatched regions (Bricks 1 and 3) are excluded from this analysis because stellar crowding limits the reliability of SFH determinations in the central regions. Foreground stars are masked and shown as white circles. The off-galaxy sky regions show obvious, unphysical variations due to the shallow imaging and background subtraction problems. The black dashed line shows the an arc of constant radius \radcut{} beyond which our quality cut excludes much of the SDSS photometry. Right: \wise{} W1-band mosaic from \citet{lang14}, with the PHAT footprint shown as the green outline. All regions in the \wise{} imaging meet our photometric quality requirement.
\label{fig:mosaics}}
\end{figure*} 

\subsection{Archival M31 Mosaics}
\label{sec:mosaics}

\subsubsection{Optical Luminosities and Colors from SDSS}
\label{sec:sdss}
We use mosaics in the $g$ and $i$ bands from \citet{tempel11}\footnote{Data: \url{http://www.aai.ee/~elmo/m31/}}, composed of imaging from SDSS \citep{york00}. The pixels are $3.96''$ on a side, a factor of 10 lower resolution than the original exposures, and the units\footnote{\url{http://www.sdss3.org/dr8/algorithms/magnitudes.php}} are nanomaggies (or nMgy). We reproject these mosaics with \texttt{Montage} to obtain the flux $f$ within each SFH pixel ($83''$ on a side) then calculate the flux in AB magnitudes following $m=22.5-2.5\log(f/\mathrm{nMgy})$.

The $i$-band mosaic that we use for our brightness map is shown in the left panel of Figure~\ref{fig:mosaics}. The PHAT footprint is overplotted as the blue outline, and the hatched rectangles show regions that were excluded from the SFH analysis due to the high stellar surface density limiting the quality of resolved-star photometry. White circles are regions masked due to the presence of foreground stars (see Section~\ref{sec:image_processing}). 

\citet{tempel11} estimated and subtracted the variable sky background from the individual exposures used to generate the mosaics. There is obvious, unphysical structure in the off-galaxy regions in the $i$-band mosaic, likely due to the difficulty of properly modeling the time-variable sky background in the drift-scan SDSS data. The effect is greater in the $g$ band because the surface brightness of M31 is lower in that filter than in the $i$ band. This low, residual background level in the mosaics can lead to inaccurate fluxes and colors, when the galaxy's surface brightness becomes comparable to the scale of the sky subtraction residuals. We identify these regions of questionable photometry by calculating the mean background level and variance within off-galaxy regions $830''$ on a side. The highest and lowest 5\% of off-galaxy pixel fluxes are excluded so that our thresholds are not biased by outliers. We then require that the flux in every SFH pixel is at least $5\sigma$ above this estimated mean background flux. This quality cut removes most regions within the PHAT footprint beyond \radcut{} from our analysis of the SDSS imaging, largely due to the shallow photometry in the $g$ filter. We show a dashed black line at \radcut{} in the left panel of Figure~\ref{fig:mosaics} to illustrate the radius beyond which our photometric quality requirement excludes much of the SDSS photometry from our analysis. Some bright, blue regions in the outer disk have high enough surface brightness to meet our quality threshold, extending the dynamic range of $g-i$ probed in this study. These 20 SFH pixels make up about 4.5\% of the surface area considered in our analysis of SDSS photometry.

\subsubsection{Mid-IR Luminosities and Colors from WISE}
\label{sec:wise}
We use W1 and W2 mosaics from \citet{lang14}\footnote{Data: http://unwise.me}, constructed from \wise{} \citep{wright10} images. The pixels in these mosaics are $2.75''$ on a side, and the units are ``Vega nMgy'' -- that is, flux units whose zero-point is 22.5 in the Vega magnitude system, not in the AB system. After reprojecting the images with \texttt{Montage} to obtain the flux $f$ in each SFH pixel, we convert to Vega magnitudes following $m=22.5-2.5\log(f/\mathrm{(Vega\,nMgy)}) + \Delta m$, where $\Delta m_\mathrm{W1}=2.655$ and $\Delta m_\mathrm{W2}=3.291$ are constant offsets \citep{willmer18}. 

The W1-band mosaic is shown in the right panel of Figure~\ref{fig:mosaics} with the PHAT footprint overplotted in green. Again, the excluded high-density, central regions are hatched, and white circles show masked foreground stars (see Section~\ref{sec:image_processing} below). The \wise{} photometry is of higher quality than the relatively shallow SDSS imaging and appears to have a more well-behaved background. The morphology of M31 at these wavelengths is also smoother than in the optical, as expected for being dominated by older RGB and AGB stars and less affected by dust. However, there is some notable small-scale structure in the $10\,\mathrm{kpc}$ star-forming ring, indicating the presence of younger, mid-IR bright massive star populations like RSGs or red HeB stars \citep[][]{melbourne12}, or possibly emission from hot dust heated by these young stellar populations. These features are typically red in \wcolor{}.

A uniform background level was estimated and subtracted from these mosaics by \citet{lang14}. We perform a similar estimate of the residual background flux level and variance as for the SDSS images (Section~\ref{sec:sdss}), and again require that the flux in each SFH pixel is at least $5\sigma$ above the background flux. All regions within the PHAT footprint meet our quality threshold and are all included in our mid-IR analysis. 

\subsubsection{Masking Foreground Stars and Measuring Luminosity  within SFH Pixels}
\label{sec:image_processing}

Foreground stars contribute to the total observed flux in a given SFH pixel. To isolate the light from the smooth stellar disk of M31 that is relevant to our \mlcmd{} calculations, we identify and mask foreground stars from the SDSS and \wise{} mosaics using \texttt{sep} \citep{barbary16}, a Python re-implementation of the commonly used \texttt{Source Extractor} tool \citep{bertin96}.

The contrast between foreground stars and the smooth M31 light is strongest in the W1 mosaic, so we identify the locations and size of foreground stars in this image and mask stars at the same coordinates in all mosaics. Using a bright star as a template for the point spread function, \texttt{sep} identifies bright sources whose fluxes are more than a user-defined level above the smooth background. We visually inspected the image to ensure that all obvious foreground stars were included in the mask, and experimented with the user-defined thresholds to verify that our final surface brightness measurements are insensitive to these choices. The foreground stars contribute fractionally less to the total light in the SDSS mosaics and have smaller angular size than in the W1 image, so we scale down the radii in the mask applied to the optical images (by factors of 2 and 1.5 in the $g$ and $i$ bands, respectively). 

We require measurements of the total luminosity within the same SFH pixels within which the ancient SFHs were determined by \citetalias{williams17} (Section~\ref{sec:ancient_sfhs}). After applying the foreground star masks, the mosaics are all reprojected to the SFH pixel scale with \texttt{Montage}, where the SFH pixels are $20-30$ times larger than the pixels in the input mosaics. The masked regions are ignored by \texttt{Montage} when computing the average flux across the SFH pixels.

\subsection{Maps of Observed Brightness and Color}
\label{sec:colors}

\begin{figure*}[!ht]
\begin{centering}
  \includegraphics[width=\linewidth]{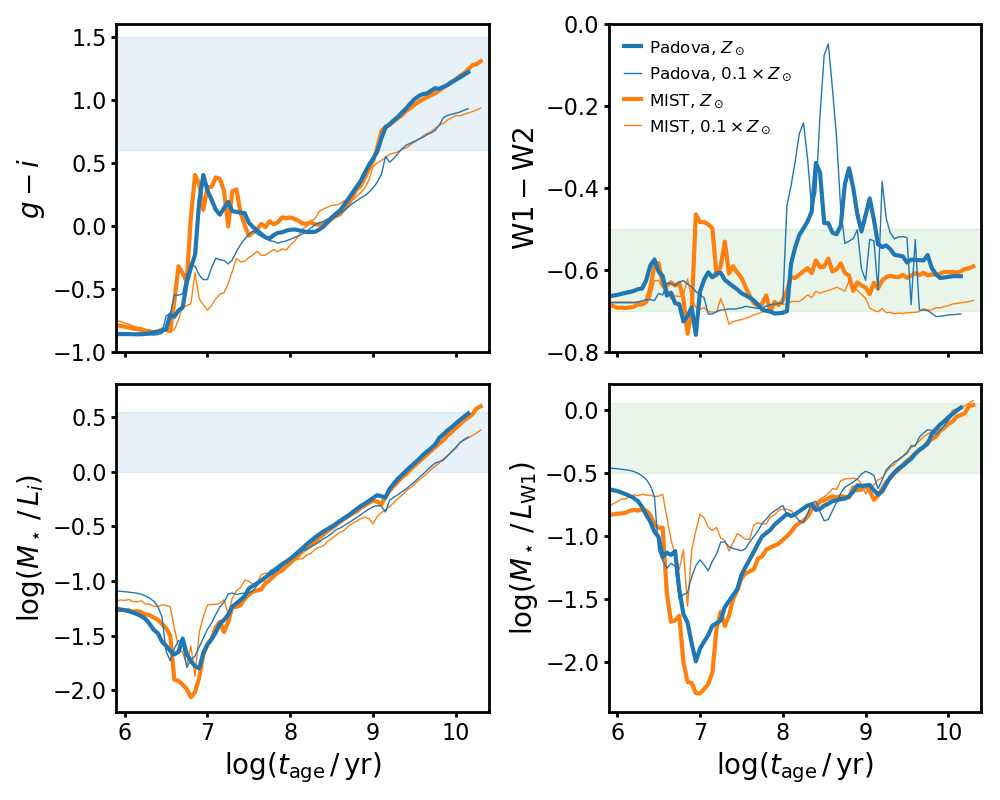}
\caption{\textbf{\textsc{Predicted colors and $\bm{M_\star/L}$ of dust-free stellar populations from stellar evolution models.}} The time evolution of optical and mid-IR colors and \ml{} for a dust-free simple stellar population, as predicted by the Padova (blue) and MIST (orange) stellar evolution models. We plot the following quantities as a function of logarithmic stellar population age: \gi{} (top left), \wcolor{} (top right), \logmli{} (bottom left), and \logmlw{} (bottom right). In all panels, thick lines indicate solar metallicity, while thin lines show one-tenth solar metallicity. The blue and green shaded regions show the range of each quantity observed in M31 in the optical and mid-IR, respectively. 
\label{fig:sps_models}}
\end{centering}
\end{figure*}

\begin{figure*}[!ht]
\begin{centering}
\minipage{0.5\textwidth}
  \includegraphics[width=\linewidth]{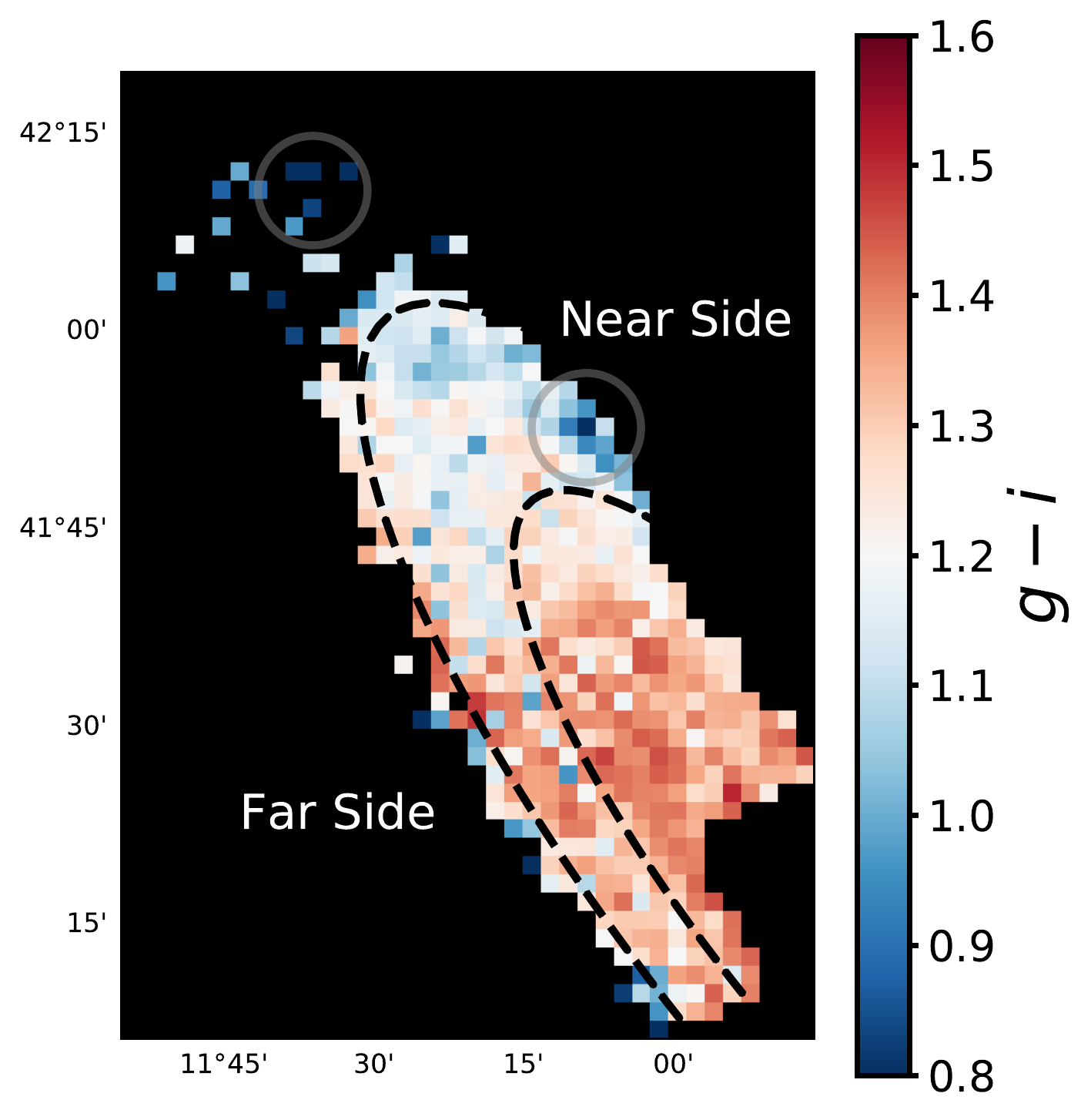}
\endminipage\hfill
\minipage{0.5\textwidth}
  \includegraphics[width=\linewidth]{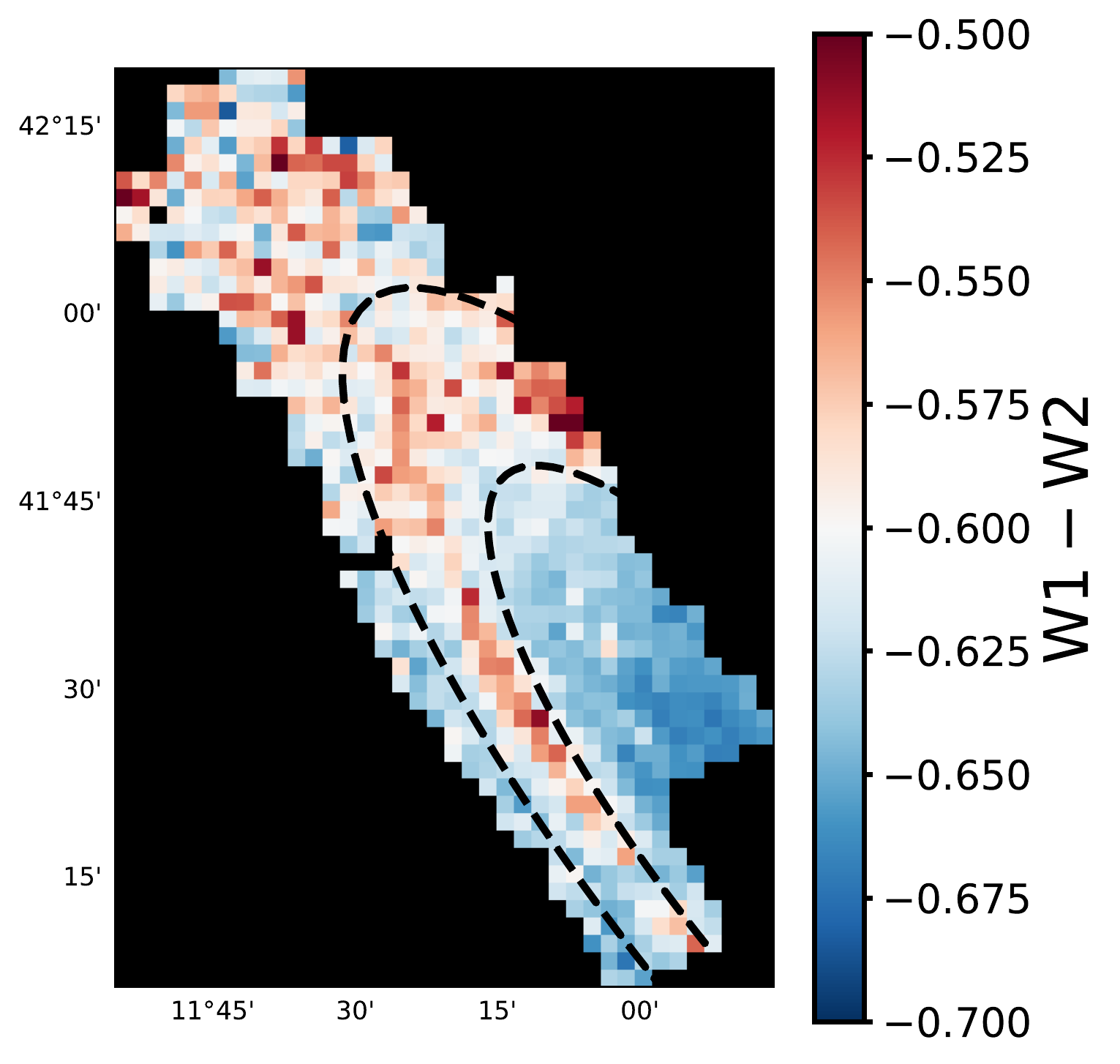}
\endminipage\hfill
\caption{\textsc{\textbf{Maps of observed $\bm{g-i}$ and $\bm{\mathrm{W}1-\mathrm{W}2}$ colors.}} Left: A map of the PHAT footprint matched to the resolution of the \citetalias{williams17} ancient SFHs, color-coded by the observed \gi{} color from SDSS. The colorbar is centered on the median \gi{} so that regions with typical colors appear light, redder regions are colored red, and bluer regions are colored blue. Many SFH pixels in the lower surface brightness outer disk are excluded by our photometric quality cut; 547 SFH pixels are used in our analysis for the SDSS filters. The dashed black lines approximately bound the 10 kpc star-forming ring, while the grey circles indicate two blue star-forming regions discussed in the text (corresponding to high \ssfr{} regions in the left panel of Figure~\ref{fig:useful_maps}). Right: Same as the left panel, but color-coding shows the \wcolor{} color from \wise{}. Again, the colorbar is centered at the median \wcolor{}, but a smaller range in color is shown than in the left panel due to the small dynamic range in observed \wcolor{}. All 778 SFH pixels meet our photometric quality requirement for the \wise{} filters.
\label{fig:colors}}
\end{centering}
\end{figure*}

Here, we present color maps across the PHAT footprint at the resolution of the $\Sigma^\mathrm{CMD}_\star$ map (Figure~\ref{fig:mass_map}). To guide our discussion of qualitative trends in these color maps, we will refer to the theoretical colors of dust-free simple stellar populations shown in the top row of Figure~\ref{fig:sps_models}. The evolution of \gi{} and \wcolor{} as a function of logarithmic age of the stellar population are shown in the top left and top right panels, respectively. The blue lines show predictions from the Padova evolutionary tracks, while orange show predictions from MIST. Solar and one-tenth solar metallicity are shown as thick and thin lines, respectively, and the range of colors observed in M31 are shown as the blue and green shaded bands. 

Figure~\ref{fig:sps_models} shows that different stellar evolutionary tracks make similar predictions for the evolution of \gi{}, but very different predictions for the \wcolor{} of intermediate age (roughly 100 Myr to few Gyr old) populations. The uncertain ingredients in modeling evolved stars \citep[e.g., mixing, convective overshooting, dredge-up events, mass-loss rates;][]{herwig05, conroy13, karakas14} have a larger impact on the mid-IR, where AGB stars provide a larger fraction of the integrated luminosity of a stellar population than in the optical. Some regions in M31 have redder optical colors than the oldest models; this discrepancy is likely due to reddening by dust and/or $\alpha$-enhancement relative to the scaled-solar evolutionary tracks.

With this theoretical context established, we now turn to the observed colors in M31 that we will use in our CMLR analysis. Figure~\ref{fig:colors} shows maps of \gi{} (left) and \wcolor{} (right). In both panels, the color ranges are chosen such that the light pixels are close to the median colors ($\left< g-i \right>_\mathrm{med} = 1.26$ and $\left< \mathrm{W1}-\mathrm{W2} \right>_\mathrm{med} = -0.61$), while SFH pixels that appear blue (red) in Figure~\ref{fig:colors} are bluer (redder) than typical. The colorbars are scaled differently to reflect the wider dynamic range in \gi{} compared to \wcolor{}. Only the SFH pixels that passed our photometric quality requirement (Section~\ref{sec:sdss}) are shown, and black dashed lines at radii of 8 and 13 kpc, approximately bounding the 10 kpc star-forming ring, are shown for reference. For comparison to other work in the literature reporting \wise{} colors in the Vega system, we note that $(\mathrm{W}1-\mathrm{W}2)_\mathrm{Vega} = (\mathrm{W}1-\mathrm{W}2)_\mathrm{AB} + 0.64$.

In the left panel of Figure~\ref{fig:colors}, there is a radial gradient in \gi{} such that the outer disk tends to be bluer than the inner regions. This would be consistent with radially decreasing stellar age and/or metallicity (top left panel of Figure~\ref{fig:sps_models}), in agreement with the findings of \citetalias{williams17}. Two clusters of blue pixels (indicated by grey circles in the left panel of Figure~\ref{fig:colors}) are associated with intensely star-forming regions: one in the outer disk, and one on the near side of the disk coinciding with the 10 kpc star-forming ring. Overall though, the star-forming rings are surprisingly not well-defined in the \gi{} map when averaged over the $83''$ scale of the SFH pixels. The lack of contrast is likely due to increased dust attenuation in the star-forming regions reddening the observed colors of intrinsically blue, young stellar populations, producing weaker color variations.

In the right panel of Figure~\ref{fig:colors}, there is a radial gradient in \wcolor{} in the opposite sense of the optical color gradient, such that the center is bluer. This may be a metallicity effect, due to CO absorption in the W2 band driving \wcolor{} bluer in higher metallicity regions \citep[e.g.,][]{meidt14}. However, different theoretical treatments of stellar atmospheres result in different predictions for the impact of metallicity on \wcolor{} (top right panel of Figure~\ref{fig:sps_models}). In addition to the overall gradient, the reddest \wcolor{} generally traces star-forming regions. Most SFH pixels redder than typical reside in 10 kpc ring or the large OB association in the outer disk that appears blue in \gi{}. This could be explained by an increased contribution to the mid-IR flux of young/intermediate age stellar populations (though again, models disagree on the time evolution of \wcolor{}; top right panel of Figure~\ref{fig:sps_models}). Because dust is co-located with ongoing star formation, it is also possible that dust emission drives redder \wcolor{} \citep[e.g.,][]{querejeta15}.

\begin{figure*}[!ht]
\begin{centering}
\minipage{0.5\textwidth}
  \includegraphics[width=\linewidth]{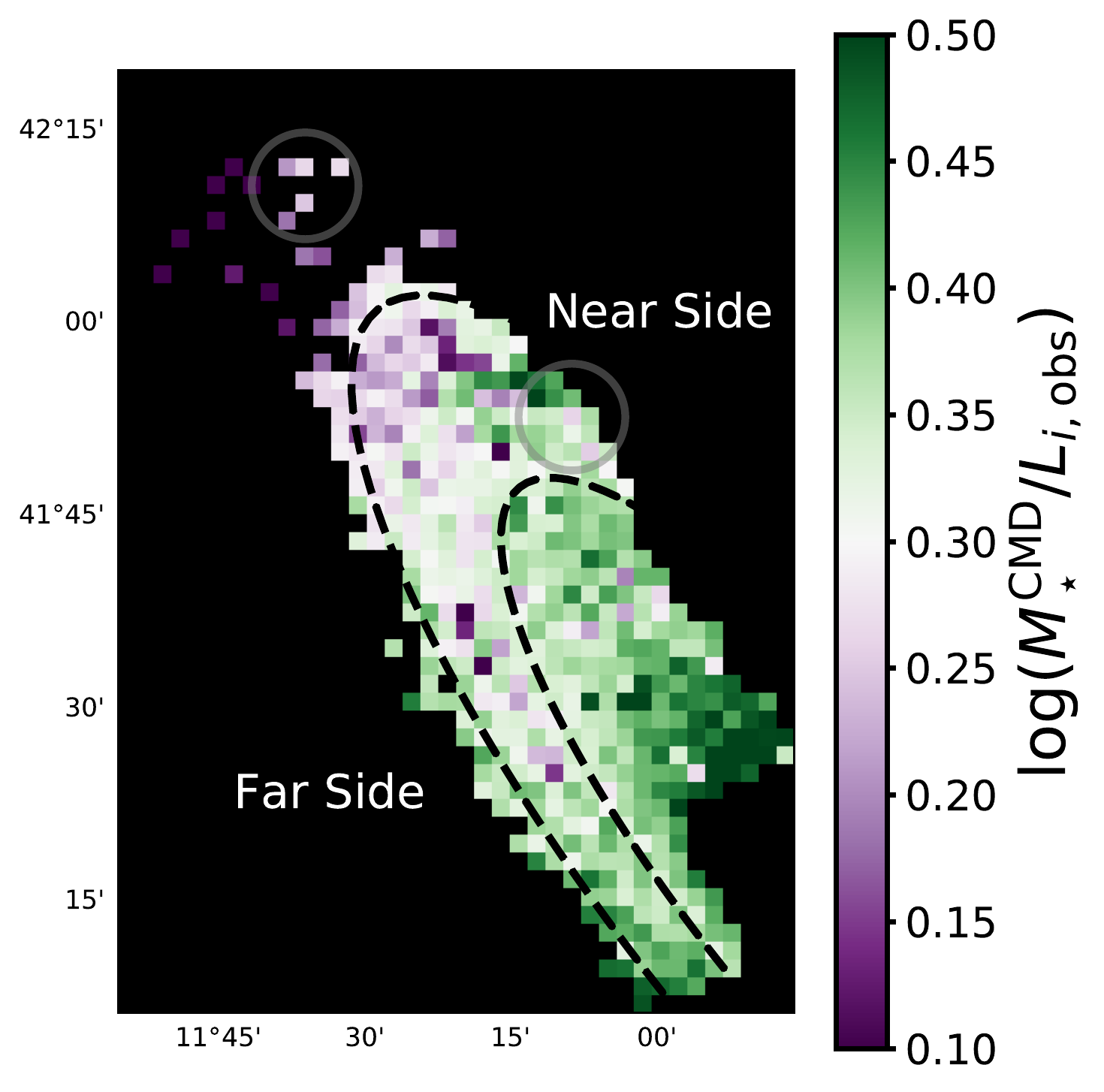}
\endminipage\hfill
\minipage{0.5\textwidth}
  \includegraphics[width=\linewidth]{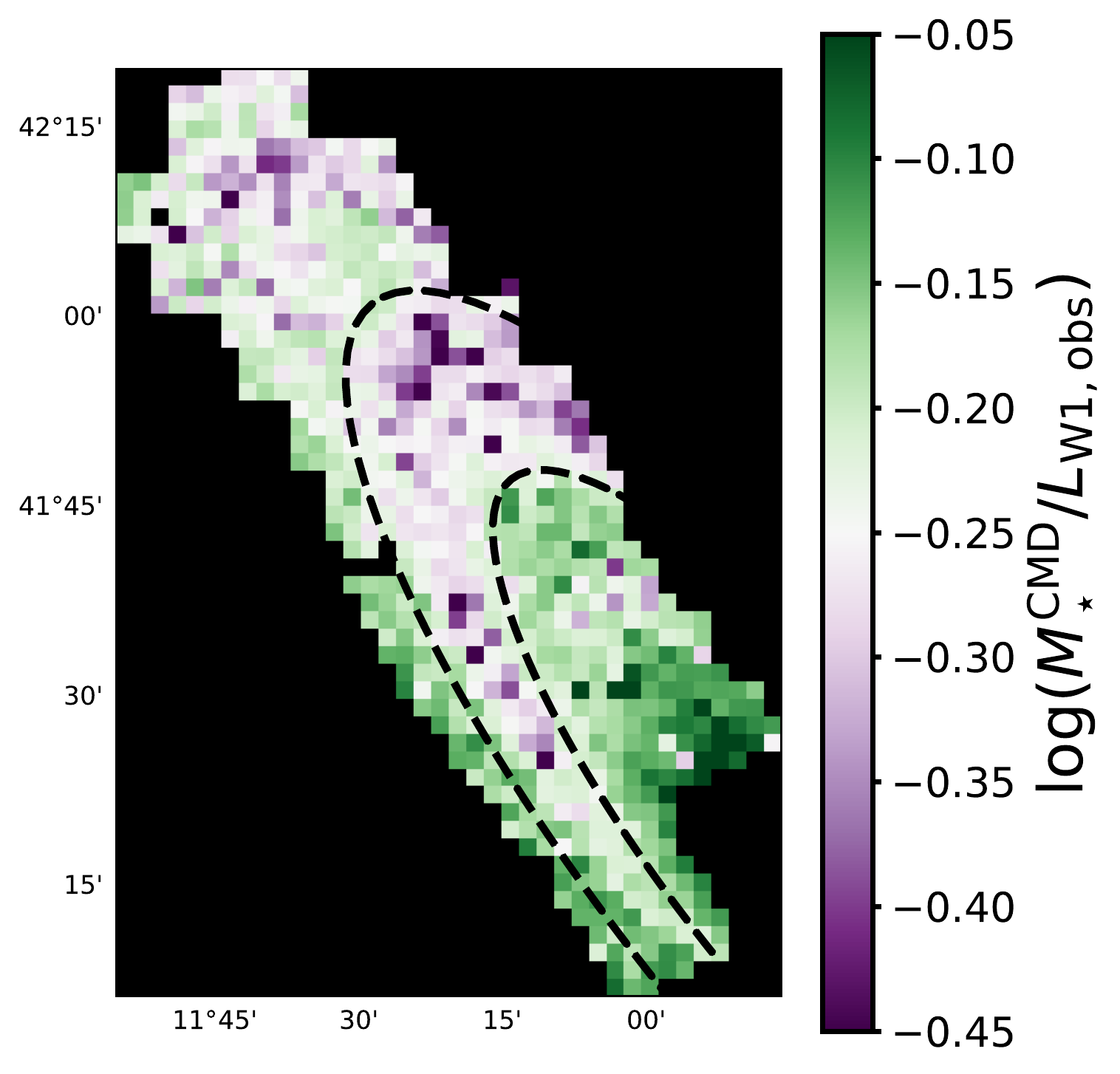}
\endminipage\hfill
\caption{\textbf{\textsc{Maps of $\bm{M^\mathrm{CMD}_\star/L_{i, \mathrm{obs}}}$ and $\bm{M^\mathrm{CMD}_\star/L_\mathrm{W1, obs}}$.}} Left: a map of the PHAT footprint at the resolution of the SFH pixels, color-coded by \logmlicmd{}. The colorbar is centered near the median \logmlicmd{} so that typical SFH pixels appear light, lower \mlicmd{} regions are purple, and higher \mlicmd{} regions are green. Again, black dashed lines approximately bounding the 10 kpc star-forming ring are shown for reference and grey circles indicate two star-forming regions discussed in the text. Right: same as the left panel, but color-coded by \logmlwcmd{}. The colorbar is centered near the median \logmlwcmd{} and spans the same range (0.4 dex) as the colorbar in the left panel, highlighting the similar spread in \mlcmd{} in the optical and IR filters.
\label{fig:ml_cmd}}
\end{centering}
\end{figure*}

\subsection{Maps of \mlcmd{} Ratios}
\label{sec:ml_cmd}

\begin{table}
\caption{Adopted Absolute Magnitudes of the Sun} 
\label{tab:abs_mags}
\begin{center}
\begin{tabular}{lcc}
Filter & Central $\lambda$ ($\mu$m) & $M_\odot$ (mag)
\\ \hline
SDSS $g$ & 0.477 & 5.11 \\
SDSS $i$ & 0.763 & 4.53 \\
\wise{} W1 & 3.4 & 5.91 \\
\wise{} W2 & 4.6 & 6.57 \\
\end{tabular}
\end{center}
\tablecomments{All magnitudes are from \citet{willmer18} and reported in the AB system.}
\end{table}

Here, we use resolution-matched surface brightness and $\Sigma^\mathrm{CMD}_\star$ maps to calculate \mlcmd{}; again, we adopt this notation to differentiate our \mlcmd{} in M31 from standard SPS-based \ml{} inference. The solar absolute magnitudes in Table~\ref{tab:abs_mags} are adopted to convert the brightness maps from magnitudes to physical luminosity units. We construct \mlcmd{} maps by dividing the stellar mass map shown in Figure~\ref{fig:mass_map} by the luminosity map in each filter.

Figure~\ref{fig:ml_cmd} shows maps of \logmlicmd{} (left) and \logmlwcmd{} (right) in the M31 disk. We focus on these filters (instead of $g$ and W2) because \ml{} in the $i$ and W1 bands are more commonly used in CMLRs in the literature (Section~\ref{sec:lit_cmlrs}). In both panels, the color bars are centered close to the median \logmlcmd{} in that filter ($\left< \log{(M^\mathrm{CMD}/L_{i, \mathrm{obs}})} \right>_\mathrm{med} = 0.35$ and $\left< \log{(M^\mathrm{CMD}/L_\mathrm{W1, obs})} \right>_\mathrm{med} = -0.22$) and span a range of 0.4 dex. \mlcmd{} in the green pixels are higher than the median (light pixels), while \mlcmd{} in the purple pixels are lower. The dashed black lines shown for reference at 8 and 13 kpc roughly bound the 10 kpc star-forming ring.

Again, we refer to the predicted time evolution of \ml{} for dust-free stellar populations shown in the bottom row of Figure~\ref{fig:sps_models} to aid in our discussion of the broad morphological features of the \mlcmd{} maps. The left panel of Figure~\ref{fig:ml_cmd} shows that \logmlicmd{} decreases with radius systematically, which would again be consistent with a radial decrease in stellar age and/or metallicity (bottom left panel of Figure~\ref{fig:sps_models}). The outer disk's  low \mlicmd{} is consistent with its blue colors in Figure~\ref{fig:colors}. On the other hand, the star-forming region at the near side of the disk in the 10 kpc ring has high \mlicmd{}, despite its blue \gi{}. This combination is not readily explained by the stellar evolutionary tracks in Figure~\ref{fig:sps_models}, nor can it be attributed to a foreground dust screen, which would drive both redder \gi{} and higher \mli{}. In Section~\ref{sec:cmlr_dust}, we show that star-dust geometry (varying \fred{}) may explain these observations. 

In both the optical and mid-IR \mlcmd{} maps in Figure~\ref{fig:ml_cmd}, two regions tend to have the highest \mlcmd{}: the inner disk and the SFH pixels that lie in the low surface brightness regions along the far side of the disk (near the bottom right corner of the map). The stellar evolutionary models in Figure~\ref{fig:sps_models} suggest that old ages are required to drive the highest \ml{} in both the optical and mid-IR. This is consistent with expectations for the inner disk, which is thought to have formed and enriched early (e.g., \citetalias{williams17}; \citealt{saglia18}). However, this is not necessarily expected for low surface brightness regions in the outer disk. It is possible that those SFH pixels may be probing the stellar halo \citep{williams12}, which could explain the high \mlcmd{} if halo stars are typically old. Indeed, the oldest mass-weighted mean stellar ages calculated from the \citetalias{williams17} SFHs ($\sim$9.5-10 Gyr) are found in this region.

In the right panel of Figure~\ref{fig:ml_cmd}, regions with lower \mlwcmd{} quite cleanly trace the star-forming and dusty rings in the M31 disk. Moreover, they appear to trace the star-forming rings more clearly than the \mlicmd{} variations in the optical, despite the widely held idea that the NIR/mid-IR is less sensitive to recent star formation. The lowest \mlwcmd{} in the PHAT footprint are consistent with predictions for few Gyr old stellar populations (bottom left panel of Figure~\ref{fig:sps_models}). We discuss the effects of recent SFH and dust on our \mlcmd{} measurements in detail in Section~\ref{sec:cmlr_scatter} below.


\section{Optical and Mid-IR Color-\ml{} Relations\label{sec:cmlrs}} 

\subsection{Characterizing the CMLRs in M31}
\label{sec:m31_cmlrs}

\begin{figure*}[!ht]
\begin{centering}
\minipage{0.5\textwidth}
  \includegraphics[width=\linewidth]{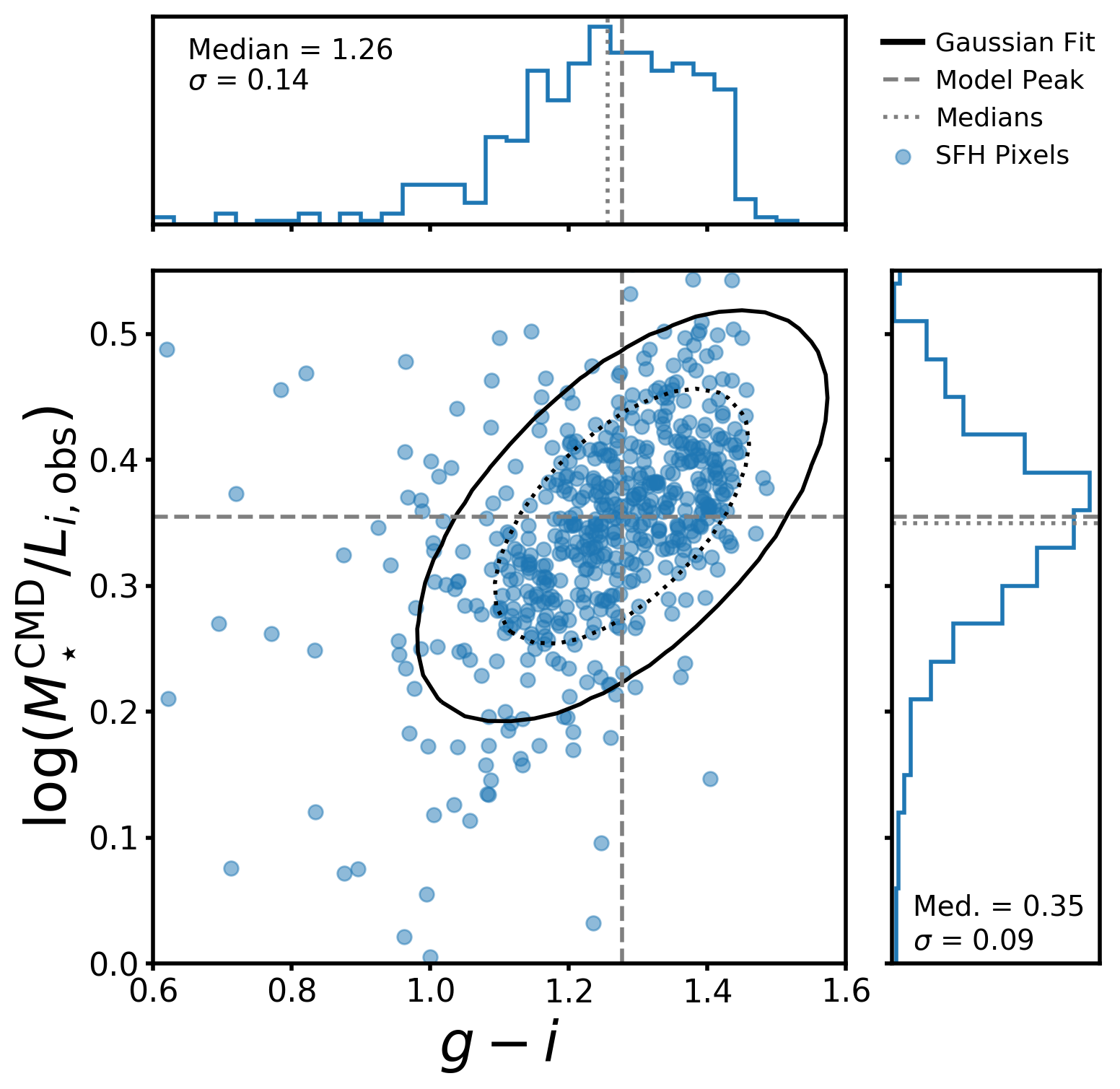}
\endminipage\hfill
\minipage{0.5\textwidth}
  \includegraphics[width=\linewidth]{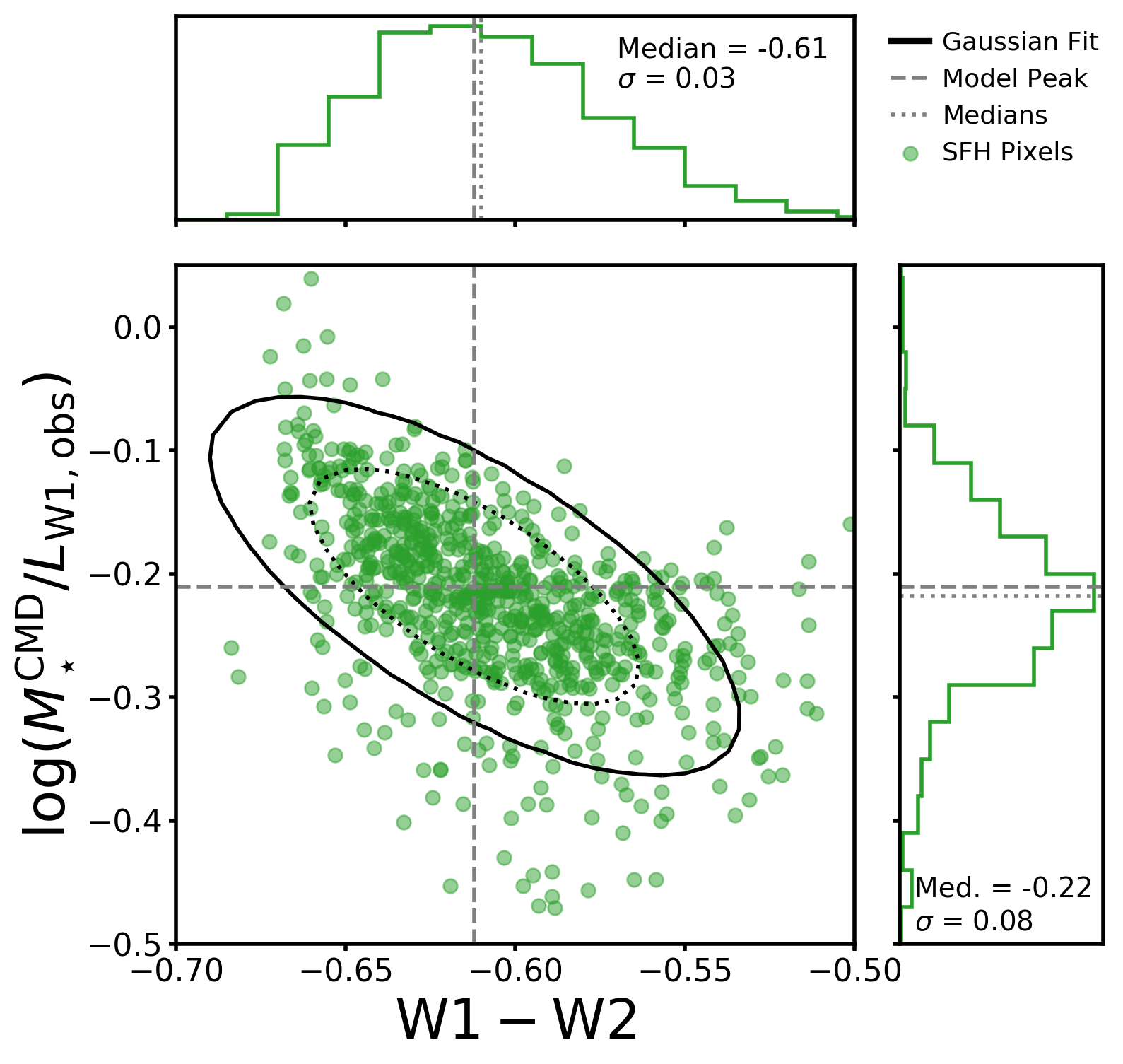}
\endminipage\hfill
\caption{\textbf{\textsc{Color--$\bm{M^\mathrm{CMD}_\star/L_\mathrm{obs}}$ relations in M31.}} Left: \logmli{} vs. \gi{}, where each SFH pixel passing our photometric quality cut is shown as a blue point, and the blue histograms to the top and right show the marginal distributions of \gi{} and \logmli{}, respectively. The black ellipses show the best-fit Gaussian model to the data, where the dotted and solid lines enclose 68\% and 95\% of the distribution, respectively. The location of the Gaussian model peak is shown by the dashed grey lines. The medians of the marginal distributions are shown as dotted grey lines, and the median and standard deviation of each marginal distribution are annotated in the histogram panels. Right: \logmlw{} vs. \wcolor{}, where all SFH pixels are shown as green points and the marginal distributions in each quantity are shown as green lines. All SFH pixels are included due to the higher quality \wise{} photometry, and all lines are analogous to those in the left panel. The vertical axes span the same range (0.55 dex), emphasizing the similar widths of the optical and mid-IR \logmlcmd{} histograms.
\label{fig:cmlr_distributions}}
\end{centering}
\end{figure*}

\begin{table*}
\caption{Parameters of Best-Fit 2D Gaussian Models}
\label{tab:params}
\begin{center}
\begin{tabular}{lcccccc}
\multicolumn{7}{c}{\bf Optical} \\
Fit Description & Slope & Intercept & Scatter about CMLR & $N_\mathrm{pixels}$ & \gi{} Peak & \logmlicmd{} Peak 
\\ \hline 
All SFH Pixels & 0.39 & -0.15 & 0.051 & 547 & 1.28 & 0.36 \\
$\log$\ssfr{} $< -11.3$ & 0.34 & -0.04 & 0.051 & 199 & 1.30 & 0.39 \\
$\log$\ssfr{} $> -11.3$ & 0.38 & -0.14 & 0.045 & 348 & 1.26 & 0.33 \\
SF, \av{} $< 1.0$ & 0.47 & -0.28 & 0.045 & 178 & 1.27 & 0.32 \\
SF, \av{} $> 1.0$ & 0.33 & -0.07 & 0.040 & 170 & 1.24 & 0.34 \\
$0.45 < f_\mathrm{red} < 0.55$ & 0.73 & -0.53 & 0.031 & 69 & 1.21 & 0.35 \\
\multicolumn{7}{c}{\bf Mid-IR} \\
Fit Description & Slope & Intercept & Scatter about CMLR & $N_\mathrm{pixels}$ & \wcolor{} Peak & \logmlw{} Peak
\\ \hline 
All SFH Pixels & -2.51 & -1.75 & 0.068 & 778 & -0.61 & -0.21 \\
$\log$\ssfr{} $< -11.3$ & -3.05 & -2.09 & 0.048 & 276 & -0.63 & -0.17 \\
$\log$\ssfr{} $> -11.3$ & -2.52 & -1.75 & 0.049 & 502 & -0.60 & -0.24 \\
SF, \av{} $< 0.8$ & -4.42 & -2.92 & 0.042 & 226 & -0.61 & -0.23 \\
SF, \av{} $> 0.8$ & -2.08 & -1.48 & 0.051 & 276 & -0.59 & -0.25 \\
\end{tabular}
\end{center}
\end{table*}

We begin our quantitative analysis of the \mlcmd{} maps in Figure~\ref{fig:ml_cmd} by comparing them to the color maps in Figure~\ref{fig:colors}.  Figure~\ref{fig:cmlr_distributions} presents the resulting relation between \logmlicmd{} and observed \gi{} (left, blue points) and between \logmlwcmd{} and observed \wcolor{} (right, green points), where each point represents a single SFH pixel. The histograms to the top and right of each scatter plot show the marginal distributions of color and \logmlcmd{}, respectively. The number of SFH pixels in each plot is different due to the photometric quality cut imposed on the SDSS data: 547 and 778 SFH pixels are shown in the left and right panels, respectively.

Clear correlations between color and \mlcmd{} are apparent in both the optical and mid-IR, as one might expect from the similar morphologies in Figures~\ref{fig:colors} and \ref{fig:ml_cmd}. We quantify these CMLRs and the scatter about them by fitting two-dimensional Gaussian models to the data. This approach has two advantages over a simple linear fit: (1) it is less sensitive to outliers, since the Gaussian model captures the covariance of the data in the most densely populated regions of color-\mlcmd{} space; and (2) the model quantifies the scatter in the distribution about the best-fit CMLR. The free parameters in our Gaussian model are the means (i.e., the location of the peak) and standard deviations along the $x$ and $y$ directions, the angle at which the direction of largest variation is rotated with respect to the horizontal, and the amplitude. The best-fit CMLR is the eigenvector pointing along the direction of greatest variance in the Gaussian model. The parameters describing the best-fit models for each sample of SFH pixels considered in this paper are given in Table~\ref{tab:params}.

It is standard practice in the CMLR literature to fit a line to the data, in contrast to the two-dimensional Gaussian model we have adopted here. However, we are hampered by the lack of published uncertainties on the SDSS surface brightness maps (discussed in Section~\ref{sec:sdss}). It is well-known that simple least-squares minimization can bias linear fits to data \citep{hogg10, cluver14}, and more sophisticated techniques typically require uncertainty information. As a check on our adopted method, we assigned our \gi{} and \mlicmd{} arbitrary, constant uncertainties and used an MCMC procedure to fit a mixture model of a line plus background level to handle outliers. We obtained similar slopes and intercepts to those we report here for our 2D Gaussian models, so we are confident that our modeling approach provides a realistic description of trends in the data.

In Figure~\ref{fig:cmlr_distributions}, the black ellipses show contours of the best-fit Gaussian models, where the dotted and solid lines enclose 68\% and 95\% of the model density, respectively. The dashed grey lines show the peak of the 2D Gaussian model fit in \mlcmd{} and color. The peak location is close to the median of the data (shown in the histogram panels as dotted grey lines for comparison). The agreement between the medians of the data and the central location of the Gaussian models indicates that the model appropriately captures the key features of the empirical relationships between color and \mlcmd{}.

The positive correlation between \mlicmd{} and observed \gi{} that we find in M31 is expected from stellar evolutionary models (left column of Figure~\ref{fig:sps_models}). Younger and lower-metallicity stellar populations have bluer \gi{} and lower \mli{}. Dust attenuation by a uniform foreground screen both decreases brightness and reddens optical colors. At mid-IR wavelengths, however, stellar evolutionary models do not agree on a predicted CMLR (right column of Figure~\ref{fig:sps_models}) due to the larger contribution of uncertain stellar evolutionary phases to the integrated light. We discuss the negative correlation between \mlwcmd{} and \wcolor{} we find in M31 in the context of other ``semi-empirical'' studies in the literature in Section~\ref{sec:lit_cmlrs}.

The scatter about these empirical CMLRs is due to a combination of measurement uncertainties and intrinsic scatter due to stellar population variations and dust. Therefore, the scatter about the best-fit CMLRs can be thought of as upper limits on the intrinsic scatter for the specific case of the M31 disk. Within the PHAT footprint, the signal-to-noise of the \wise{} photometry is typically 100 or better (formal uncertainty only, not including systematic uncertainty in the background estimation; \citealt{lang14}). We do not have measurement uncertainties for the SDSS mosaics, which are clearly dominated by sky subtraction systematics and not photon counting. Our attempt to limit the effect of these systematics by imposing a signal-to-noise ratio of at least 5 should leave the photometry good to within $\sim$20\%. However, some SFH pixels in the upper left corner of the scatter plot in the left panel of Figure~\ref{fig:cmlr_distributions} (\gi{}$\,\lesssim 0.9$ and \logmli{}$\,\gtrsim 0.3$) are obvious outliers. These SFH pixels tend to lie along the low surface brightness, far side of the disk, and have anomalously high $g$-band luminosities, enabling them to pass our quality cut. 

Typical uncertainties (random + systematic) on the formed \mstar{} from \citetalias{williams17} are $\sim20$\%. We can assess the achieved level of uncertainty by looking at the smoothness of the \mstar{} map. M31's stellar population is dominated by old stars, which should be well-mixed, and thus should show smooth variation across adjacent SFH pixels. No smoothness was imposed on the \mstar{} formed in adjacent SFH pixels in their CMD modeling, and thus small deviations from smoothness can be seen in the \mstar{} map shown in Figure~\ref{fig:mass_map}. Some of the scatter to low \ml{} can be attributed to regions where the CMD-based \mstar{} is more than 20\% lower than the \mstar{} in neighboring SFH pixels. However, scatter in the CMD-based \mstar{} does not dominate the total scatter; flux measurement uncertainties and/or intrinsic scatter are more important.

The optical and mid-IR \mlcmd{} distributions have remarkably similar standard deviations: 0.09 dex in \logmlicmd{}, and 0.08 dex in \logmlwcmd{}. It is common to think of stellar population variations as driving larger \ml{} variations in the optical than in near/mid-IR filters. However, the presence of young stellar populations and dust emission can strongly affect both NIR and mid-IR \ml{} \citep[e.g.,][]{melbourne12, querejeta15}. The measurements presented here demonstrate that the variations in optical and mid-IR \mlcmd{} are comparable in M31, a late-type massive spiral with low-level, ongoing star formation. 

We now compare the scatter about the optical and mid-IR best-fit CMLRs in M31 (reported in the right column of Table~\ref{tab:params}) to the observed scatter in the \logmlcmd{} distributions. If the scatter about the CMLR is lower, that indicates that using color information enables a more precise \mstar{} estimate. The Gaussian model scatter perpendicular to the best-fit optical CMLR is 0.05 dex, a reduction from the 0.09 dex spread in the observed \logmlicmd{} distribution. This is consistent with a rich literature showing that color information improves \mstar{} estimates using optical data \citep[e.g.,][and many others]{bell01}. The scatter about the mid-IR CMLR is 0.07 dex, a much smaller reduction compared to the 0.08 dex spread in the observed \logmlwcmd{} distribution.However, this is a $\sim 25\%$ reduction in variance, compared to a $\sim 70\%$ reduction in variance for the optical CMLR. Knowledge of \wcolor{} does provide information about \mlwcmd{}, but the smaller reduction in variance suggests that using a CMLR in the mid-IR may only modestly improve \mstar{} estimates above adopting a constant \mlw{}. We discuss this point further in Section~\ref{sec:lessons}.

\subsection{Comparison to Literature CMLRs}
\label{sec:lit_cmlrs}

Here, we compare the CMLRs in M31 to CMLRs previously reported in the literature for SDSS and \wise{} filters. The \mcmd{} map that we use to construct the M31 CMLRs was inferred from modeling CMDs of resolved stellar populations, and is therefore independent of the uncertainty introduced by the treatment of SFH, dust, and evolved stellar populations in SPS modeling of integrated light. The M31 CMLRs provide a critical check on the ability of SPS model-based CMLRs to capture the causes of \ml{} and color variations in real galaxies. 

\subsubsection{Optical: \ml$_i$ vs. \gi{}}

\begin{figure*}[!ht]
\begin{centering}
\minipage{0.5\textwidth}
  \includegraphics[width=\linewidth]{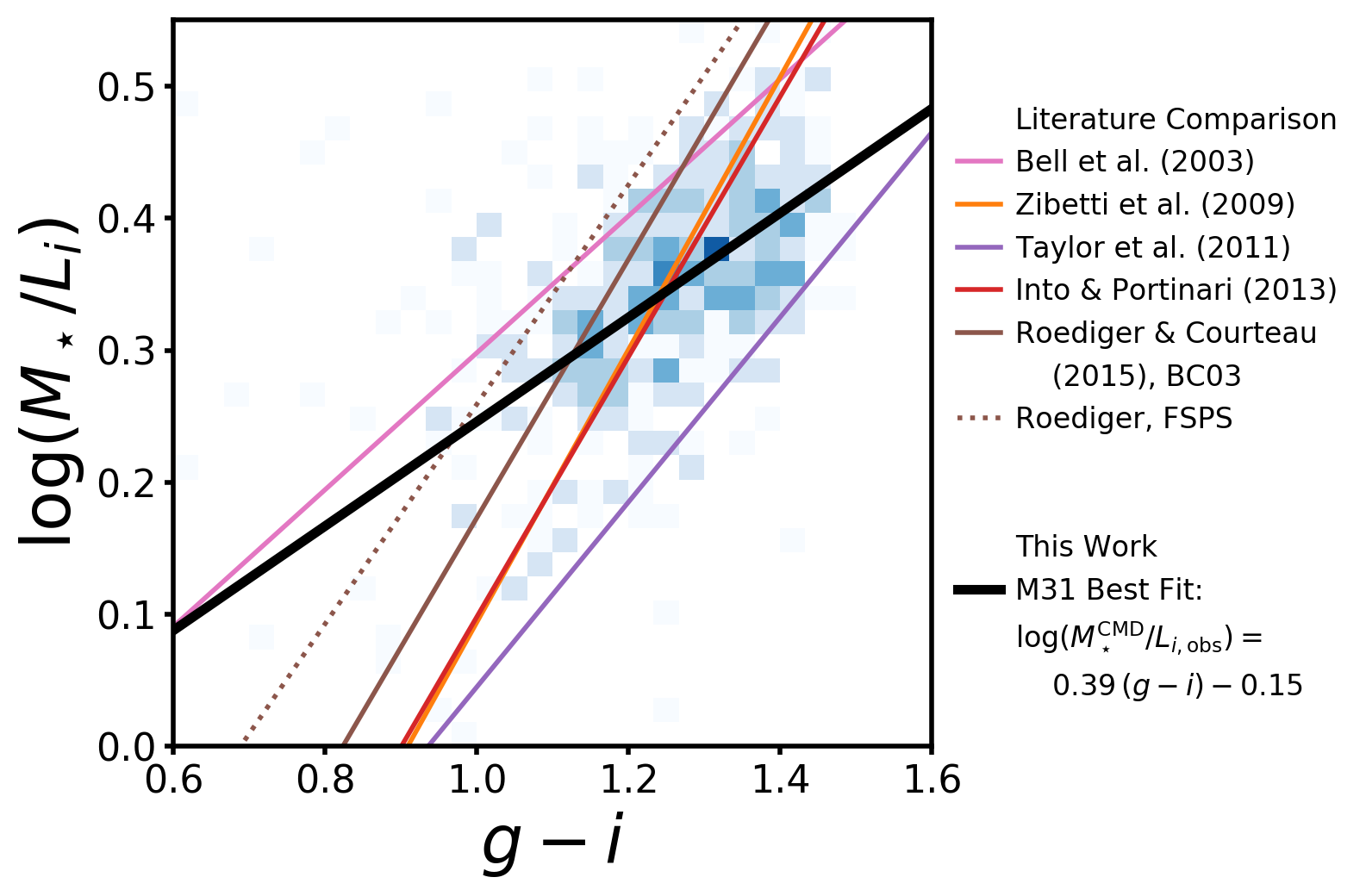}
\endminipage\hfill
\minipage{0.5\textwidth}
  \includegraphics[width=\linewidth]{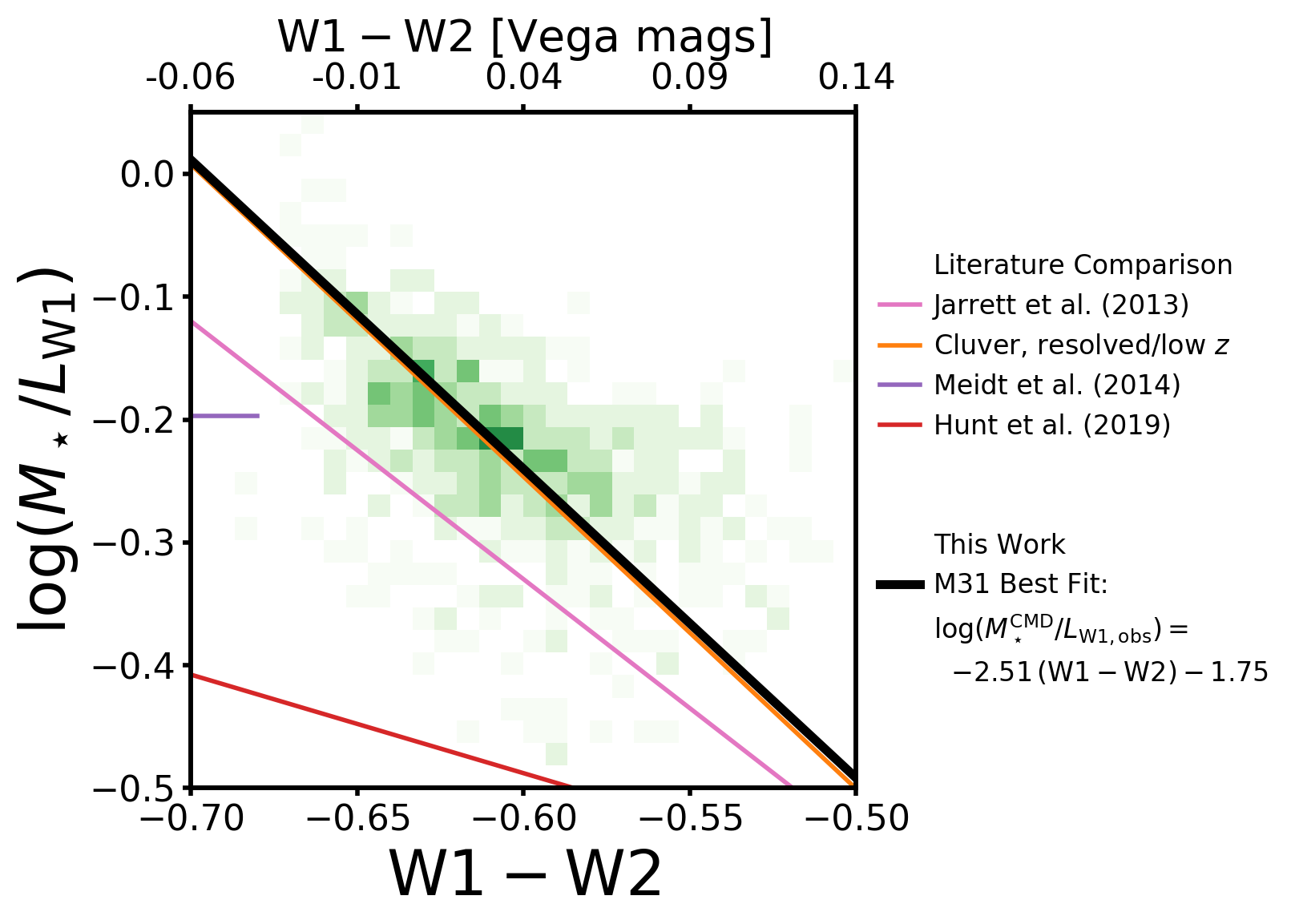}
\endminipage\hfill
\caption{\textsc{\textbf{Comparing the M31 Color--$\bm{M^\mathrm{CMD}_\star/L_\mathrm{obs}}$ Relations to Literature CMLRs.}} Left: a two-dimensional histogram in \gi{} and \logmlicmd{} is shown in blue, where more SFH pixels reside in darker bins. The best-fit CMLR for M31 is shown as the solid black line, while the colored lines show SPS model-based CMLRs reported in various papers (citations in the legend). Right: a two-dimensional histogram in \mlwcmd{} and \wcolor{} is shown in green, where darker bins contain more SFH pixels. The best-fit CMLR for M31 is shown as the black line, and the orange, pink, and red lines show semi-empirical CMLRs for galaxies detected by \wise{}. The purple line shows a constant, SPS model-based \logml{}$_{3.6 \mu m}$ advocated for use with old stellar populations only. The various CMLRs span a range of both slope and normalization (despite all being scaled to a common \citealt{kroupa01} IMF) in both the optical and mid-IR. 
\label{fig:lit_cmlrs}}
\end{centering}
\end{figure*}

In the left panel of Figure~\ref{fig:lit_cmlrs}, we show several \logmli{} vs. \gi{} CMLRs from the literature as colored lines, where all relations have been scaled to a common \citet{kroupa01} IMF (the constant offsets we use are given in Table~\ref{tab:imfs}). Our M31 CMLR is shown as the black line, calculated as the eigenvector along the direction of maximum variance in our best-fit Gaussian model. A 2D histogram of the data from the left panel of Figure~\ref{fig:cmlr_distributions} is shown as the blue shading, where darker blue indicates that more SFH pixels fall into a given region of color-\ml{} space.

Each literature \ml{}$_i$ vs. \gi{} relation is fit to a different library of SPS model predictions. These libraries are constructed by varying parameters describing the SFH, dust, and metallicity, all of which impact color and \ml{}, in such a way that aims to capture the range present in the real galaxy population. The variety of slopes and normalizations of these literature CMLRs is driven by the different choices made in generating the SPS models to which each relation was fit, including the parameterization/complexity of the SFH, the treatment of dust (or if dust is included at all), and the adoption of different isochrone sets and treatment of uncertain phases of stellar evolution \citep[e.g,][]{gallazzi09, pforr12, roediger15}. We discuss these issues further in Section~\ref{sec:discussion}.

Clearly, the M31 \mlicmd{} vs. \gi{} relation has a flatter slope than all of the theoretical CMLRs, but is closest to the calibrations of \citet{bell03} and \citet{taylor11}. The slope of SPS-based CMLRs is strongly influenced by the priors on the SFH, and in particular on the allowed strength of recent bursts. Model libraries that include more active recent SFHs tend to have lower \ml{} at bluer optical colors than libraries with more quiescent SFH priors \citep[e.g.,][]{de-jong07, roediger15}. The M31 CMLR is more consistent with the SPS libraries that have fewer recent bursts of star formation.

Another possible driver of the optical CMLR slope is whether purely theoretical SPS libraries or models fit to real data are used. Both \citet{bell03} and \citet{taylor11} used their model SPS libraries to fit a large sample of observed optical-NIR galaxy SEDs, and then fit CMLRs to the subset of models that described the population real galaxies. \citet{taylor11} point out that when no observations are incorporated in the best-fit CMLR, the slope and normalization are dictated only by the priors and choices made in assembling the library of SPS models. The ``semi-empirical'' methods may be at least partially responsible for the flatter CMLR slopes than were found for the CMLRs fit to purely theoretical SPS model libraries by \citet{zibetti09}, \citet{into13}, and \citet{roediger15}. 

\subsubsection{Mid-IR: \ml$_{\mathrm{W}1}$ vs. \wcolor{}}

The right panel of Figure~\ref{fig:lit_cmlrs} shows \logmlw{} vs. \wcolor{} CMLRs from the literature as solid colored lines (all scaled to a common \citealt{kroupa01} IMF), while our M31 CMLR is shown as the solid black line. A 2D histogram of the data from the right panel of  Figure~\ref{fig:cmlr_distributions} is shown as the green shading, where darker green bins contain more SFH pixels.

We compare to several ``semi-empirical'' CMLRs in the mid-IR, where \mstar{} is inferred using SPS-based techniques but the CMLR is constructed with observed, rather than modeled, W1 luminosity and \wcolor{}. \citet{jarrett13} and \citet{cluver14} both constructed their relations by using literature CMLRs to estimate \mstar{} from photometry (NIR and optical, respectively), then dividing their \mstar{} by observed W1 luminosity for the same galaxies. Similarly, \citet{hunt19} inferred \mstar{} by fitting UV through far-IR photometry for a sample of  star-forming galaxies from the KINGFISH survey \citep{dale17} with the CIGALE code \citep{noll09}, which simultaneously models the stellar emission, dust attenuation, and reprocessed dust emission. All of these relations are dependent on SPS models as the underlying theoretical tool that calibrated the \mstar{} estimates, but the colors and luminosities are observed quantities (not SPS-based). 

We also compare to a theoretical prediction from \citet{meidt14}, who used SPS models calibrated to the observed NIR and mid-IR colors of old, giant stars to show that a constant \ml{}$_{3.6}$ can be applied to \textit{old stellar populations only} to estimate \mstar{} within $\sim0.1$ dex. We assume that \mlw{} and \ml{}$_{3.6}$ are equivalent \citep[which is true to within a few percent;][]{jarrett13}, and only show the \citet{meidt14} constant \ml{}$_{3.6}$ for blue \wcolor{} colors that are expected for old stellar populations. As demonstrated in the right panels of Figure~\ref{fig:sps_models}, theoretical predictions for mid-IR colors and \ml{} vary widely among different stellar evolutionary tracks, so SPS-based mid-IR CMLRs are not considered reliable for estimating \mstar{} in young/intermediate-age stellar populations.

The slope of the mid-IR CMLR in M31 is in good agreement with the CMLR slopes reported by \citet{jarrett13} and \citet{cluver14}, and remarkably, the relation from the latter paper fit to nearby, resolved galaxies is almost the same as the M31 CMLR. The \citet{hunt19} relation has a substantially lower \mlw{} normalization and shallower slope, possibly attributable to the fact that many galaxies in their sample are highly star-forming; we discuss this discrepancy further in Section~\ref{sec:mass_uncertainty}. The \citet{meidt14} constant \ml{}$_{3.6}$ is quite close to typical \mlwcmd{} values for blue \wcolor{} SFH pixels in M31.


\section{Drivers of Color--\mlcmd{} Relation Slope and Scatter\label{sec:cmlr_scatter}}

Here, we explore the structure in the residuals about the best-fit CMLRs fit to our observed colors and \mlcmd{} for individual SFH pixels in M31. From here forward, $\Delta$\logmlcmd{} refers to the residual in \logmlcmd{} after subtracting off the best-fit CMLR (reported in Table~\ref{tab:params}). We then analyze how the recent SFH and dust content and geometry affect the normalization and slope of CMLRs in the optical and mid-IR. 

\subsection{Effect of Recent SFH on CMLRs\label{sec:cmlr_recent_sfh}}

\begin{figure*}[!ht]
\begin{centering}
\minipage{0.5\textwidth}
  \includegraphics[width=\linewidth]{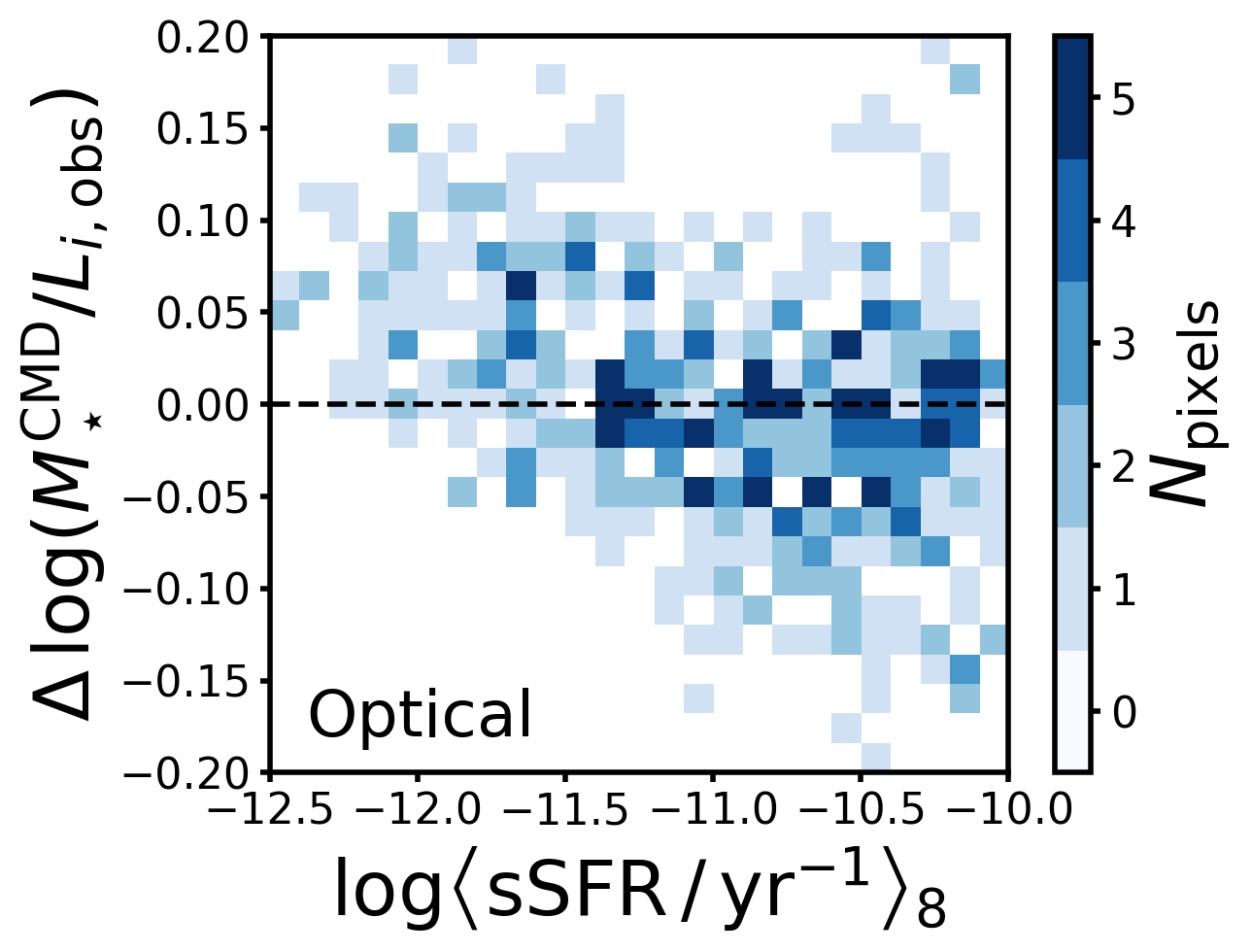}
\endminipage\hfill
\minipage{0.5\textwidth}
  \includegraphics[width=\linewidth]{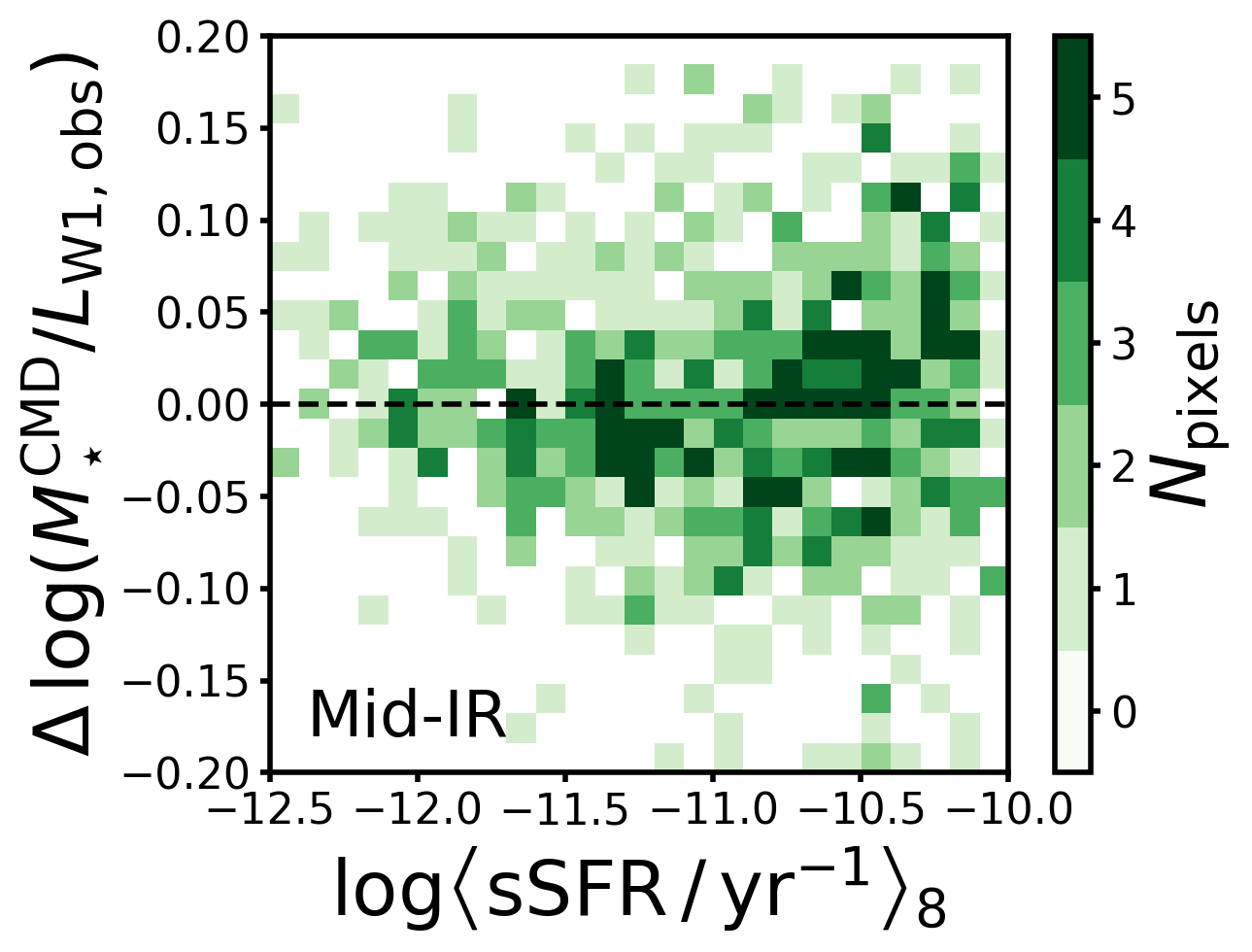}
\endminipage\hfill
\caption{\textsc{\textbf{Correlations Between SFH and M31 Color--$\bm{M^\mathrm{CMD}_\star/L_\mathrm{obs}}$ Relation Residuals.}} Number of SFH pixels in bins of $\log$\ssfr{} and offset from the best-fit CMLR, $\Delta$\logmlcmd{} (i.e., the residual after subtracting the best-fit CMLR from the \logmlcmd{} in individual SFH pixels). The left panel shows the offset from the optical \mlicmd{} vs. \gi{} relation in blue, while the right panel shows the offset from the mid-IR \mlwcmd{} vs. \wcolor{} relation in green; darker colors indicate more populated bins. The dashed horizontal line in each panel shows zero offset from the best-fit CMLR in M31.
\label{fig:cmlr_scatter_sfh}}
\end{centering}
\end{figure*}

\begin{figure*}[!ht]
\begin{centering}
  \includegraphics[width=0.75\linewidth]{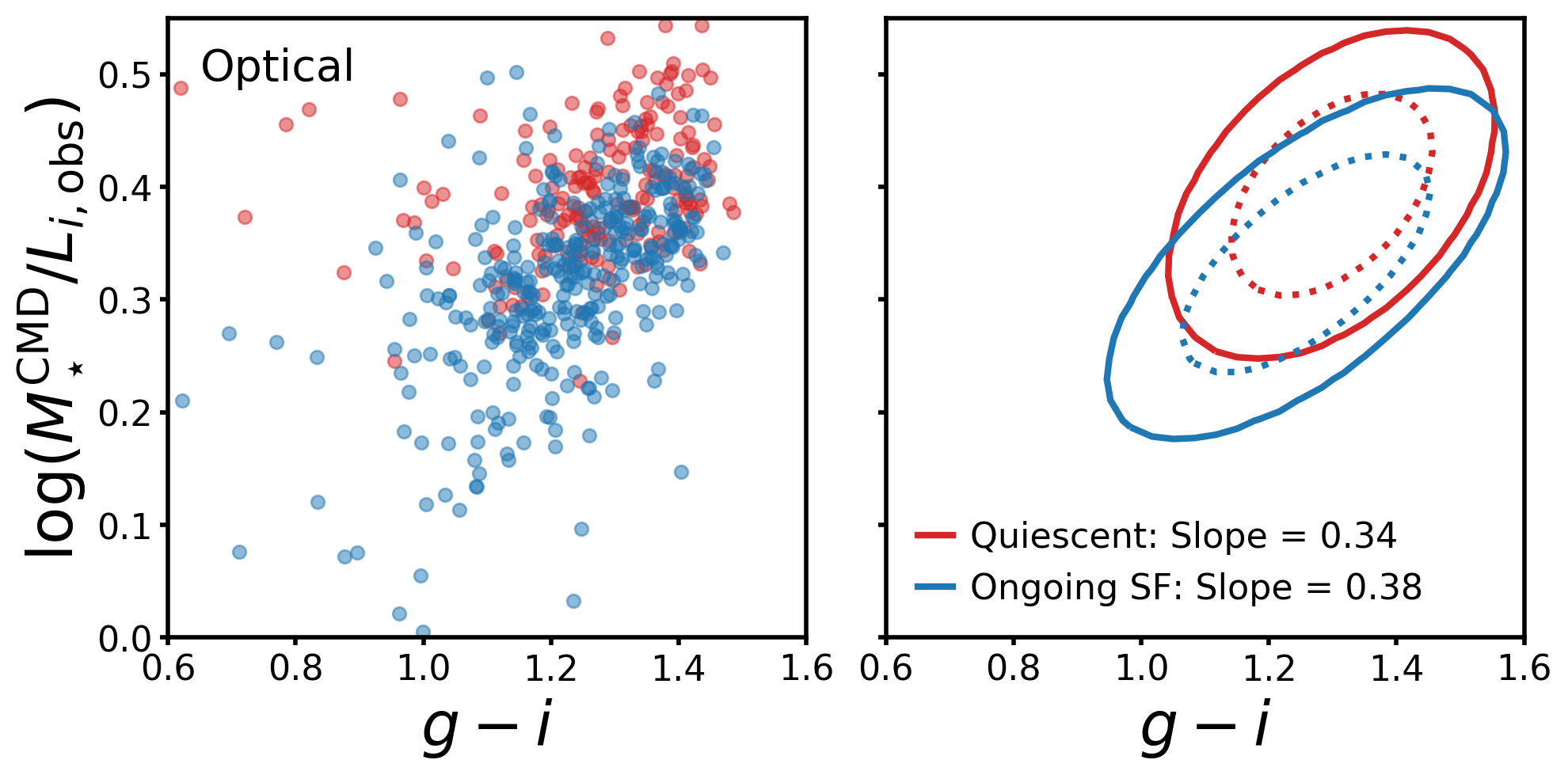}
  \includegraphics[width=0.75\linewidth]{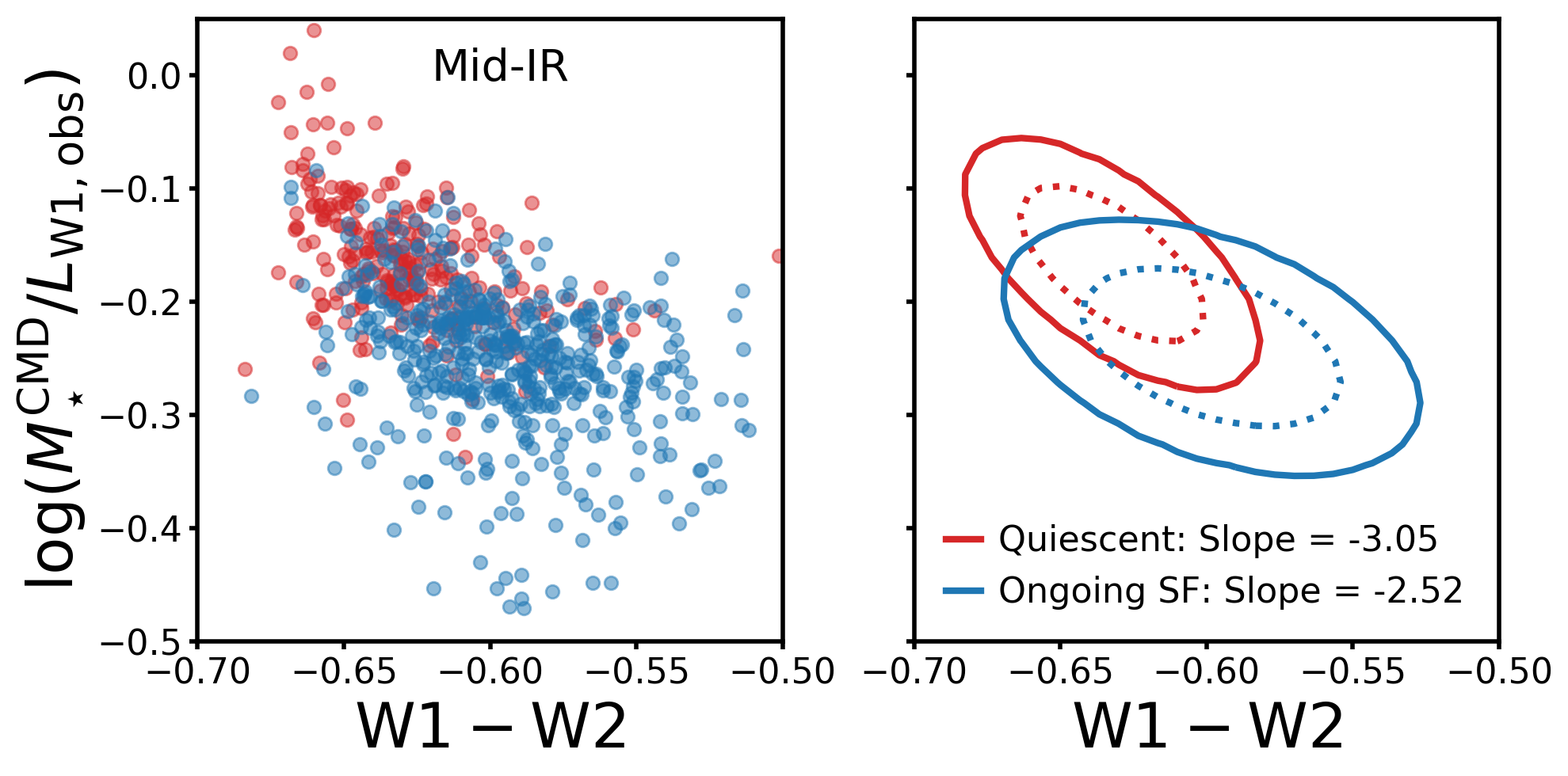}
\caption{\textbf{\textsc{Older stellar populations have higher $\bm{M^\mathrm{CMD}_\star/L_\mathrm{obs}}$ at fixed color.}} Top row: \logmlicmd{} as a function of \gi{}, where red and blue colors indicate sub-samples of the SFH pixels in M31 defined as ``quiescent'' (red, $\log$\ssfr{} $< -11.3$) or ``star-forming'' (blue). The left panel shows a scatter plot of the data, while the right panel shows Gaussian models fit to each sub-sample. The dotted and solid ellipses contain 68\% and 95\% of the model density, respectively. Bottom row: same as the top row, but for \logmlwcmd{} as a function of \wcolor{}. In both the optical and mid-IR, the quiescent and ongoing SF regions overlap in color, but the quiescent SFH pixels are offset to higher \mlcmd{} at a given color. 
\label{fig:cmlr_split_sfh}}
\end{centering}
\end{figure*}

We begin our analysis of the drivers of scatter about the M31 CMLRs by searching for correlations with the average sSFR over the past 10$^8$ yr, \ssfr{}. Figure~\ref{fig:cmlr_scatter_sfh} shows two-dimensional histograms where SFH pixels are binned in a grid of $\log$\ssfr{} and $\Delta$\logmlcmd{}. The left and right panels show the optical and mid-IR mass-to-light ratio residuals, $\Delta$\logmlicmd{} (blue) and $\Delta$\logmlwcmd{} (green), respectively, as a function of $\log$\ssfr{}. Darker colors indicate more SFH pixels per 2D bin, and the dashed black lines at $\Delta$\logmlcmd{} $=0$ indicate no offset from the best-fit CMLR.

In the optical (left panel), quiescent regions with $\log$\ssfr{} $\lesssim -11.3$ have their \mlicmd{} systematically over-estimated by the best-fit CMLR. In the mid-IR (right panel), regions with ongoing SF tend to have their \mlwcmd{} under-estimated, but with increased scatter. The distribution of SFH pixels is no longer well-approximated by a linear fit at redder \wcolor{} (see the apparent plateau around \logmlwcmd{} $\sim-0.2$ in Figure~\ref{fig:cmlr_distributions}). 

Clearly, the star-forming and quiescent regions of M31 are not simultaneously well described by the same best-fit CMLR. To understand how regions with different SFHs behave in color-\mlcmd{} space, we fit 2D Gaussian models to ``quiescent'' and ``star-forming'' (SF) subsamples of SFH pixels. We split the SFH pixels at $\log$\ssfr{} $= -11.3$, the apparent threshold below which quiescent regions have their \mlicmd{} underestimated by the best-fit optical CMLR (left panel of Figure~\ref{fig:cmlr_scatter_sfh}). 

Figure~\ref{fig:cmlr_split_sfh} presents the best-fit optical (top row) and mid-IR (bottom row) Gaussian models for the SF (blue) and quiescent (red) subsamples of SFH pixels. The data are shown in the left column and the Gaussian models fit to those data are shown in the right column. The dotted and solid ellipses contain 68\% and 95\% of the model density, and the slopes of the best-fit CMLRs (i.e., the eigenvectors along the direction of maximum variation in the Gaussian models) are reported in the legends. Again, the parameters of the best-fit model for each subset of SFH pixels are given in Table~\ref{tab:params}.

The top right panel of Figure~\ref{fig:cmlr_split_sfh} shows that the slopes of the best-fit optical CMLRs to the quiescent and SF subsamples are similar. At a fixed \gi{}, the \mlicmd{} of quiescent regions are higher than for SF regions, so varying SFH does \textit{not} necessarily move stellar populations along a linear CMLR in the optical. The SF regions span a wider range of \gi{}, encompassing the full range covered by the quiescent sample. This large overlap in \gi{} between the SF and quiescent regions highlights the fact that optical color is not necessarily a useful proxy for age. These findings are consistent with previous work showing that increased recent star formation can drive optical \ml{} low at a given color \citep[e.g.,][]{bell03, roediger15}. 

We see a similar effect in the mid-IR (bottom right panel of Figure~\ref{fig:cmlr_split_sfh}): the quiescent regions are offset to higher \mlwcmd{} than the SF regions and cover a narrower range in \wcolor{}. The relationship between \logmlwcmd{} and \wcolor{} is both steeper and tighter for the quiescent SFH pixels. While the SF regions are clearly offset to lower \mlwcmd{} than the quiescent regions, within the SF sample there is only a weak CMLR with large scatter. This result suggests that a mid-IR CMLR is of limited utility for \mstar{} inference, since the \wcolor{} does not provide much information about \mlwcmd{} (similar to the findings of \citealt{eskew12} for \spitzer{} photometry). We return to this point in Section~\ref{sec:lessons}.

\subsection{Effect of Dust on CMLRs\label{sec:cmlr_dust}}

\begin{figure*}[!ht]
\begin{centering}
\minipage{0.5\textwidth}
  \includegraphics[width=\linewidth]{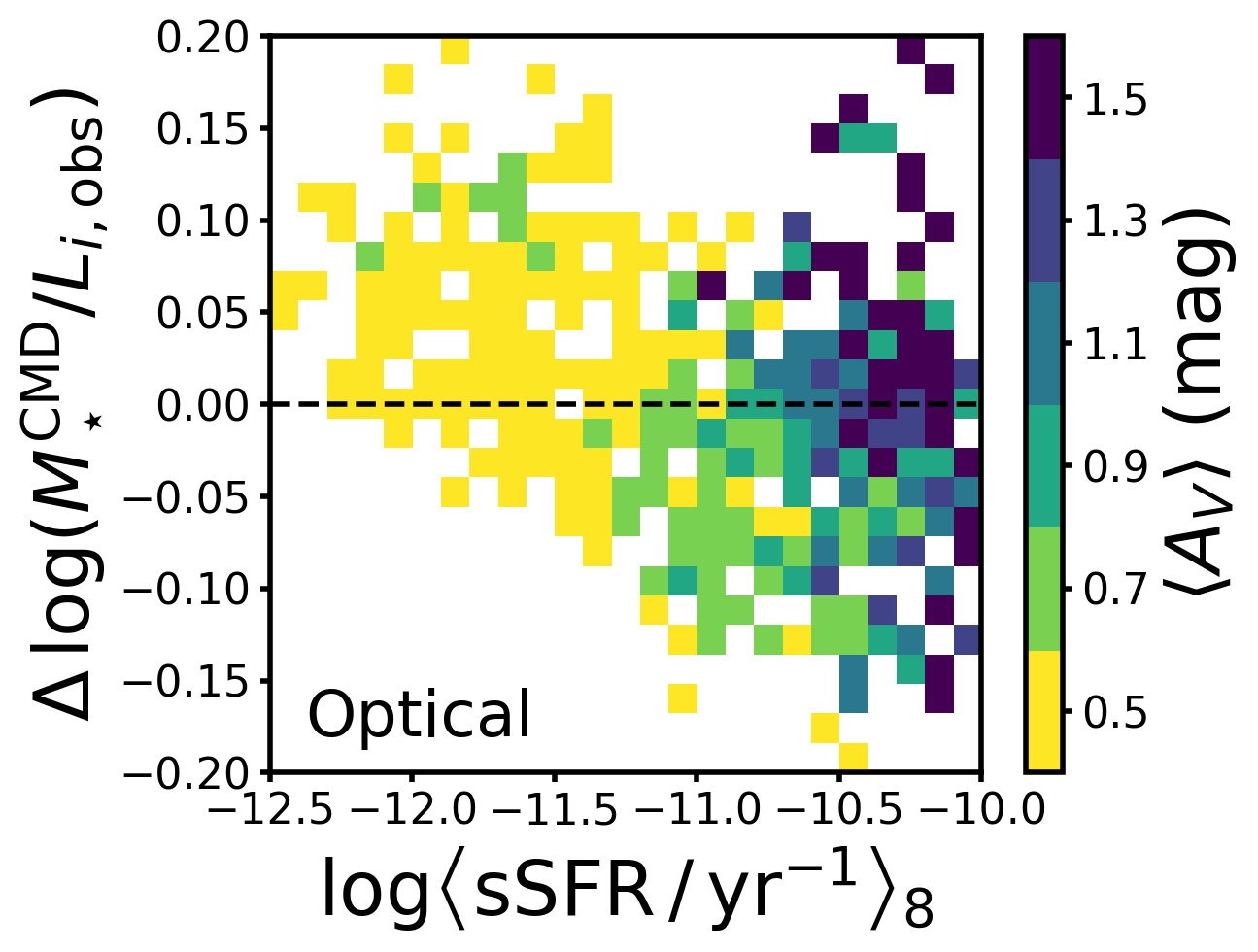}
\endminipage\hfill
\minipage{0.5\textwidth}
  \includegraphics[width=\linewidth]{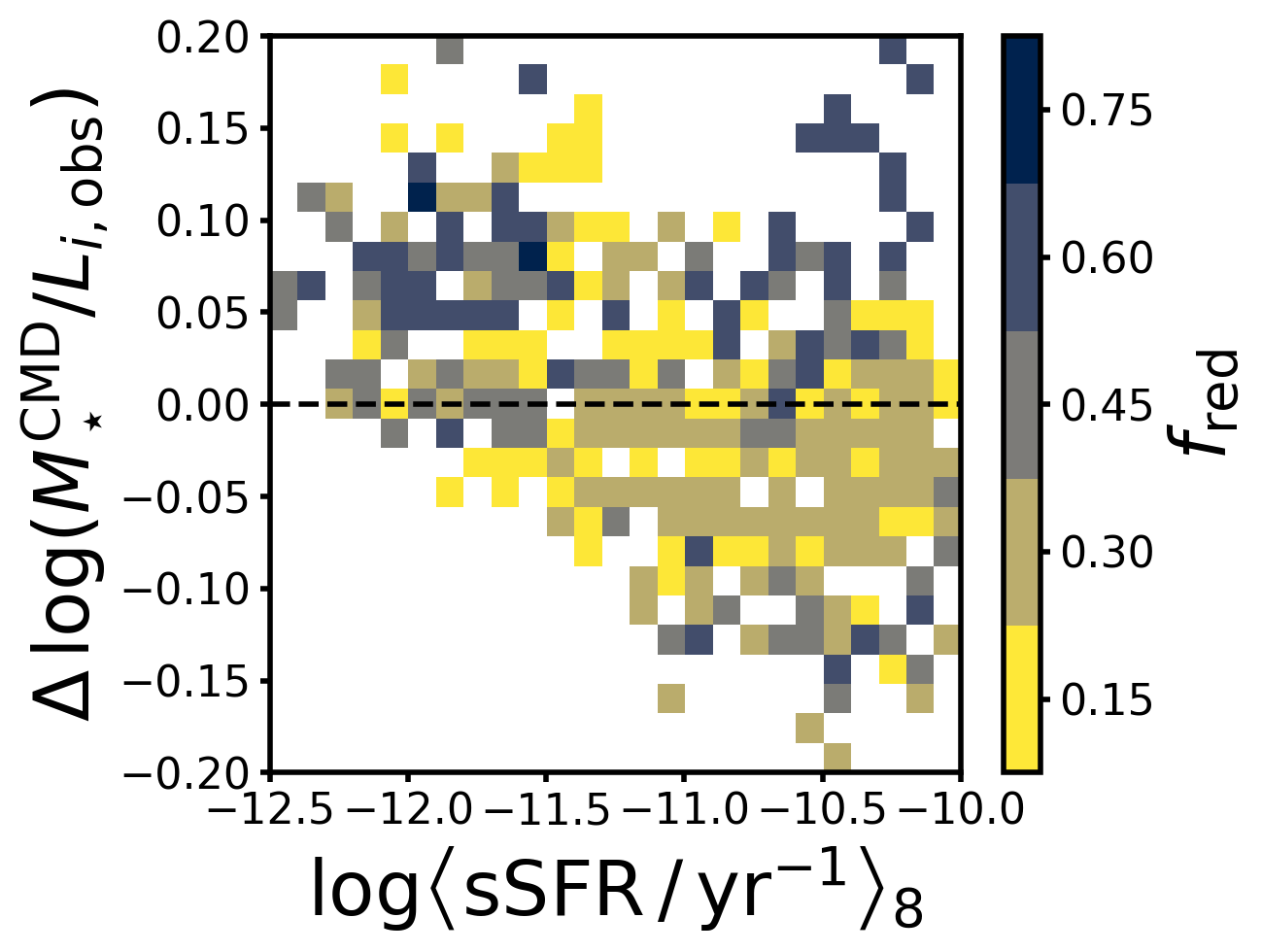}
\endminipage\hfill
\minipage{0.5\textwidth}
  \includegraphics[width=\linewidth]{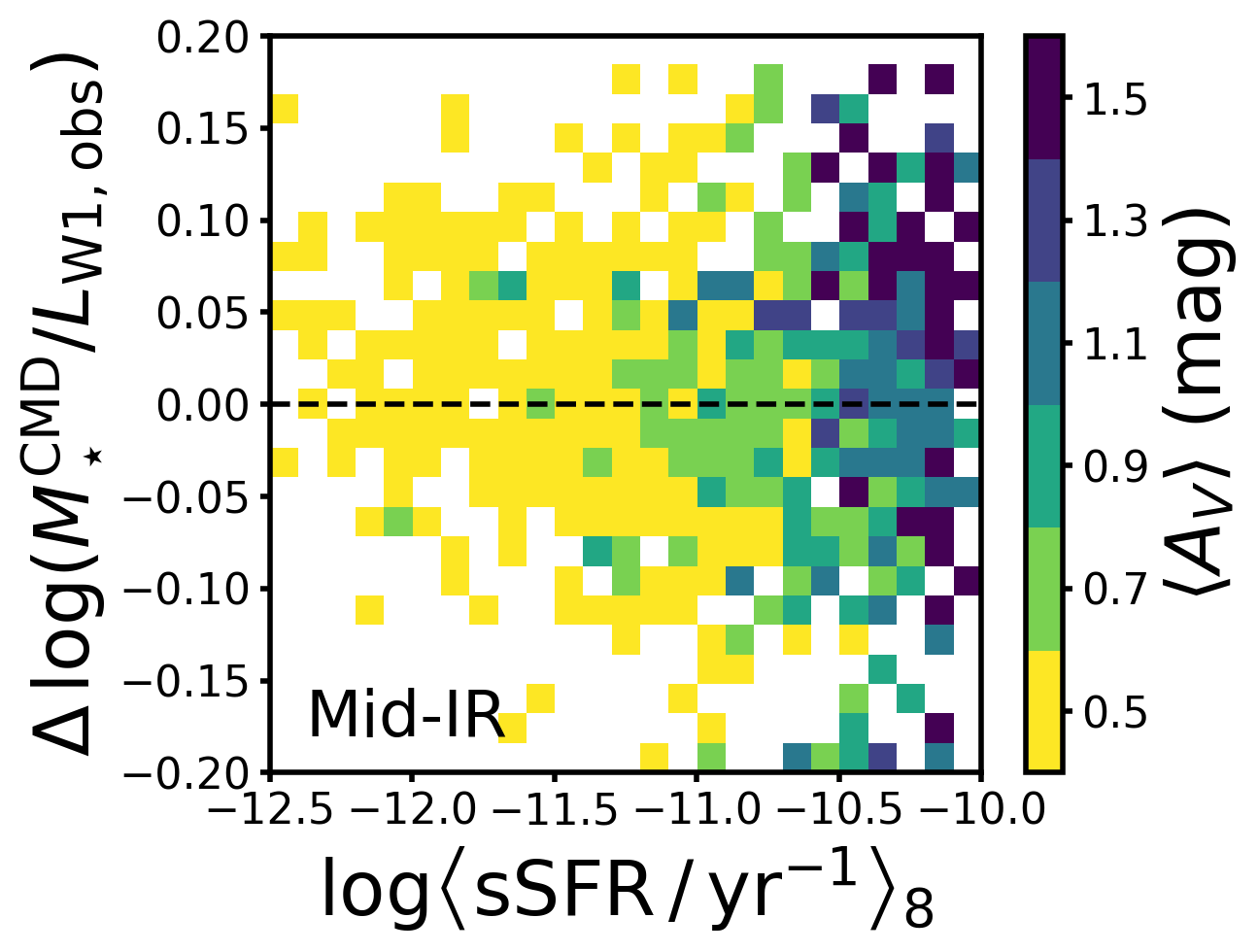}
\endminipage\hfill
\minipage{0.5\textwidth}
  \includegraphics[width=\linewidth]{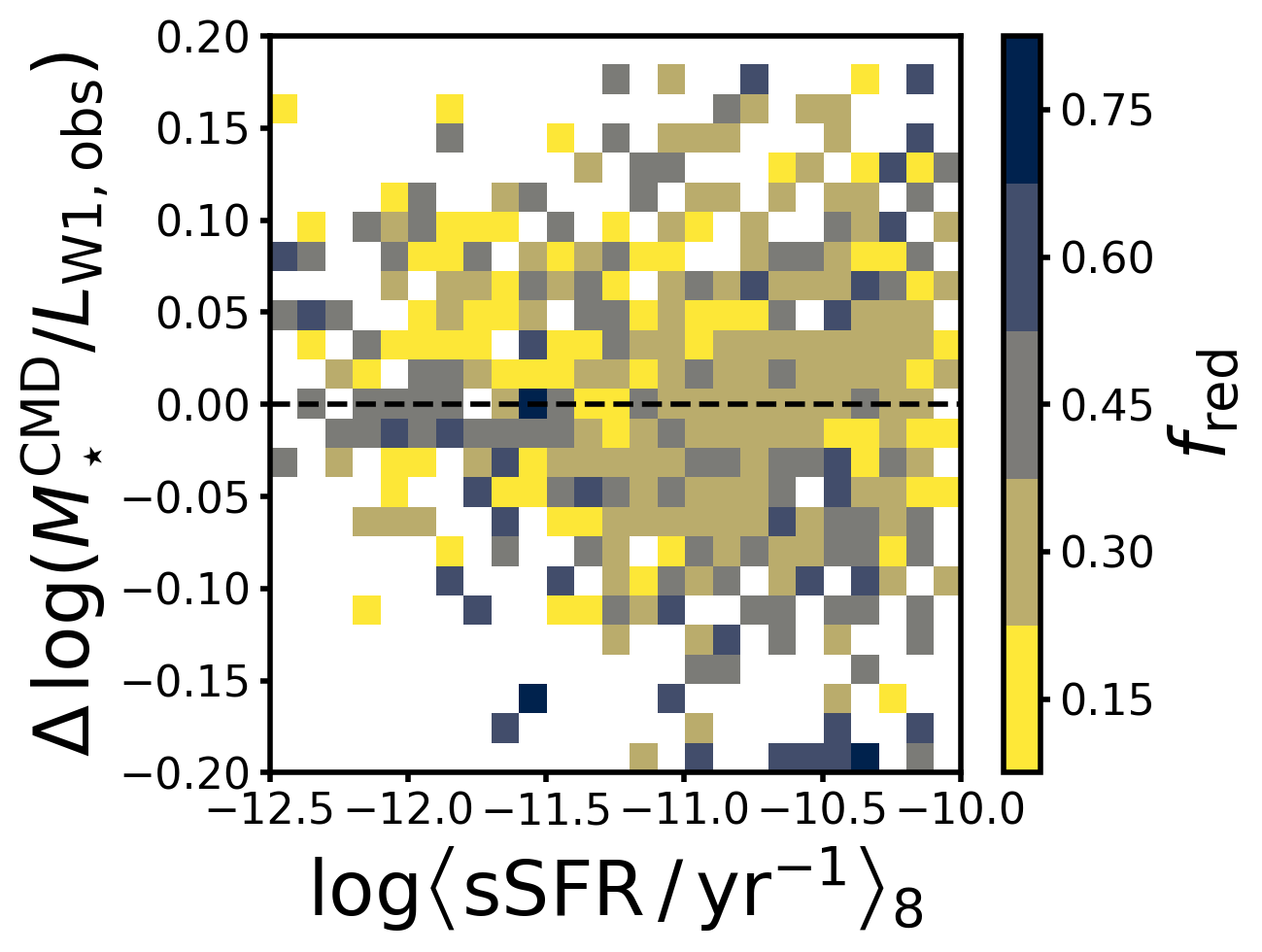}
\endminipage\hfill
\caption{\textsc{\textbf{Correlations Between Dust Parameters and M31 Color--$\bm{M^\mathrm{CMD}_\star/L_\mathrm{obs}}$ Relation Residuals.}} The same 2D histograms as in Figure~\ref{fig:cmlr_scatter_sfh} for the optical (top row) and mid-IR (bottom row), but now color-coded by the median \meanav{} (left column) and \fred{} (right column) of the SFH pixels in each bin. Darker pixels indicate higher typical dust content (left column) and higher fraction of old stars behind the dust layer (right column). The dashed horizontal line in each panel shows zero offset from the best-fit CMLR.
\label{fig:cmlr_scatter_dust}}
\end{centering}
\end{figure*}

\begin{figure*}[!ht]
\begin{centering}
  \includegraphics[width=0.75\linewidth]{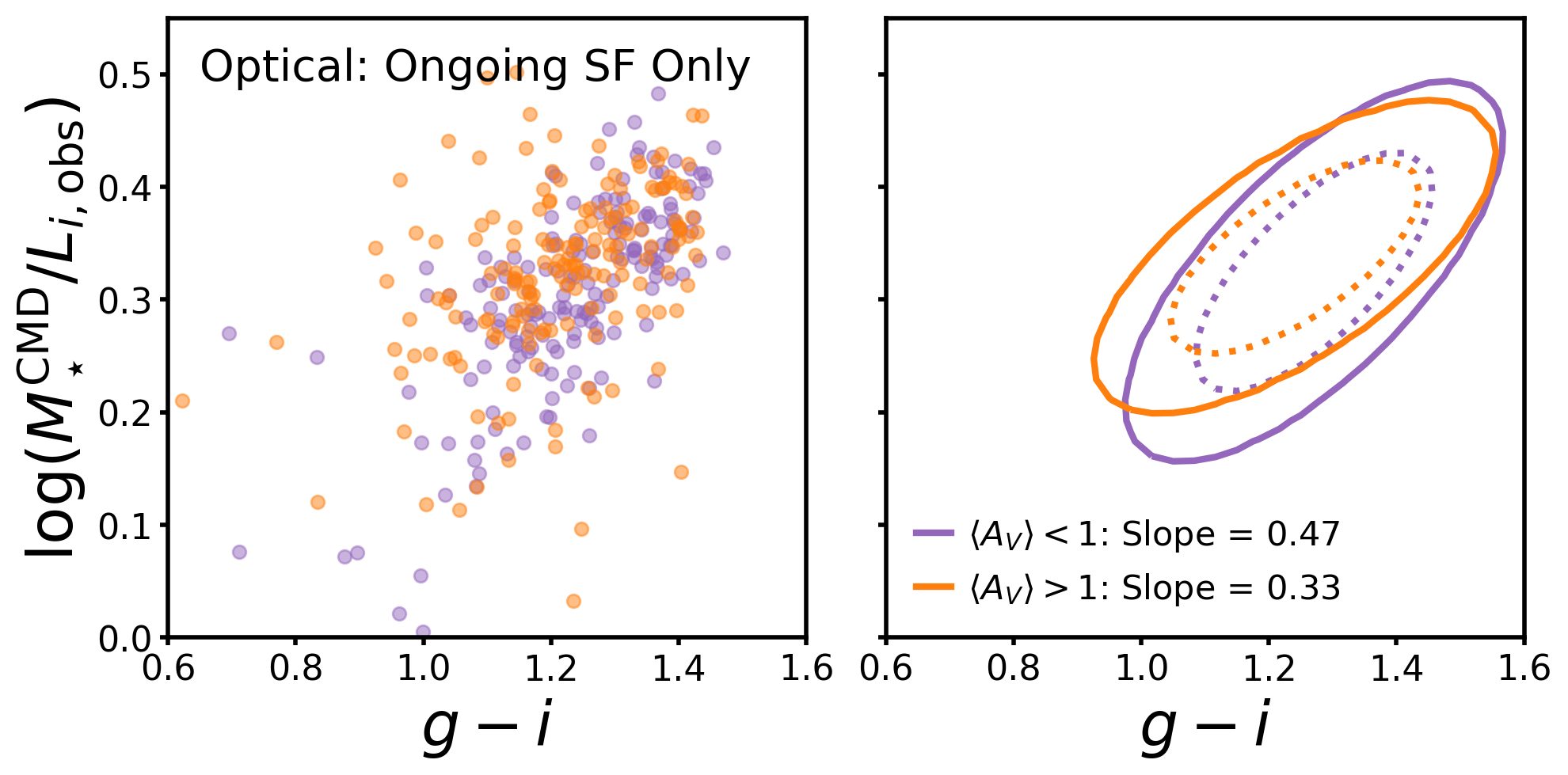}
  \includegraphics[width=0.75\linewidth]{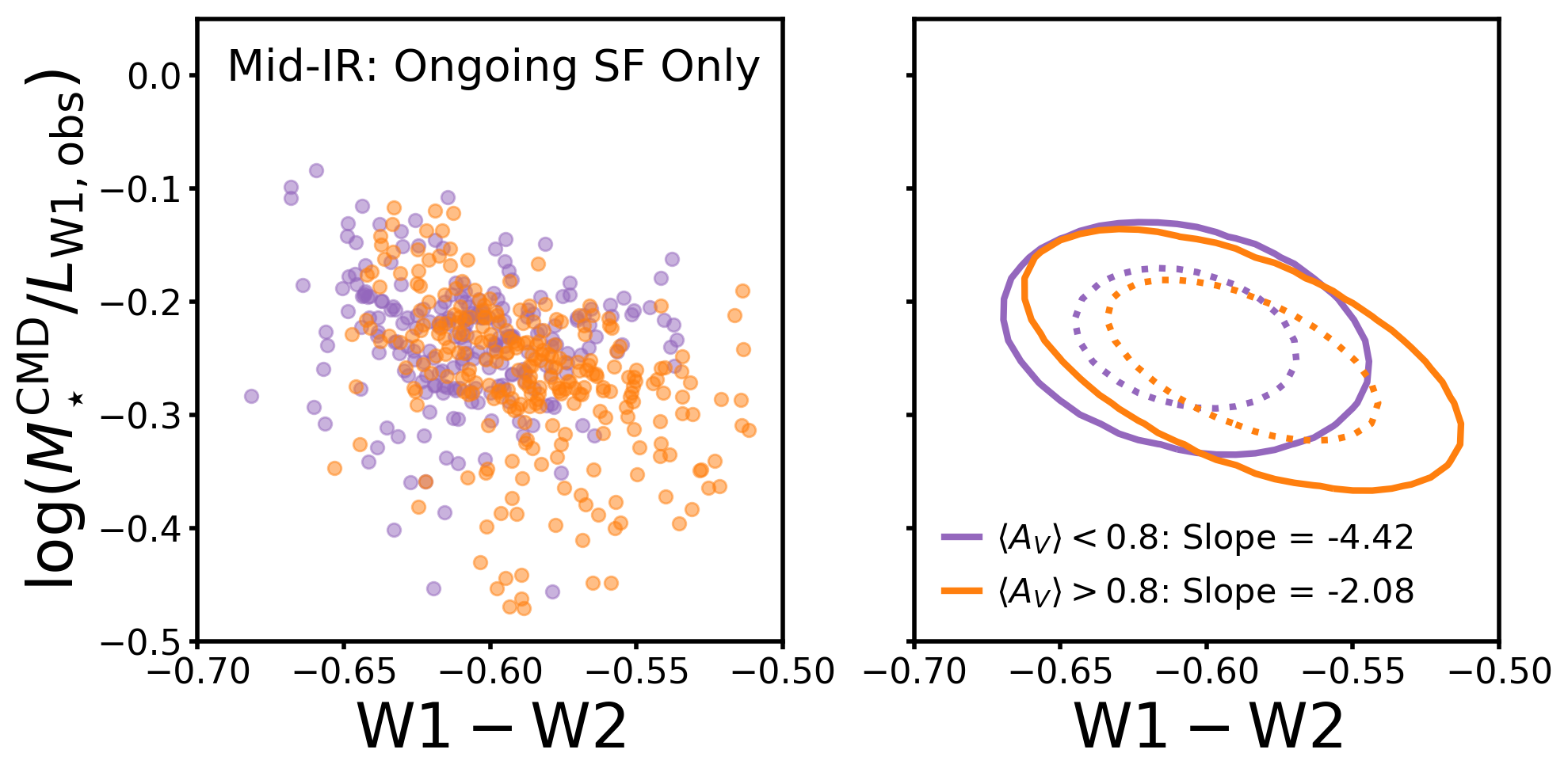}
\caption{\textbf{\textsc{Dust changes the CMLR slope for star-forming regions in M31.}} Same as Figure~\ref{fig:cmlr_split_sfh}, except showing only the SFH pixels with ongoing star formation (shown as blue points in Figure~\ref{fig:cmlr_split_sfh}, $\log$\ssfr{} $> -11.3$). These samples are then split at the median \meanav{}: 1.0 for the optical sample (where fewer SFH pixels are included due to the shallower SDSS photometry), and 0.8 for the mid-IR. Orange and purple colors indicate high- and low-dust regions, respectively. The shape of the best-fit Gaussian models is different for star-forming regions with low and high dust content in both the optical and mid-IR, demonstrating that dust does not simply move regions within a single galaxy along the best-fit CMLRs.
\label{fig:cmlr_split_dust}}
\end{centering}
\end{figure*}

\begin{figure*}[!ht]
\begin{centering}
  \includegraphics[width=0.75\linewidth]{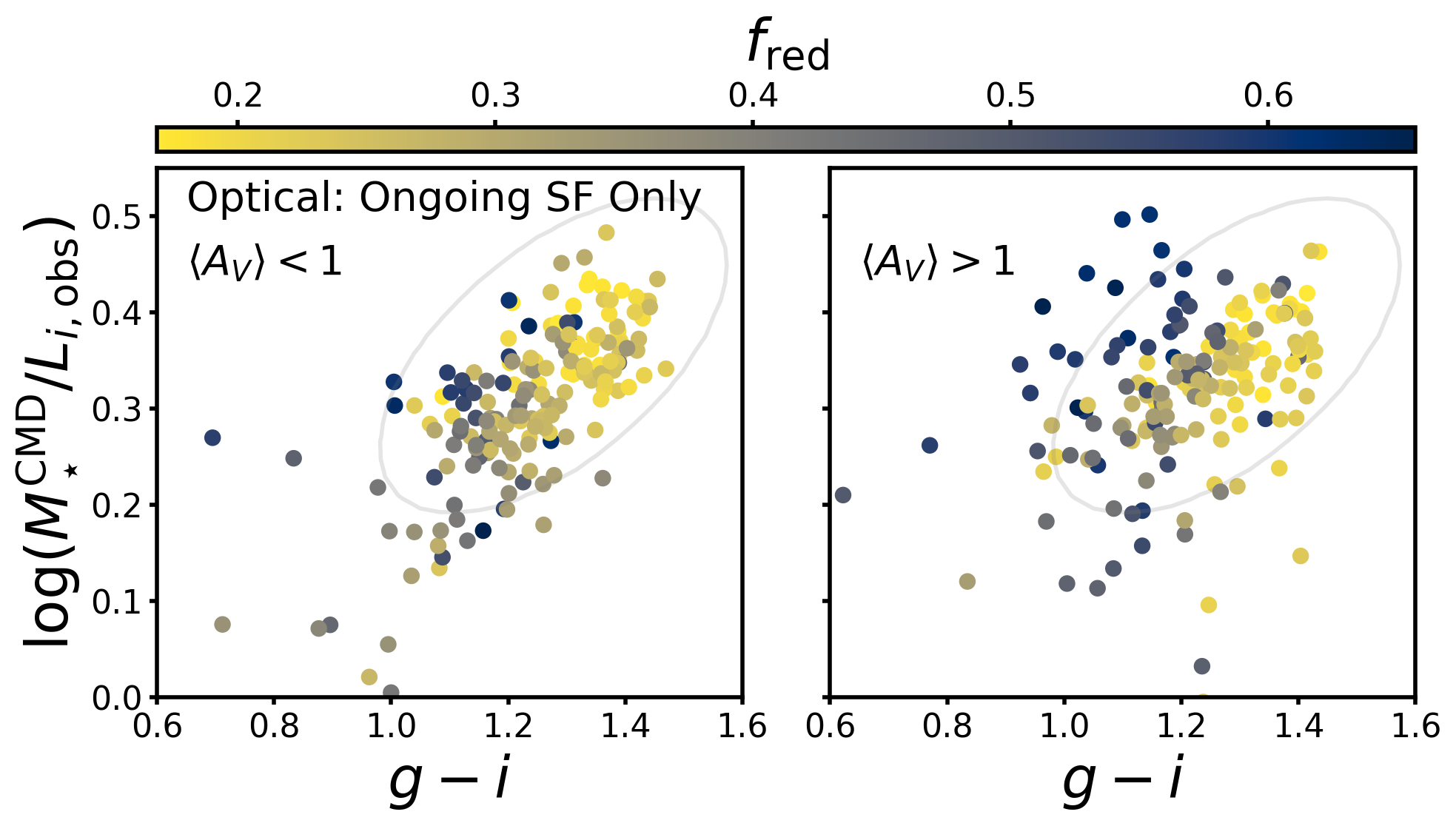}
\caption{\textbf{\textsc{Reddened fraction drives scatter in $\bm{M^\mathrm{CMD}_\star/L_{i, \mathrm{obs}}}$ at blue $\bm{g-i}$ and flattens the optical CMLR.}} \logmlicmd{} vs. \gi{} for SFH pixels with ongoing star formation (shown as blue points in Figure~\ref{fig:cmlr_split_sfh}, $\log$\ssfr{} $> -11.3$). The left and right panels show the low- and high-\meanav{} subsamples, respectively, and the points in both panels are color-coded by \fred{}. For reference, the grey ellipses enclose 95\% of the density in the best-fit Gaussian model for all SFH pixels shown in the left panel of Figure~\ref{fig:cmlr_distributions}. Higher \fred{} correlates with bluer \gi{}, and in the high-\av{} regime (right panel), the scatter in \logmlicmd{} at a given \gi{} is larger for the high-\fred{}, blue SFH pixels. The combined effects of high dust content, ongoing star formation, and the variation the relative geometry of dust and old stars across a highly inclined thick galaxy disk tend to flatten the relationship between \mlicmd{} and \gi{}.
\label{fig:fred}}
\end{centering}
\end{figure*}

\begin{figure}[!ht]
\begin{centering}
  \includegraphics[width=\linewidth]{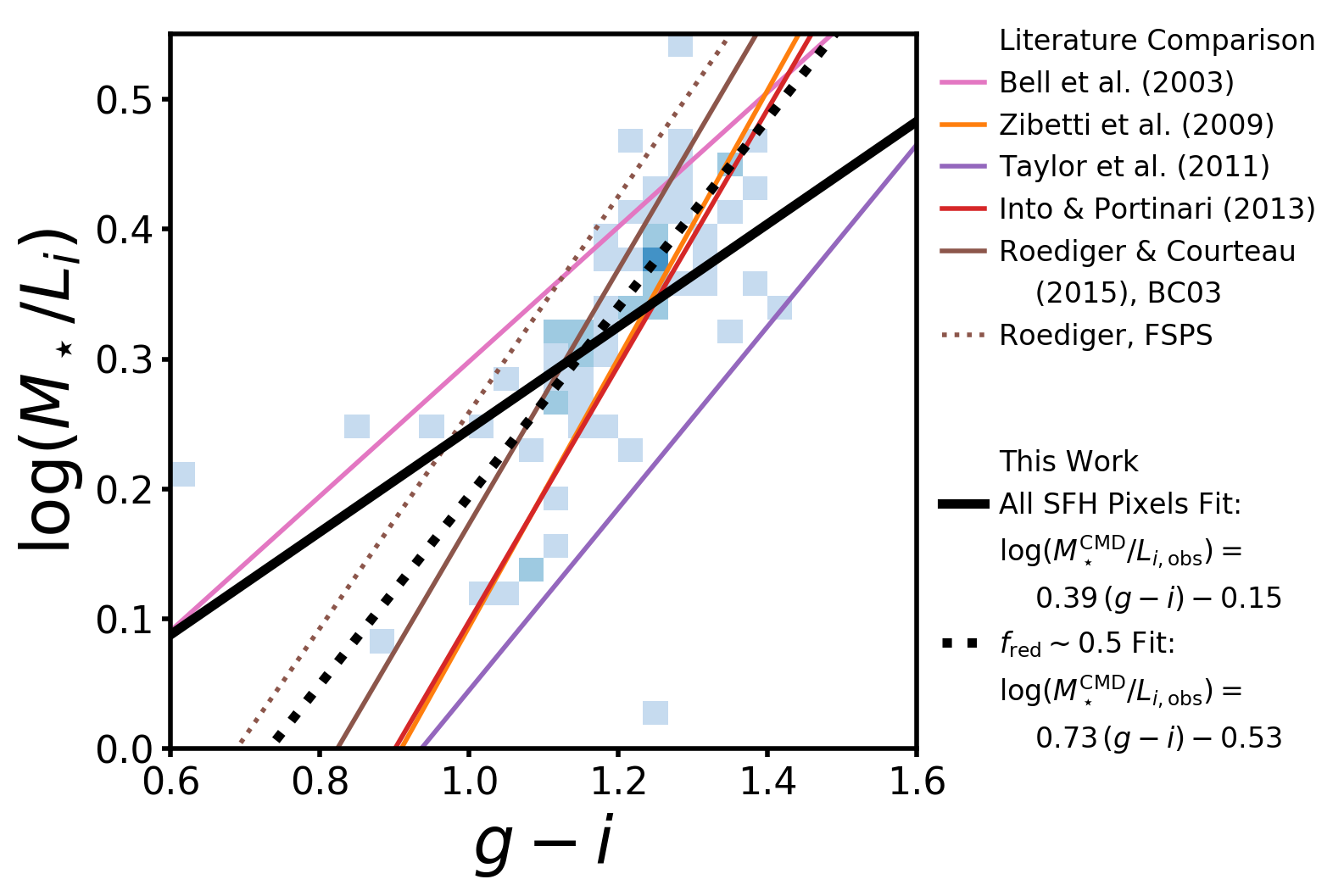}
\caption{\textbf{\textsc{SFH pixels with $\bm{f_\mathrm{red} \sim 0.5}$ follow a steeper optical CMLR, similar to SPS model predictions.}} A reproduction of the literature \logmli{} vs. \gi{} relations shown in the left panel of Figure~\ref{fig:lit_cmlrs}, along with the CMLR fit to SFH pixels with $0.45 < f_\mathrm{red} < 0.55$ (black dotted line). Now, the underlying two-dimensional histogram shows only the $f_\mathrm{red} \sim 0.5$ subset of SFH pixels. The CMLR fit to $f_\mathrm{red} \sim 0.5$ regions only is steeper than that for the full sample of SFH pixels in M31 (black solid line), and is more similar to literature CMLRs fit to SPS model libraries.
\label{fig:fred_cmlr}}
\end{centering}
\end{figure}

We now search for correlations between the \mlcmd{} residuals and dust content and geometry, beyond the correlations with SFH explored in Section~\ref{sec:cmlr_recent_sfh}. Figure~\ref{fig:cmlr_scatter_dust} shows the same 2-D histograms of SFH pixels in bins of $\log$\ssfr{} and $\Delta$\logmlicmd{} (top row) or $\Delta$\logmlwcmd{} (bottom row) from Figure~\ref{fig:cmlr_scatter_sfh}, but now color-coded by the median \meanav{} (left column) and \fred{} (right column) in each bin. \meanav{} is tightly correlated with \ssfr{} because dust is colocated with ongoing star formation. In both the optical and mid-IR, the scatter in $\Delta$\logmlcmd{} is larger for regions with high \ssfr{}. The top row of Figure~\ref{fig:cmlr_scatter_dust} shows that regions with high $\Delta$\logmlicmd{} (where \mlicmd{} has been underestimated by the best-fit CMLR) tend to have both higher \meanav{} and higher \fred{}. No such trends between the residuals and either \meanav{} or \fred{} are obvious for the mid-IR (bottom row). We discuss in more detail the effects of dust content and star-dust geometry in turn.

\subsubsection{Dust Content: \meanav{} \label{sec:dust_content}}

To clarify how the presence of dust affects the relationship between \ml{} and color, we again split the SFH pixels into subsamples defined by their dust content. Because dust is colocated with ongoing star formation, splitting the full sample of SFH pixels into high- and low-dust regions would effectively be making an age selection. We therefore restrict the remainder of this analysis to the SF regions (with $\log$\ssfr{} $> -11.3$, shown in blue in Figure~\ref{fig:cmlr_split_sfh}). We divide the SF regions into high- and low-dust subsamples defined by the median \meanav{} (which scales linearly with $\Sigma_\mathrm{dust}$) and fit 2D Gaussian models to the distributions in color-\mlcmd{} space. Because some regions in the outer disk were excluded from the SDSS imaging due to our photometric quality cut (Section~\ref{sec:sdss}), the median \meanav{} is different in the optical ($\left< A_V\right>_\mathrm{med} =1.0$) and mid-IR ($\left< A_V\right>_\mathrm{med} =0.8$). Figure~\ref{fig:cmlr_split_dust} shows the 2D Gaussian models fit to our high- and low-dust subsamples in orange and purple, respectively. This figure is analogous to Figure~\ref{fig:cmlr_split_sfh}, but the data shown and to which the models are fit are restricted to the SF sample ($\log$\ssfr{} $> -11.3$). Parameters of the best-fit models are given in Table~\ref{tab:params}.

The top right panel of Figure~\ref{fig:cmlr_split_dust} shows that the high-dust regions follow a flatter optical CMLR than the low-dust regions, with a tendency towards higher \mlicmd{} at a given color than less dusty regions. Interestingly though, the characteristic \gi{} ranges of the two subsamples are nearly the same. This suggests that in the dustiest regions of M31 (i.e., in the star-forming rings), dust removes $i$-band light in a way that does not produce the amount of reddening in \gi{} expected from a uniform foreground dust screen. This could potentially be explained by a clumpier dust distribution in those regions that allows blue light to escape through dust-free channels, flattening the effective attenuation law \citep[e.g.,][]{calzetti00}. However, the upper right panel of Figure~\ref{fig:cmlr_scatter_dust} hints that in the SFH pixels with highest \ssfr{}, \fred{} is correlated with higher \mlicmd{}. We discuss the influence of \fred{} on optical colors and \mlicmd{} in Section~\ref{sec:dust_geometry}.

We now turn to the impact of dust on the mid-IR, where dust emission may contribute to the \wcolor{} color and effective \mlw{} \citep[e.g.,][]{querejeta15}. The bottom right panel of Figure~\ref{fig:cmlr_split_dust} shows the best-fit Gaussian models to the low- and high-dust subsamples as purple and orange ellipses, respectively. The range of \wcolor{} and \mlwcmd{} spanned by the high-dust sample covers the entire range of the low-dust sample, and extends to redder \wcolor{} and lower \mlwcmd{}. The high-dust regions are also those with the highest \ssfr{}, in the star-forming rings of M31. Because the dust is colocated with the most intense star formation, we cannot determine whether the redder \wcolor{} and lower \mlwcmd{} is driven by young stellar populations or dust emission. Given the large ratio of scatter in \mlwcmd{} to dynamic range in \wcolor{} seen in the bottom row of Figure 13, knowledge of \wcolor{} can only modestly improve \mlwcmd{} estimates over adopting a constant value for the star-forming regions in M31.

\subsubsection{Star-Dust Geometry: \fred{} \label{sec:dust_geometry}}

Star-dust geometry varies across galaxies in many ways that could affect observed colors and \ml{}: clumpiness, prominent dust lanes, etc. Here, we analyze the effect on the optical CMLR of variation in a specific type of star-dust geometry: the fraction of old stars behind a relatively thin dust layer, \fred{}. We focus only on the optical CMLR because star-dust geometry should affect only dust attenuation, which is negligible at mid-IR wavelengths. Indeed, the bottom right panel of Figure~\ref{fig:cmlr_scatter_dust} shows no correlation between $\Delta$\logmlwcmd{} and \fred{}. In contrast, the upper right panel hints that SFH pixels with higher \fred{} tend to have their \mlicmd{} underestimated by the best-fit CMLR.

Figure~\ref{fig:fred} shows \logmlicmd{} vs. \gi{} for the low- and high-dust, SF subsamples in the left and right panels, respectively. Each point represents a single SFH pixel and the color-coding shows the median \fred{} within that pixel. In the low-dust regime, higher \fred{} points tend to be bluer, but they do not appear to deviate from the overall trend between \logmlicmd{} and \gi{}. In the high-dust regime, however, high \fred{} pixels are bluer and show an increased scatter in \mlicmd{} for a fixed color. Counterintuitively, regions where a larger fraction of old stars lie behind the dust layer have bluer \gi{}. We speculate that this is due to the star-dust geometry dictating the relative contribution of old and young stellar populations to the total light. In dusty regions with low \fred{}, the old stars experience little dust attenuation while the young stars are embedded in the dust layer, resulting in redder \gi{} dominated by the light from old, red stars. In high \fred{} regions, however, the light from old stars is more attenuated and therefore contributes less to the total light than in low \fred{} regions, driving \gi{} bluer due to the increased contribution of the young stars.

Strong variation in \fred{} occurs for inclined spirals with thick stellar disks. In galaxies that are less highly inclined or lack thick disks, \fred{} is close to 0.5 everywhere in a galaxy, and this is exactly true in the case of face-on disks. We test whether the \fred{} variation is responsible for the shallow slope (relative to SPS-based predictions) of the optical CMLR in the disk of M31 by identifying a sample of 69 SFH pixels where $0.45 < f_\mathrm{red} < 0.55$, with no restriction on \ssfr{}, and fitting a 2D Gaussian model to the distribution of those points in \gi{} vs. \mlicmd{} space (best-fit parameters given in Table~\ref{tab:params}). The resulting CMLR is shown in Figure~\ref{fig:fred_cmlr} as the dotted black line, while the original CMLR fit to the full sample of SFH pixels is reproduced as the solid black line. Clearly, this is far steeper than the CMLR fit to all SFH pixels in M31 and is more similar to the steeper SPS-based CMLRs, which are shown for reference as the thin colored lines in Figure~\ref{fig:fred_cmlr}. 

\textit{The takeaway from this exploration is that the geometry of old stars relative to the dust in a galaxy can strongly affect the slope of the true relationship between \mlicmd{} and \gi{} in inclined galaxies with thick stellar disks.} This effect is not captured by SPS models, which typically approximate the effects of dust with a uniform foreground screen model (sometimes including additional attenuation by birth cloud dust for young stellar populations). The results shown here imply that possible variation in \fred{} should be accounted for in SPS models used to infer \mstar{} maps from spatially resolved optical light for inclined spiral galaxies.


\section{Discussion}\label{sec:discussion}

We have used CMD-based \mstar{} to construct optical and mid-IR \mlcmd{} and CMLRs in M31, compared them to previously reported CMLRs in the literature, and analyzed the effect of SFH and dust content and geometry on the slope and normalization of our CMLRs. Here, we discuss the implications of our results for estimating \mstar{} in other galaxies.

\subsection{Lessons for Spatially Resolved \mstar{} Inference}\label{sec:lessons}

The \mcmd{} that we use to map \mlcmd{} in M31 were inferred from modeling resolved star CMDs, providing a complementary measurement to the \ml{} predicted by SPS models of integrated light. Our main goal in this work was to test the performance of SPS-based CMLRs by comparing against the empirical relations in M31. However, because these \mlcmd{} are only measured within a single galaxy, we must be cautious in extrapolating our findings to lessons for \mstar{} inference in other situations.The M31 results certainly inform the interpretation of SPS-based, resolved \mstar{} measurements within highly inclined galaxies with thick stellar disks and low-level star formation. Our results suggest that variation in \fred{} within such galaxies (Section~\ref{sec:cmlr_dust}) affects observed optical color and \ml{} in a manner not reproduced by current SPS models, but such effects may average out when considering only the integrated light of entire galaxies. The impact of SFH on CMLR normalization (Section~\ref{sec:cmlr_recent_sfh}), however, is applicable to both resolved and integrated \mstar{} inference; we discuss this further in Section~\ref{sec:mass_uncertainty}.

In Section~\ref{sec:lit_cmlrs}, we show that the optical CMLR fit to all SFH pixels within the PHAT footprint has a flatter slope than SPS-based CMLRs fit to libraries of predicted color and \ml{}. The slope of the adopted CMLR affects the inferred distribution of \mstar{}, either across a population of galaxies spanning a range of colors, or within individual galaxies with color gradients. Accurate maps of the \mstar{} distribution within galaxies are particularly important for dynamical studies. Recently, \citet{nguyen19} fit SPS models to optical spectroscopy of the inner regions of low-mass galaxies to infer \ml{}, with the goal of mapping \mstar{} to look for dynamical signatures of supermassive black holes in these galaxies. Interestingly, they found different CMLR slopes in each of the four galaxies they studied, suggesting that no ``one size fits all'' CMLR can be used for the precise \mstar{} mapping required for such dynamical modeling efforts. The relatively flat optical CMLR that we find in M31 supports this conclusion. 

We recommend taking into account galaxy morphology and inclination when using SPS models or SPS-based CMLRs to construct spatially resolved \mstar{} maps from optical colors, particularly if those galaxies are inclined and may harbor thick disks and therefore strong variation in \fred{}. This finding takes on additional importance with the advent of massive, optical IFU surveys, e.g., CALIFA \citep{sanchez12}, SAMI \citep{bryant15}, and MaNGA \citep{bundy15}. Though advanced spectral fitting techniques are being used to fit SPS models to the spatially resolved SEDs of galaxies in these surveys, SPS models do not typically account for possible variation in the relative geometry of old stellar populations and dust, like that in M31. Allowing for varying \fred{} in SPS models would improve future efforts to map \mstar{} using spatially resolved optical light.

In the mid-IR, SPS models cannot at present appropriately model the light from young stellar populations \citep[e.g.,][]{peletier12}. Short-lived, luminous phases of stellar evolution such as core He burning and TP-AGB are notoriously difficult to model, and different approaches result in very different predictions for the time evolution of mid-IR \ml{} and colors (for example, the right column of Figure~\ref{fig:sps_models}). While, in principle, ``semi-empirical'' CMLRs could be used to sidestep the need for SPS models to explain \wcolor{} colors, we found in Section~\ref{sec:cmlr_scatter} that incorporating \wcolor{} information can only modestly improve \mlw{} estimates \citep[in agreement with][]{eskew12}, especially for SFH pixels with ongoing star formation. 

This finding suggests that, in the case of spatially resolved observations with high enough signal-to-noise, the best approach to inferring the distribution of \mstar{} from mid-IR light is the technique of subtracting off ``contaminating'' light from regions with red \wcolor{} \citep[as pioneered by][]{meidt12} to isolate the ancient stellar light, then using an SPS-based \mlw{} appropriate for old stellar populations to convert the luminosity map to \mstar{}. If this is not possible (either for the case of integrated light or due to poor data quality), we suggest that using two \mlw{} values, typical for each old and young stellar populations, would be a practical approach to estimating \mstar{}. \wcolor{} can be used to approximately determine whether the stellar population is quiescent or star-forming, adopting a threshold around \wcolor{} $\sim -0.62$ (where bluer colors indicate quiescent populations). While the significant overlap in the colors of star forming and non star forming pixels makes it difficult to choose a single boundary color, $-0.62$ marks close to the center of the overlap, and places the SFH pixels in the 10 kpc star-forming ring in M31 squarely in the star forming regime. The peak values of \logmlw{} and scatter perpendicular to our best-fit CMLRs given in Table~\ref{tab:params} for our quiescent and star-forming samples would be appropriate estimates of \logmlw{} and uncertainty for this purpose. However, even the highest \ssfr{} values in M31 are modest, and so \mlwcmd{} estimates based on M31 data should not be used in highly star-forming (regions within) galaxies. 

It is well-established that redder mid-IR colors correlate with young stellar populations and with the presence of dust heated by starlight, but the relative importance of these to determining \ml{} and color remains unclear. The polycyclic aromatic hydrocarbon (PAH) emission feature at 3.3 $\mu$m is often pointed to as a potential driver of low \mlw{} due to dust heating by young stellar populations. Substantial flux in that emission line would drive \mlw{} low and \wcolor{} bluer. However, regions with lower \mlwcmd{} in M31 tend to have redder \wcolor{} (albeit with large scatter), suggesting that the 3.3 $\mu$m PAH emission is not the dominant driver of low \mlw{}. The SED of hot dust is expected to have red \wcolor{} \citep{querejeta15}, so variation in the relative importance of PAH emission and the overall SED shape likely contributes to the range of colors observed in dusty, star-forming regions. Given the present incomplete understanding of the relative contributions of young stellar populations and dust emission to observed \wcolor{} and \mlw{}, we point out that including mid-IR flux in full SED fitting could potentially bias results and should be treated with caution.

Finally, we find in Section~\ref{sec:m31_cmlrs} that the spread in \logmlicmd{} and \logmlwcmd{} are comparable across the disk of M31 (0.08 and  0.09 dex, respectively). This is in opposition to the common idea that optical \ml{} are more sensitive to recent SFH, and therefore are more variable than and mid-IR \ml{}. We speculate that the variation in optical and mid-IR \ml{} would also be comparable within other relatively early-type, massive spiral galaxies with low-level, ongoing star formation. 

\subsection{Drivers of Uncertainty in Absolute \ml{}\label{sec:mass_uncertainty}}

As discussed in Section~\ref{sec:mass_norm}, we have attempted to put all CMLRs considered in Section~\ref{sec:lit_cmlrs} on the same absolute \mstar{} scale by correcting for differences in the adopted IMF. Yet, Figure~\ref{fig:lit_cmlrs} shows that substantial differences in the normalization of various CMLRs remain, spanning $\sim$0.25 dex and $\sim$0.4 dex in the optical and mid-IR, respectively. Even after accounting for the choice of IMF,  systematic differences at the factor of $\sim$2 (0.3 dex) level among \mstar{} inferences using different SED modeling techniques are acknowledged in the literature \citep[e.g.,][]{courteau14, mcgaugh14, hunt19}, but the causes of these offsets have not been definitively identified. Here, we discuss sources of offsets among various \mstar{} inference methods.

Most likely, the discrepancies among SPS-based \mstar{} are due to a combination of the SFH priors and stellar evolution models used in the various SPS codes. The features of the SFH that most strongly affect inferred \ml{} are the time at which star formation began and the allowed magnitude and timing of bursts \citep{gallazzi09, roediger15}. Earlier star formation allows for higher \mstar{} without requiring a large change in brightness, while recent bursts tend to drive down \ml{} due to the increased brightness of young stellar populations. \citet{bell01} showed that a 1-2 Gyr old burst of star formation can lower the observed \ml{} at a fixed optical color. This is similar to our finding that quiescent regions in M31 are offset to higher \mli{} at fixed \gi{} in Section~\ref{sec:cmlr_recent_sfh}. The priors imposed on the onset of star formation and the burstiness of the recent SFH in SPS model libraries can result in different predictions for both the slope and \mstar{} normalization of optical CMLRs.

It is also well-known that stellar evolutionary models struggle to simultaneously explain observations in the optical and near-IR \citep[e.g.,][]{taylor11, mcgaugh14}, and there remain large discrepancies in the predicted \ml{} at red/infrared wavelengths among SPS models that adopt different treatments of luminous stellar evolutionary phases \citep[e.g.,][]{conroy13}. The short timescales of the most luminous phases of stellar evolution lead to strong time variability in the NIR, posing a formidable challenge to stellar evolutionary models \citep[e.g.,][]{melbourne12}. The discrepancies among various stellar evolutionary models can also drive different \ml{} predictions at fixed color, with more pronounced differences in redder filters \citep{roediger15}. 

The normalization of the \wise{} CMLR found by \citet{hunt19} using the CIGALE code to model the UV through IR SED is far lower than either our CMLR in M31 or other relations in the literature (right panel of Figure~\ref{fig:lit_cmlrs}). Their KINGFISH galaxy sample is composed of mostly star-forming galaxies, which do tend to have lower \mlw{} \citep[][Section~\ref{sec:cmlr_recent_sfh}]{eskew12, querejeta15}. However, their choice of \mstar{} inference technique may also contribute to the low \ml{}. Intriguingly, \citet{buat19} used CIGALE to model the SEDs of a sample of dusty $z \sim 2$ galaxies, and for the subset of their galaxy sample where the physical extent of the stars and dust emission was similar, they infer systematically lower \mstar{} for their fits to stellar + dust emission than from the fits to the stellar continuum only.

In principle, fitting the full SED with a model that simultaneously captures stellar and dust emission should give more robust \mstar{} measurements because the far-IR emission can be used to break the dust-age degeneracy in the optical. The dust mass inferred from the far-IR emission constrains the amount of dust extinction that can remove light in the optical and UV, and therefore how much intrinsic stellar emission is allowed. However, it is not clear that commonly used dust emission models can accurately predict dust extinction, given measured dust emission. The \citet{draine07} dust models that are used in CIGALE SED fitting have been shown to predict a factor of $\sim2.5$ higher extinction than observed for a given amount of dust emission \citep{dalcanton15, planck16-dust}. If the SED model applied too much dust extinction, then the inferred dust-free stellar emission would be too blue, potentially leading to stellar age and \mstar{} both being biased low. This possibility highlights the challenges in simultaneously modeling stellar emission, dust extinction, and dust emission, and reminds us that differences among the wavelength baselines of observations used to infer \mstar{} likely contribute to the well-known discrepancies among \ml{} in the literature.

Dynamical measurements can also be used as a complementary \mstar{} inference technique to SPS models, with the caveat that dark matter can introduce substantial uncertainty. \citet{de-jong07} reviewed the available dynamical constraints on galaxy \mstar{} at that time, and showed that they agree to within $\sim$0.4 dex. They concluded that SPS-based \ml{} that adopt a \citet{chabrier03} or \citet{kroupa01} IMF are broadly consistent with dynamical measurements. More recently, \citet{martinsson13} used the vertical velocity dispersion in a sample of galaxy disks from the DiskMass survey to estimate near-IR \ml{}. These authors found a \ml{} that is roughly a factor of 2 lower than most SPS-based \ml{} \citep{mcgaugh14}, but consistent with the \ml{} found by \citet{hunt19}. On the other hand, \citet{nguyen19} found that their \ml{} inferred from fitting SPS models adopting a \citet{kroupa01} IMF to optical spectra of four low-mass galaxies were consistent with dynamical constraints. We reiterate that \textit{all} \mstar{} estimates, including our CMD-based measurements in M31, are systematically uncertain. At present, the best approach is to account for the possible offsets among various \mstar{} inference techniques when comparing results across different studies.


\section{Conclusions\label{sec:conclusions}}

In this paper, we construct linear relations between observed optical and mid-IR colors and \logmlcmd{} for spatially resolved stellar populations in the massive spiral galaxy M31. \mcmd{} is calculated from CMD-based SFHs, which were inferred from PHAT resolved-star photometry independently of the integrated light.  $L_\mathrm{obs}$ is measured from SDSS and \wise{} surface brightness maps. These \mlcmd{} vs. color relations in M31 are constructed in a fundamentally different manner from standard CMLRs that are fit to libraries of SPS models. Our key conclusions are:

\begin{enumerate}

\item We find comparable spread in the observed \mlicmd{} and \mlwcmd{} distributions across M31, contrary to the common idea that optical \ml{} are more sensitive to variation in SFH. This finding may extend to other early-type spirals experiencing low levels of star formation (Section~\ref{sec:m31_cmlrs}, Figures~\ref{fig:ml_cmd} and \ref{fig:cmlr_distributions}).

\item We fit 2D Gaussian models to the distribution of SFH pixels in optical and mid-IR color--\mlcmd{} space, and calculate linear CMLRs as the eigenvector pointing along the direction of maximum variance (Table~\ref{tab:params}). We compare these CMLRs in M31 against previous results in the literature, finding good agreement with previous ``semi-empirical'' relations in the mid-IR, but a shallower CMLR slope than predicted by most SPS models in the optical (Section~\ref{sec:lit_cmlrs}, Figure~\ref{fig:lit_cmlrs}).

\item The residuals about both the optical and mid-IR CMLRs correlate with the recent SFH inferred from PHAT CMDs. Quiescent stellar populations are systematically offset to higher \mlicmd{} at fixed \gi{} in the optical, and to both higher \mlwcmd{} and bluer \wcolor{} in the mid-IR. Star-forming and quiescent regions overlap in color in both the optical and mid-IR (Section~\ref{sec:cmlr_recent_sfh}, Figures~\ref{fig:cmlr_scatter_sfh} and \ref{fig:cmlr_split_sfh}).

\item We show that the strong variation in the fraction of old stars behind the dust layer, \fred{}, in M31 results in dusty, star-forming regions following a flatter CMLR than low-dust regions. This effect is not captured by SPS models, and is important to account for in studies of spatially resolved \mstar{} using optical data (Sections~\ref{sec:cmlr_dust} and \ref{sec:lessons}, Figures~\ref{fig:cmlr_scatter_dust}, \ref{fig:cmlr_split_dust}, \ref{fig:fred}, and \ref{fig:fred_cmlr}).

\item We find that using a mid-IR CMLR to estimate \mlw{} can provide only modest improvement over adopting a constant \mlw{}. We therefore recommend against using a linear CMLR to estimate \mlw{} for star-forming galaxies, and instead advocate for removing ``contaminating'' light from young stars and/or dust emission for spatially resolved data with high enough signal-to-noise. If those conditions are not met, we recommend using \wcolor{} to select an appropriate \mlw{} that is typical of either old or young stellar populations (Sections~\ref{sec:cmlr_scatter} and \ref{sec:lessons},  Figures~\ref{fig:cmlr_split_sfh} and ~\ref{fig:cmlr_split_dust}).

\item In Appendix~\ref{sec:appendix}, we report the slope, intercept, and scatter about CMLRs fit to observed colors and \mlcmd{} in M31 for various combinations of SDSS filters (Figure~\ref{fig:cmlrs_appendix}, Table~\ref{tab:params_appendix}).
\end{enumerate}

\acknowledgements
We thank Alexia Lewis, Anil Seth, Jessica Werk, Kristen McQuinn, Dustin Lang, Elmo Tempel, and Leslie Hunt for helpful conversations and for making data publicly available. OGT is supported by an NSF Graduate Research Fellowship under grant DGE-1256082, and was supported in part by NSF IGERT grant DGE-1258485. This research has extensively used NASA's Astrophysics Data System and the arXiv preprint server. This research has made use of Montage. It is funded by the National Science Foundation under Grant Number ACI-1440620, and was previously funded by the National Aeronautics and Space Administration's Earth Science Technology Office, Computation Technologies Project, under Cooperative Agreement Number NCC5-626 between NASA and the California Institute of Technology. 

This work was supported by the Space Telescope Science Institute through GO-12058. This work is based on observations made with the NASA/ESA Hubble Space Telescope, obtained from the data archive at the Space Telescope Science Institute. STScI is operated by the Association of Universities for Research in Astronomy, Inc. under NASA contract NAS 5-26555.

Funding for the SDSS and SDSS-II has been provided by the Alfred P. Sloan Foundation, the Participating Institutions, the National Science Foundation, the U.S. Department of Energy, the National Aeronautics and Space Administration, the Japanese Monbukagakusho, the Max Planck Society, and the Higher Education Funding Council for England. The SDSS Web Site is http://www.sdss.org/. The SDSS is managed by the Astrophysical Research Consortium for the Participating Institutions. The Participating Institutions are the American Museum of Natural History, Astrophysical Institute Potsdam, University of Basel, University of Cambridge, Case Western Reserve University, University of Chicago, Drexel University, Fermilab, the Institute for Advanced Study, the Japan Participation Group, Johns Hopkins University, the Joint Institute for Nuclear Astrophysics, the Kavli Institute for Particle Astrophysics and Cosmology, the Korean Scientist Group, the Chinese Academy of Sciences (LAMOST), Los Alamos National Laboratory, the Max-Planck-Institute for Astronomy (MPIA), the Max-Planck-Institute for Astrophysics (MPA), New Mexico State University, Ohio State University, University of Pittsburgh, University of Portsmouth, Princeton University, the United States Naval Observatory, and the University of Washington.

This publication makes use of data products from the Wide-field Infrared Survey Explorer, which is a joint project of the University of California, Los Angeles, and the Jet Propulsion Laboratory/California Institute of Technology, funded by the National Aeronautics and Space Administration.

\facilities{\textit{HST}, SDSS, \wise{}}
\software{iPython \citep{ipython}, Astropy \citep{astropy}, matplotlib \citep{matplotlib}, NumPy \citep{numpy}, FSPS \citep{conroy09, conroy10a}, python-fsps \citep{foreman-mackey14}, HDF5 \citep{hdf5}, Montage \citep{berriman03, jacob10}}, sep \citep{bertin96, barbary16}

\appendix
\section{Color-\mlcmd{} Relations for All SDSS Filters}\label{sec:appendix}

\begin{figure*}[!ht]
\begin{centering}
  \includegraphics[width=\linewidth]{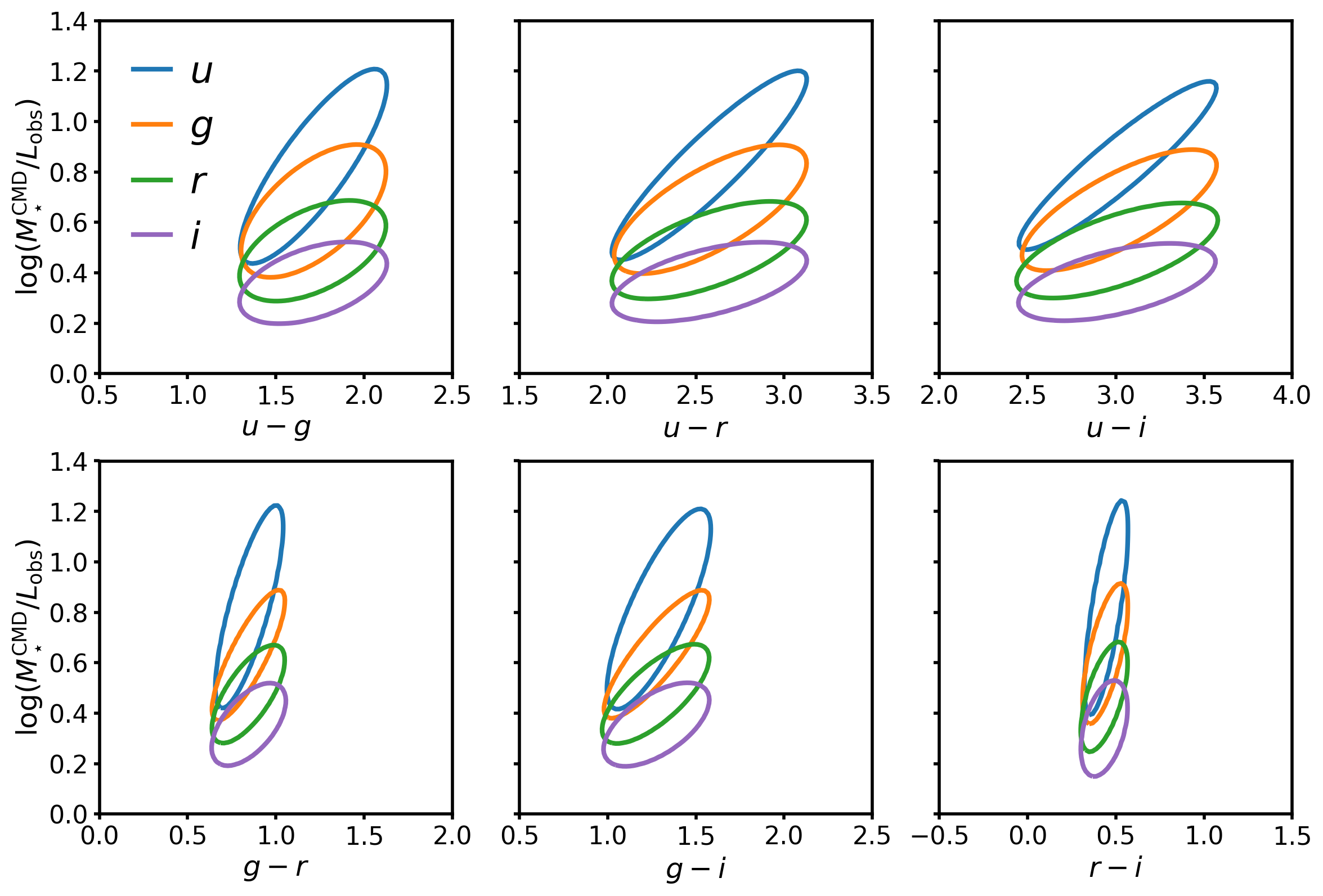}
\caption{\textbf{\textsc{Summary of M31 CMLRs for SDSS filters.}} Ellipses containing 95\% of the density of 2D Gaussian models fit to all combinations of \logmlcmd{} vs.\ color constructed from SDSS surface photometry of M31. Each panel shows the best fit relation for \logmlcmd{} in the four $ugri$ filters, color-coded as shown in the top left panel, plotted against a single color. The color axis spans the same range (though centered at different colors) in all panels, such that the observed colors in M31 covering smaller dynamic ranges correspond to ellipses that appear narrower in the horizontal direction. The parameters for each of these CMLRs are reported in Table~\ref{tab:params_appendix}.
\label{fig:cmlrs_appendix}}
\end{centering}
\end{figure*}

We restricted our analysis of optical CMLRs to the \logmli{} vs. \gi{} relation in the main text of this paper for clarity. In this Appendix, we report the best-fit parameters of optical-NIR CMLRs fit to M31 data for other combinations of SDSS filters. The CMLRs reported here may be useful for comparison or for applications involving M31-like galaxies, but we caution that these CMLRs are not expected to be generically applicable to arbitrary galaxy types for \mstar{} inference.

The $ugriz$ mosaics were all constructed by \citet{tempel11}, and we execute foreground star masking, quality thresholding, and reprojection to the scale of PHAT SFH pixels as described in Section~\ref{sec:sdss} above. We exclude any colors and \mlcmd{} constructed using the $z$-band surface photometry because the high sky background in the NIR rendered much of the data in that filter unreliable. \mlcmd{} in each SDSS filter is equal to the ratio of CMD-based $\Sigma^\mathrm{CMD}_\star$ (Section~\ref{sec:mass_map}) to the surface brightness across the PHAT footprint. We fit two-dimensional Gaussian models to the distributions of SFH pixels in color-\logmlcmd{} planes as described in Section~\ref{sec:m31_cmlrs}, and calculated the best-fit CMLRs as the eigenvectors along the direction of maximum variance in each Gaussian model. 

Figure~\ref{fig:cmlrs_appendix} presents a visual summary of the best-fit 2D Gaussian models to the various combinations of observed M31 colors and \logmlcmd{} in SDSS filters. Each panel shows \logmlcmd{} in the $ugri$ filters as a function of a single color, where the color of the ellipse indicates the filter in which \logmlcmd{} is measured. The ellipses enclose 95\% of the best-fit model density, and the color axes all span a range of 2.0 mag to enable visual comparison of the dynamic range of the various SDSS colors observed in M31. Table~\ref{tab:params_appendix} presents the slope and intercept of each best-fit CMLR (i.e., the line along the direction of maximum variance in the 2D Gaussian model) in M31, as well as the scatter about the best-fit CMLR and the location of the Gaussian model peak.

\begin{table*}
\caption{Parameters of Best-Fit 2D Gaussian Models and CMLRs for All SDSS Filters}
\label{tab:params_appendix}
\begin{center}
\begin{tabular}{lcccccc}
Color & $M_\star^\mathrm{CMD}/L_\mathrm{obs}$ Filter & Slope & Intercept & Scatter about CMLR & Color Peak & $\log{(M_\star^\mathrm{CMD}/L_\mathrm{obs})}$ Peak
\\ \hline 
  & $u$ & 0.92 & -0.75 & 0.222 & 1.71 & 0.82 \\
$u-g$ & $g$ & 0.51 & -0.22 & 0.079 & 1.71 & 0.65 \\
  & $r$ & 0.30 & -0.02 & 0.175 & 1.71 & 0.49 \\
  & $i$ & 0.21 & 0.01 & 0.057 & 1.71 & 0.36 \\\hline
  & $u$ & 0.66 & -0.88 & 0.268 & 2.58 & 0.83 \\
$u-r$ & $g$ & 0.38 & -0.32 & 0.069 & 2.58 & 0.65 \\
  & $r$ & 0.24 & -0.13 & 0.231 & 2.58 & 0.49 \\
  & $i$ & 0.17 & -0.06 & 0.053 & 2.58 & 0.36 \\\hline
  & $u$ & 0.57 & -0.89 & 0.045 & 3.01 & 0.83 \\
$u-i$ & $g$ & 0.35 & -0.41 & 0.061 & 3.02 & 0.65 \\
  & $r$ & 0.23 & -0.19 & 0.057 & 3.01 & 0.49 \\
  & $i$ & 0.16 & -0.11 & 0.052 & 3.01 & 0.36 \\\hline
  & $u$ & 2.48 & -1.29 & 0.044 & 0.85 & 0.82 \\
$g-r$ & $g$ & 1.35 & -0.51 & 0.130 & 0.84 & 0.63 \\
  & $r$ & 0.97 & -0.34 & 0.044 & 0.84 & 0.48 \\
  & $i$ & 0.70 & -0.24 & 0.048 & 0.85 & 0.36 \\\hline
  & $u$ & 1.41 & -1.01 & 0.193 & 1.29 & 0.81 \\
$g-i$ & $g$ & 0.83 & -0.42 & 0.155 & 1.28 & 0.63 \\
  & $r$ & 0.56 & -0.24 & 0.140 & 1.27 & 0.48 \\
  & $i$ & 0.39 & -0.15 & 0.130 & 1.28 & 0.36 \\\hline
  & $u$ & 4.46 & -1.17 & 0.033 & 0.45 & 0.82 \\
$r-i$ & $g$ & 2.84 & -0.61 & 0.120 & 0.44 & 0.64 \\
  & $r$ & 2.16 & -0.47 & 0.096 & 0.43 & 0.47 \\
  & $i$ & 2.19 & -0.61 & 0.045 & 0.43 & 0.34 \\
\end{tabular}
\end{center}
\end{table*}

\end{document}